\documentclass[a4paper]{article}
\usepackage{geometry}
\usepackage[table]{xcolor}
\usepackage[utf8]{inputenc}
\usepackage{amsmath}
\DeclareMathOperator*{\argmax}{arg\,max}

\usepackage[capitalise]{cleveref}
\usepackage{graphicx}
\usepackage{subcaption}
\usepackage{booktabs}
\usepackage{natbib}
\usepackage{multirow}

\begin{document}

\title{Bounded rationality for relaxing best response and mutual consistency: The Quantal Hierarchy model of decision-making}

\author{
Benjamin Patrick Evans\footnote{benjamin.evans@sydney.edu.au} \and
Mikhail Prokopenko
}

\date{\footnotesize Centre for Complex Systems, The University of Sydney, Sydney, NSW {2006}, Australia}

\maketitle

\begin{abstract}
While game theory has been transformative for decision-making, the assumptions made can be overly restrictive in certain instances. In this work, we investigate some of the underlying assumptions of rationality, such as mutual consistency and best response, and consider ways to relax these assumptions using concepts from level-$k$ reasoning and quantal response equilibrium (QRE) respectively. Specifically, we propose an information-theoretic two-parameter model called the Quantal Hierarchy model, which can relax both mutual consistency and best response while still approximating level-$k$, QRE, or typical Nash equilibrium behavior in the limiting cases. The model is based on a recursive form of the variational free energy principle, representing higher-order reasoning as (pseudo) sequential decision-making in extensive-form game tree. {This representation enables us to treat simultaneous games in a similar manner to sequential games, where reasoning resources deplete throughout the game-tree.} Bounds in player processing abilities are captured as information costs, where future branches of reasoning are discounted, implying a hierarchy of players where lower-level players have fewer processing resources. We demonstrate the effectiveness of the Quantal Hierarchy model in several canonical economic games, {both simultaneous and sequential}, using out-of-sample modelling.
\end{abstract}

\section{Introduction}

A crucial assumption made in Game Theory is that all players behave perfectly rationally. Nash Equilibrium \citep{nash1951non} is a key concept that arises based on all players being rational and assuming other players will also be rational, requiring correct and consistent beliefs amongst players. However, despite being the traditional economic assumption, perfect rationality is incompatible with many human processing tasks, with models of limited rationality better matching human behaviour than fully rational models \citep{gachter2004behavioral, camerer2010behavioural, gabaix2006costly}. Further, if an opponent is irrational, then it would be rational for the subject to play ``irrationally" \citep{raiffa1957games}. That is, ``[Nash equilibrium] is consistent with perfect foresight and perfect rationality only if both players play along with it. But there are no rational grounds for supposing they will" \citep{koppl2002all}. These assumptions of perfect foresight and rationality often lead to contradictions and paradoxes \citep{cournot1838recherches, morgenstern1928wirtschaftsprognose, hoppe1997certainty, glasner2020hayek}. 

Alternate formulations have been proposed that relax some of these assumptions and model boundedly rational players, better approximating actual human behaviour and avoiding some of these paradoxes \citep{simon1976substantive}. For example, relaxing mutual consistency allows players to form different beliefs of other players \citep{stahl1995players, camerer2004cognitive} avoiding the infinite self-referential higher-order reasoning which emerges as the result of interaction between rational players \citep{knudsen1993rationality} (I have a model of my opponent who has a model of me $\dots$ \textit{ad infinitum} \citep{morgenstern1935vollkommene}) and non-computability
of best response functions \citep{rabin1957effective, koppl2002all}. The ability to break at various points in the higher-order reasoning chain can be considered as ``partial self-reference" \citep{lofgren1990partiality, mackie1971can2}. Importantly, rather than implicating negation \citep{prokopenko2019self}, this type of self-reference represents higher-order reasoning as a logically non-contradictory chain of recursion (reasoning about reasoning...). Hence, bounded rationality arises from the ability to break at various points in the chain, discarding further branches. ``Breaking" the chain on an otherwise potentially infinite regress of reasoning about reasoning can be seen as the players limitations in information processing, determining when to end the recursion.

Another example of bounded rationality based on information processing constraints is the relaxation of the best response assumption of players, which allows for erroneous play, with deviations from the best response governed by a resource parameter \citep{haile2008empirical, goeree2005regular}.

In this work, we adopt an information-theoretic perspective on reasoning (decision-making). By enforcing potential constraints on information processing,  we are able to relax both mutual consistency and best response, and hence, players do not necessarily act perfectly rational. The proposed approach provides an information-theoretic foundation for level-$k$ reasoning and a generalised extension where players can make errors at each of the $k$ levels.

Specifically, in this paper, we focus on three main aspects:

\begin{itemize}
    \item Players reasoning abilities decrease throughout recursion in a wide variety of games, motivating an increasing error rate at deeper levels of recursion. That is, it becomes more and more difficult to reason about reasoning about reasoning $\dots$ (necessitating a relaxation in best response decisions).    
    \item Finite higher-order reasoning can be captured by discounting future chains of recursion and ultimately discarding branches once resources run out. This representation introduces an implicit hierarchy of players, where a player assumes they have a higher level of processing abilities than other players, motivating relaxation of mutual consistency.
    \item Existing game-theoretic models can be explained and recovered in limiting cases of the proposed approach. This fills an important gap between methods relaxing best response and methods relaxing mutual consistency.
\end{itemize}

The proposed approach features only two parameters, $\beta$ and $\gamma$, where $\beta$ quantifies relaxation of players' best response, and $\gamma$ governs relaxation of mutual consistency between players. In the limit, $\beta \to \infty$ best response can be recovered, and in the limit $\gamma=1$ mutual consistency can be recovered. Equilibrium behaviour is recovered in the limit of both $\beta \to \infty, \gamma=1$. For other values of $\beta$ and $\gamma$, interesting out-of-equilibrium behaviour can be modelled which concurs with experimental data, and furthermore, in repeated games that converge to Nash equilibrium, player learning can be captured in the model with increases in $\beta$ and $\gamma$. Importantly, we also show how fitted values of $\beta$ and $\gamma$ also generalise well to out-of-sample data.

The remainder of the paper is organised as follows. \cref{secBackground} analyses bounded rationality in the context of decision-making and game theory, \cref{secBoundedDecisions} introduces information-theoretic bounded rational decision-making, \cref{secMethod} extends the idea to capture higher-order reasoning in game theory. We then use canonical examples highlighting the use of information-constrained players in addressing bounded rational behaviour in games in \cref{secExamples}. We draw conclusions and outline future work in \cref{secConclusions}.

\section{Background and Motivation}\label{secBackground}

While a widespread assumption in economics, perfect rationality is incompatible with the observed behaviour in many experimental settings, motivating the use of bounded rationality \citep{camerer2011behavioral}. Bounded rationality offers an alternative perspective, by acknowledging that players may not have a perfect model of each other or may not play perfectly rationally. In this section, we explore some common approaches to modelling bounded rational decision-making.

\subsection{Mutual Consistency}

Equilibrium models assume mutual consistency of beliefs and choices \citep{camerer2003models, camerer2003behavioral}, however, this is often violated in experimental settings \citep{polonio2019testing}  where ``differences in belief arise from the different iterations of strategic thinking that players perform" \citep{chong2005ch}.

Level-$k$ reasoning \citep{stahl1995players} is one attempt at incorporating bounded rationality by relaxing mutual consistency,  where players are bound to some level $k$ of reasoning. A player assumes that other players are reasoning at a lower level than themselves, for example, due to over-confidence. This relaxes the mutual consistency assumption, as it implicitly assumes other players are not as advanced as themselves. Players at level $0$ are not assumed to perform any information processing, and simply choose uniformly over actions (i.e., a Laplacian assumption due to the principle of insufficient reason), although alternate level-$0$ configurations can be considered \citep{wright2019level}. Level-$1$ players then exploit these level-$0$ players and act based on this. Likewise, level-$2$ players act based on other players being level-$1$, and so on and so forth for level-$k$ players acting as if the other players are at level-$(k-1)$. Various extensions have also been proposed \citep{levin2019bridging}.

A similar approach to level-$k$ is that of Cognitive Hierarchies (CH) \citep{camerer2004cognitive}, where again it is assumed other players have lower reasoning abilities. However, rather than assuming that the other players are at $k-1$, players can be distributed across the $k$ levels of cognition according to Poisson distribution with mean and variance $\tau$. The validation of the Poisson distribution has been provided in \cite{chong2005ch}, where an unconstrained general distribution offered only marginal improvement. Again, various extensions have been proposed \citep{CHONG2016257, koriyama2021inclusive} and there are many examples of successful applications of such depth-limited reasoning in literature, for example, \cite{goldfarb2011thinks}.

Endogenous depth of reasoning (EDR) is a similar approach to level-$k$ and CH, but it separates the player's cognitive bounds from their beliefs of their opponent's reasoning \citep{alaoui2016endogenous}. EDR captures player reasoning as if they are following a cost-benefit analysis \citep{alaoui2022cost}, with cognitive abilities (costs) and payoffs (benefits).

One fundamental similarity across these methods is that they all maintain best-response. That is, they best respond based on the lower-level play assumptions. The following section introduces methods that instead maintain mutual consistency, but relax the best-response assumption.

\subsection{Best response}
Alternate approaches assume that a player may make errors when deciding which strategy to play, rather than playing perfectly rationally. That is, they relax the best response assumption. Quantal Response Equilibrium \citep{mckelvey1995quantal, mckelvey1998quantal} (QRE) is a well-known example, where rather than choosing the best response with certainty, players choose noisily based on the payoff of the strategies and a resource parameter controlling this sensitivity. Another method of capturing this erroneous play is Noisy Introspection \citep{goeree2004model}. Utility proportional beliefs \citep{bach2014utility} is another method that relaxes the best response assumption, where the authors note that ``possibly, the requirement that only rational choices are considered and zero probability is assigned to any irrational choice is too strong and does not reflect how real world [players] reason", giving merit to the relaxation of best-response. By allowing for errors in decision-making, these methods offer a more realistic perspective on how individuals make choices. 

\subsection{Infinite-Regress}

When considering reasoning about reasoning, infinite regress can emerge \cite{knudsen1993rationality}. The problem of infinite regress can be formulated as a sequence $$A, f(A), f(f(A)), f(f(f(A))) $$
where the player is first confronted with an initial choice of an action $a$ from the set of actions $A$, or a computation $f(A)$. If the action is chosen, the decision process is complete. If, instead, the computation is chosen, the outcome of $f(A)$ must be calculated, and the player is then faced with another choice between the obtained result or yet another computation. This process is repeated until the player chooses the result, as opposed to performing an additional computation. \cite{lipman1991decide} investigates whether such sequence converges to a fixed point. A similar approach by \cite{mertens1985formulation} formally approximates Harsanyi's infinite hierarchies of beliefs \citep{harsanyi1967games, harsanyi1968games} with finite states.

\subsection{Contributions of our work}

In contrast to existing works, we propose an information-theoretic approach to higher-order reasoning, where each level or hierarchy corresponds to additional information processing for the player. While \cite{alaoui2016endogenous} capture the trade-off between additional reasoning and payoff, by measuring the first intersection of the players' payoff improvement (from $k$ to $k+1$), and the cost of performing this additional reasoning $c(k+1)$. This cost function $c$ has to be determined (for all $c(k)$), for example, with maximum likelihood estimation to estimate the average cost of performing this extra level of reasoning \citep{alaoui2022cost}. In contrast, we propose capturing this trade-off with information processing costs by using the Kullback-Leibler divergence to constrain the overall change in action probabilities at each stage of reasoning.

Our approach constrains the overall amount of information processing available to the players, leading to potential errors at each stage of reasoning, which is not present in existing level-$k$ type approaches. By doing so, we establish a foundation for ``breaking" the chain of higher-order reasoning based on the depletion of players' information processing resources.

This results in a principled information-theoretic explanation for decision-making in games involving higher-order reasoning. Best response is relaxed with $\beta < \infty$ (linking to Quantal Response Equilibrium) and mutual consistency is relaxed  with $\gamma < 1$ (linking to level-$k$ type models). Best response and mutual consistency are recovered with $\beta \to \infty$ and $\gamma=1$. This contributes to the existing literature on game theory and decision-making by adopting an information-theoretic perspective on bounded rationality, quantified by information processing abilities. A key benefit of the proposed approach is while level-$k$ models relax mutual consistency, but retain best response, and QRE models relax best response but retain mutual consistency \citep{chong2005ch}, the proposed approach is able to relax either assumption through the introduction of two tunable parameters. We apply this model to various games, demonstrating the usefulness of the proposed approach for capturing human behaviour when compared to these existing approaches.

\section{Technical Preliminaries: Information-Theoretic Bounded Rationality}\label{secBoundedDecisions}

Information theory provides a natural way to reason about limitations in player cognition, as it abstracts away specific types of costs \citep{wolpert2006information,harre2021information}. This means that we can assume the existence of cognitive limitations without speculating about the underlying behavioural foundations. As pointed out by \cite{sims2003implications}, an information-theoretic treatment may not be desirable for a psychologist, as this does not give insights to where the costs arise from, however, for an economist, reasoning about optimisation rather than specific psychological details may be preferable, for example, in the Shannon model \citep{caplin2019rational}. Such models have seen considerable success in a variety of areas for information processing, for example, embodied intelligence  \citep{polani2007information}, self-organisation \citep{ay2012information}
and adaptive decision-making of agents \citep{tishby2011information} based on the information bottleneck formalism \citep{tishby2000information}.

In this work, we adopt the information-theoretic representation of bounded rational decision-making proposed by \cite{ortega2013thermodynamics}, which has been further developed in \citep{ortega2011information, braun2014information, ortega2016human, gottwald2019bounded}. This approach has a solid theoretical foundation based on the (negative) free energy principle and has been successfully applied to several tasks \citep{evans2021maximum}. We begin by providing an overview of single-step decisions and then sequential decisions, before discussing extensions for capturing the relationship between processing limitations and higher-order reasoning.

\subsection{Single-step Decisions}

A boundedly-rational decision maker who is choosing an action $a \in A$ with payoff $U[a]$ is assumed to follow the following negative free energy difference when moving from a prior belief $p[a]$, e.g., a default action, to a (posterior) choice $f[a]$, given by:

\begin{equation}
\label{eqOrtega}
- \Delta F[f[a]]  = \sum_{a \in A} f[a] U[a] - \frac{1}{\beta} \sum_{a \in A} f[a] \log \left( \frac{f[a]}{p[a]}\right)
\end{equation} 

The first term represents the expected payoff, while the second term represents a cost of information acquisition that is regularised by the parameter $\beta$. Formally, the second term quantifies information acquisition as the Kullback-Leibler (KL) divergence from the prior belief $p[a]$. Parameter $\beta$, therefore, serves as the resource allowance for a decision-maker.

By taking the first order conditions of \cref{eqOrtega} and solving for the decision function  $f[a]$, we obtain the equilibrium distribution:
\begin{equation}\label{eqSingleDecision}
f[a] = \frac{1}{Z} p[a] e^{\beta U[a]}
\end{equation}
where $Z=\sum_{a' \in A}p[a'] e^{\beta U[a', x]}$ is the partition function. This representation is equivalent to the logit function (softmax) commonly used in QRE models, and relates to control costs derived in economic literature \citep{stahl1990entropy, mattsson2002probabilistic}. 

The parameter $\beta$ serves as a resource allowance for the decision-maker, modulating the cost of information acquisition from the prior belief. Low values of $\beta$ correspond to high costs of information acquisition (and high error play), while as $\beta \to \infty$, information becomes essentially free to acquire, and the perfectly rational \textit{homo economicus} player is recovered.

\subsection{Sequential Decisions}\label{secSeq}

This free energy definition can be extended to sequential decision-making by considering a recursive negative free energy difference, as described in \citep{ortega2013thermodynamics, ortega2014generalized}.  This corresponds to a nested variational problem, and involves introducing new inverse temperature parameters $\beta_k$ for $K$ sequential decisions, which allows for different reasoning at different depths of recursion to choose a sequence of $K$ actions $a_{\leq K}$.

Therefore, for sequential decision-making, \cref{eqOrtega} can be represented as:

\begin{equation}
    - \Delta F[f]  = \sum_{a_{\leq K}} f[a_{\leq K}] \sum_{k=1}^{K} \left( U[a_k \mid a_{<k}] - \frac{1}{\beta_{k}} \log  \frac{f[a_k \mid a_{<k}]}{p[a_k \mid a_{<k}]} \right)
\end{equation}
where $a_{<k}$ abbreviates the history $a_o,..,a_{k-1}$ of decisions.  We can expand the sum:
\begin{equation}\label{eqExpansion}
\begin{split}
=
\sum_{a_1} f[a_1] \biggl( U[a_1] - \frac{1}{\beta_1} f[a_1] \log \frac{f[a_1]}{p[a_1]} + \\ 
\sum_{a_2} f[a_2 \mid a_1] \biggl( U[a_2 \mid a_1] - \frac{1}{\beta_2} f[a_2 \mid a_1] \log \frac{f[a_2 \mid a_1]}{p[a_2 \mid a_1]} + \\
\dots \\ +
\sum_{a_K} f[a_K \mid a_{<K}] \biggl( U[a_K \mid a_{<K}] - \frac{1}{\beta_K} f[a_K \mid a_{<K}] \log \frac{f[a_K \mid a_{<K}]}{p[a_K \mid a_{<K}]}
\biggr) \biggr)
\end{split}
\end{equation} 
which we can see as first choosing an action $a_1$ at $k=1$, while considering that in order to choose this action, we must consider the future stages by analysing the result at $k=2$ given the choice $a_1$, and so forth. To compute this, we can solve the innermost sum first:
\begin{equation}\label{eqOriginalInnerSum}
\begin{split}
f[a_K \mid a_{<K}] &= \frac{1}{Z_K} p[a_K \mid a_{<K}] e ^ {\beta_K U[a_K \mid a_{<K}]}
\end{split}
\end{equation}
which recovers \cref{eqSingleDecision} with the introduction of conditioning on decision histories. This represents the base-case for recursion. For steps where $k < K$, we get the following equilibrium solution for sequential decisions:

\begin{equation}\label{eqMultiDecision}
\begin{split}
f[a_k \mid a_{<k}] &= \frac{1}{Z_k} p[a_k \mid a_{<k}] \exp\left(\beta_k (U[a_k \mid a_{<k}] + \frac{1}{\beta_{k+1}}  \log Z_{k+1})\right) \\
&= \frac{1}{Z_k} \underbrace{p[a_k \mid a_{<k}]}_{\text{Prior Belief}} \times \underbrace{Z_{k+1} ^ {\beta_k / \beta_{k+1}}}_{\text{Future Contribution}} \times \underbrace{e ^{\beta_k U[a_k \mid a_{<k}]}}_{\text{Current Payoff}}\\
\end{split}
\end{equation} 
where decisions are now dependent on the history of decisions as well as on a recursive component based on the future contribution of each decision. Here
\begin{equation}\label{eqSequentialZ}
    Z_{k} = \sum_{a_k}  p[a_k \mid a_{<k}] Z_{k+1}^{\beta_k / \beta_{k+1}} e ^{\beta_k U[a_k \mid a_{<k}]}
\end{equation}
where $Z_{K}=1$, i.e., the base-case for recursion at the final level.

\subsection{Extension}
With this formal and generic treatment of information-processing costs for sequential decision-making, it is desirable to use this to capture bounded rational reasoning of an agent making sequential \textit{or} simultaneous decisions in games. By representing reasoning as an extensive-form game tree, these two can be captured in a similar manner. Simultaneous decisions can be treated as if they are pseudo-sequential decisions, considering possible repercussions for various choices.

To model higher-order reasoning, we extend the information-theoretic formulation for sequential decisions discussed in \cref{secSeq}. This involves representing reasoning as a (pseudo) sequence of decisions, where each decision corresponds to a "level" of reasoning. At each level, players may play incorrectly, producing a level-$k$ play that is modulated by $\beta_k$. A high $\beta_k$ corresponds to an exact level-$k$ thinker, while a low $\beta_k$ corresponds to an error-prone level-$k$ thinker. We can formalise this chain of reasoning as an extensive-form game tree, where at the root node, a player is faced with a decision to choose from a set of available options $A$ or perform additional processing $f(A)$ to acquire new information on the beliefs and repercussions of each choice. If the player chooses not to process additional information, each branch is terminated early, and the player makes a decision based solely on the immediately available information. However, if the player chooses to process additional information, each action branch is (potentially) extended to analyze the possible repercussions of taking an action, and a higher level thinker can examine these repercussions. This process continues until the player runs out of processing resources or, in a finite problem, converges to a solution. For example, in the $p$-beauty contest analyzed in later sections, convergence occurs once the guess hits 0.

{This extensive-form game tree representation means reasoning about simultaneous decisions can be treated in the same manner as decisions in sequential games. However, the key issue here is the requirement of \cref{eqMultiDecision} to have $K$ information processing parameters for each step or level of reasoning. In order to analyse the purpose of these parameters, we} consider a simple case, setting $\beta_k=\beta$ for all $k$, giving discrete control \citep{braun2011path}:

\begin{equation}\label{eqMultiDecisionSimple}
f[a_k  \mid  x, a_{<k}] = \frac{1}{Z_k} p[a_k  \mid a_{<k}] Z_{k+1} e^{\beta U[a_k, x \mid a_{<k}]}
\end{equation}
which clarifies that the boundedness applies to the computation of payoff $U$, but the depth of reasoning (or recursion-depth) is dependent on the length of the sequence (or level) $K$, not on $\beta$. To represent higher-order reasoning more succinctly, it would be desirable to implicitly base the sequence length $K$ on the resource parameter $\beta$ rather than keeping them separate. This would allow us to treat recursive reasoning in information-theoretic terms. For instance, when $\beta \to \infty$, $K \to \infty$, which captures (potential) infinite-regress, while $\beta=0$ would imply $K=0$, leaving the player with no information processing abilities.

{In order to achieve this, $\beta$ must be reduced at each level $k$. This reduction in $\beta$ at each level is related to the relaxation of mutual consistency, reflecting how a player perceives other (lower-level) players' reasoning about the problem. In the next section, we outline our proposed approach for modelling this. This extension allows us to reason directly about resource constraints instead of sequential level-$k$ thinking.}

\section{Proposed Approach}\label{secMethod}

{The proposed approach implicitly enforces the assumption that, in sequential games, it becomes more difficult to reason to later stages in the game (e.g. in chess, it is difficult to reason 5 steps-ahead), and likewise in simultaneous games, it is difficult to perform the level of higher-order reasoning required to arrive at the equilibrium solution. The further a player mentally tries to reason, the more likely an error is to occur, as the processing resources deplete. This assumption is captured under the proposed model in a generic information-theoretic sense.}

\begin{figure}[ht]
    \centering
    \includegraphics[width=.6\textwidth]{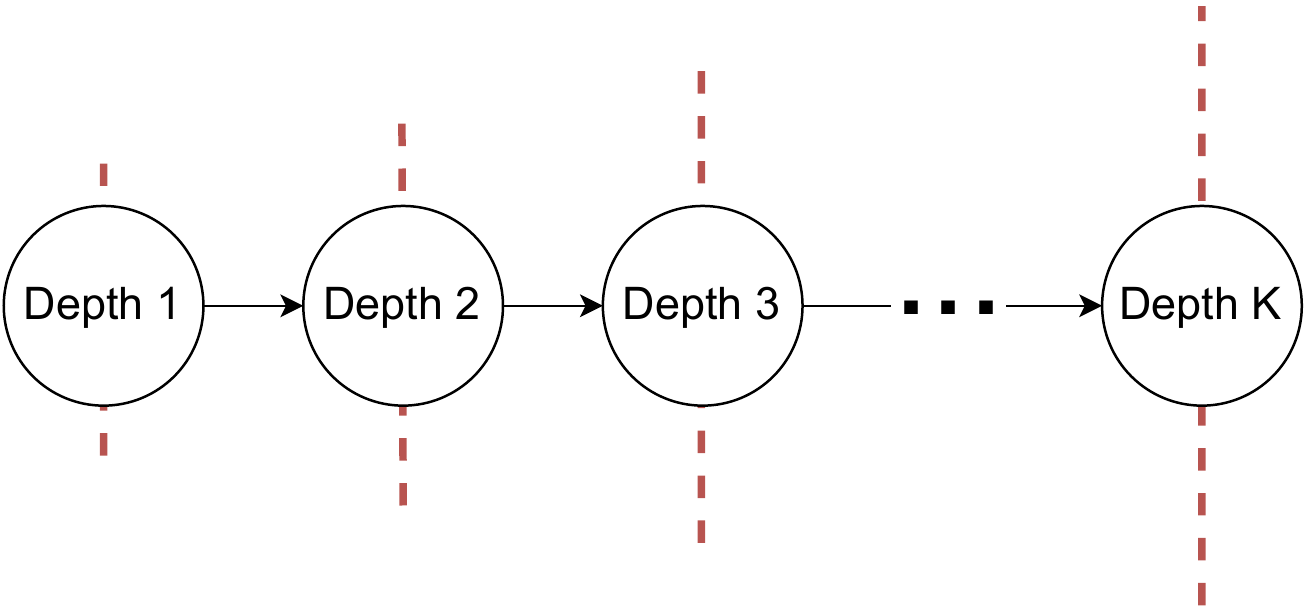}
    \caption{The effect of discounting resources over time. In the beginning, the player has $\beta$ resources and may make some error (red bars). Parameter $\gamma$ controls how this error grows over time and, implicitly, how the player believes lower-level players will respond. A low $\gamma$ means the errors increase drastically at each level, assuming opponents with much lower reasoning abilities. As $\beta\gamma^{k} \to 0$, the noise increases, and the recursion eventually stops once the utilities become indistinguishable (i.e., the player can not reason any deeper).}
    \label{figErrorBetaGamma}
\end{figure}

The key concept is that it becomes more difficult to reason about reasoning, that is the further one tries to explore through the extensive-form tree. This overall process is visualised \cref{figErrorBetaGamma}, where the players reasoning error increases throughout the steps of reasoning. This representation can be thought of as a hierarchy of players noisily responding to lower level players, {or as a sequential decision with increasing noise at each step}. Once a player's resources deplete, the later depths simply echo the prior beliefs as the noise obstructs the payoffs.

\subsection{Quantal Hierarchy Model}
{To model higher-order reasoning with processing costs, we propose a more flexible and succinct approach than simply setting a maximum depth $K$ and corresponding parameters $\beta_k$ in a pseudo-sequential decision-making task.} Instead, we introduce an overall information processing parameter $\beta$, which captures the available information processing resources for a player, and a discount parameter $\gamma \in [0,1]$, which controls the reduction in player rationality throughout the reasoning chain. This approach allows for heterogeneous bounds on player reasoning, relaxing the assumptions of best response and mutual consistency. We refer to this approach as the Quantal Hierarchy (QH) model, as it shares formal similarities with Quantal Response Equilibrium and Cognitive Hierarchy models, as discussed in previous sections.

\subsubsection{Formulation}
{We represent reasoning as information-constrained sequential decision-making.} The proposed formulation features only two parameters, $\beta$ and $\gamma \in [0,1]$ (as opposed to the vector $\beta_k$ and number of levels $K$). Reasoning resources are then set as $$\beta_k = \beta\gamma^{k}$$i.e., $\beta_k$ is $\beta$ discounted based on $\gamma$ and the current depth of reasoning. This can be represented by the following recursive free energy difference:
\begin{equation}
- \Delta F[f]  = \sum_{a_{\leq K}} f[a_{\leq K}] \sum_{k=0}^{\infty} \left( U[a_k \mid a_{<k}] - \frac{1}{\beta \gamma ^{k}} \log  \frac{f[a_k \mid a_{<k}]}{p[a_k \mid a_{<k}]} \right)
\end{equation}
where we have represented the sequence as an infinite-sum. The sum converges due to the discount parameter as the later (inner-most sums) simply echo the prior beliefs once the player's computational resources are exhausted. This formalisation draws parallels with the reinforcement learning (RL) methods, where such representations are common for reasoning about future states for the player. In RL, $\gamma$ is used to discount future timesteps. Here, $\gamma$ is used to discount future chains of reasoning about reasoning (i.e., the depth of recursion is governed by $\beta$ discounted by $\gamma$), and to represent the limited resources that we believe other players have. We have represented this as in infinite sum with $k \to \infty$, however, in various problems (such as sequential games) $K$ can be assumed to be finite.  The solution for the decision function $f[a_k \mid a_{<k}]$ with the discounted $\beta$ becomes:

\begin{equation}\label{eqDiscountDecision}
  f[a_k \mid a_{<k}] =
\begin{cases}
    \frac{1}{Z_k} \underbrace{p[a_k \mid a_{<k}]}_{\text{Prior Belief}},              & \text{if } { \beta \gamma ^{k} \approx 0}
    \\
    \\
    \frac{1}{Z_k} \underbrace{p[a_k \mid a_{<k}]}_{\text{Prior Belief}} \times \underbrace{e ^{\beta U[a_k \mid a_{<k}]}}_{\text{Current Payoff}},              & \text{if } \gamma = 0
    \\
    \\
    \frac{1}{Z_k} \underbrace{p[a_k \mid a_{<k}]}_{\text{Prior Belief}} \times \underbrace{Z_{k+1} ^ {1 / \gamma}}_{\text{Future Contribution}} \times \underbrace{e ^{\beta\gamma^{k} U[a_k \mid a_{<k}]}}_{\text{Current Payoff}}, & \text{otherwise }
    \\
    \\
\end{cases}
\end{equation}
which aims to capture decisions based on the beliefs of other players reasoning at later stages. With the assumption of a discount rate $\gamma$, we can see the recursion is now depth-bound based on the (discounted) resource parameter. {Once $\beta \gamma ^{k} \to 0$} \footnote{e.g. when  $\beta \gamma ^{k} < \epsilon$ where $\epsilon$ is some small enough term where the payoffs become indistinguishable, here, $\epsilon=10^{-8}$}, the recursion will stop since the result will simply echo the prior belief as no focus will be placed on the payoff, becoming the base case for recursion (the ``naive" player). This means that in the new implicit limit $K$ \footnote{in contrast to level-$k$, here $K$ indicates the level with the lowest resources} (denoted by the case where $\beta_k \approx 0$), reasoning about the future provides no new information, which recovers the original form of \cref{eqOriginalInnerSum} (when $\gamma=1$) with the introduction of conditioning on histories.

With the proposed representation, what was previously thought of in the context of sequential decisions, can be extended and modified to think about hierarchies of beliefs. The informationally constrained players ``break" the chain of reasoning due to depleting their cognitive or computational capabilities as bounded by $\beta$ and $\gamma$. Formally, this is represented as a (potentially) infinite-sum that converges based on $\gamma$.  
By discounting future computation in a chain of recursion, we can approximate higher-order reasoning, where players become increasingly limited as the reasoning chain progresses, making it more challenging to reason about reasoning.

\subsubsection{Parameters}

\textbf{Resource Parameter} The resource parameter $\beta$ quantifies the amount of processing a player can perform. Perfect utility maximisation behaviour is recovered with  $\beta \to \infty$. With  $\beta<\infty$, players are assumed to be limited in computational resources and must now balance the trade-off between their computation cost and payoff. With $\beta \to 0$, players have no processing resources, and choose based on their prior beliefs (default actions). Anti-rational (or adversarial) play can be modelled with $\beta \to -\infty$.

\textbf{Discount Parameter}
The discount parameter $\gamma$ quantifies other players mental processing abilities in terms of level-$k$ thinking. A high $\gamma$ assumes other players play at a relatively similar cognitive level, whereas a low $\gamma$ assumes other players have less playing ability. With $\gamma < 1$, \cref{eqDiscountDecision} is guaranteed to converge to a finite sequence of decisions, where the ability to process information decreases the further we get through the sequence. This captures the \textit{belief} about play at later stages of reasoning, where other players are assumed to be less rational (and thus, more noisy) as governed by $\gamma$. The case $\gamma < 1$ implicitly relaxes mutual consistency as lower-level thinkers are then governed by a lower resource parameter, and allows for players to believe that players at later nodes behave more noisily. In the case of otherwise infinite regress (where backward induction can not be used), the limited foresight approach proposed converges to a finite approximation of the sequence by relaxing the assumption of mutual consistency. In tractable problems, we recover an approximation of backward induction where the player performing such induction may make errors at each step due to limited computational processing abilities. With $\gamma=1$, we recover mutual consistency, as we assume other players are just as rational as ourselves, and in the special case with uniform prior beliefs and $\gamma=1$, we collapse to a logit form of agent QRE \citep{mckelvey1998quantal, turocy2010computing}. The special case for $\gamma=0$ exists in \cref{eqDiscountDecision} as in this case, no future processing will be performed. 

In the limit, with $\beta \to \infty, \gamma=1$, perfect backward induction is recovered (see \cref{secBackwardsInduction}). Crucially, the proposed QH model allows for a general representation, relaxing the perfect rationality assumption (with $\beta < \infty, \gamma < 1$), which can model out-of-equilibrium behaviour compatible with observed experimental data. We explore the role of these parameter values in more detail in the following section.

\begin{figure}[!ht]
    \centering
    \includegraphics[width=.5\textwidth]{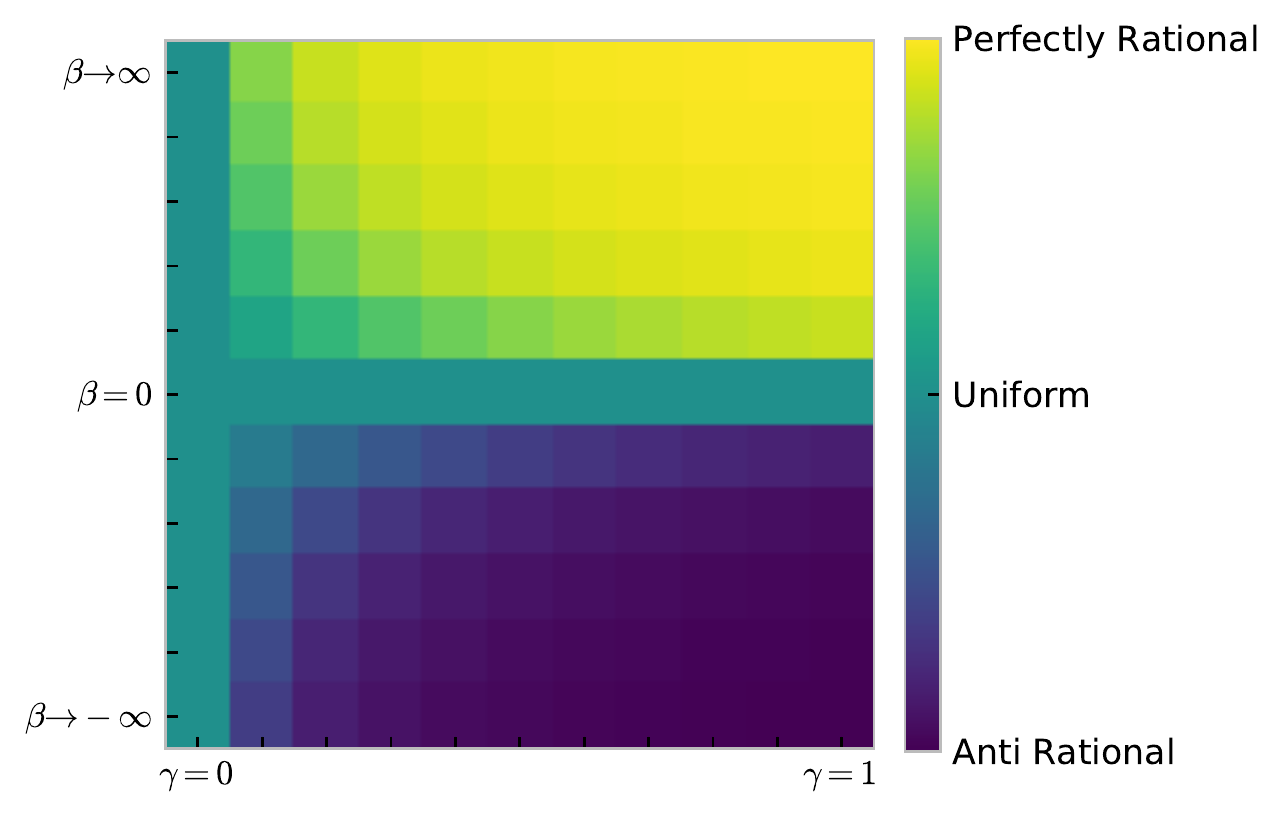}
    \caption{A heatmap visualising the resulting players expected payoff based on the values of $\beta$ and $\gamma$. The yellow colour represents the maximal expected payoff, and the purple colour represents the minimum expected payoff. When either parameter is $0$, the result is a random choice amongst the actions.
    }
    \label{figParamHeatmap}
\end{figure}

\textbf{Parameter Interactions} In \cref{figParamHeatmap} we visualise how $\beta$ and $\gamma$ interact in a general setting. For $\beta \to \infty$ and $\gamma=1$, we approach payoff maximisation behaviour, i.e., the perfectly rational (Nash Equilibrium) player is recovered. For $\beta \to -\infty$ and $\gamma=1$, payoff minimisation behaviour (an adversarial player) is recovered. In between, we can see how $\gamma$ adjusts $\beta$. It is these values in between random play ($\beta=0$) and perfect payoff maximisation behaviour which are particularly interesting, as they give rise to out-of-equilibrium behaviour not predicted by traditional methods.

%
%

\subsection{Explanation}\label{secSimpleExample}

We work through a generic example of the QH model on an extensive-form game tree. A player is given decision-making resources, governed by $\beta$, to make a decision $f$. At each stage of reasoning $k$, the player's resources are discounted by $\gamma$. Ultimately, the player's resources become depleted (at $K$, i.e., once $\beta\gamma^k \approx 0$), and the game tree is considered terminated, and the naive player chooses based on their prior belief (which we assume to be uniform). This decision is then propagated backwards, and becomes the prior belief at the higher stage of reasoning $p[a_{K-1}]$, where the player has processing resources $\beta\gamma^{K-1}$, and hence, noisily responds to the lower-level play based on these resources. This process is continued, with noisy responses from the lower-level thinker captured by the resource constraint. Finally, once all results have been recursively propagated, the higher-level players' decision is made (which may still be noisy, as captured by $\beta$). That is, their decision is made recursively, starting from the most basic level of player reasoning (the naive player) and reasoning upwards.

\subsubsection{Basic Example}\label{secExample}
To help illustrate the proposed approach, we will use a simplified version of the Ultimatum game (see \cref{secAppendixBargaining}) as a specific example. In this game, a player must decide what percentage of a pie to take. We assume that there are uniform priors among the available options at each stage.

At the first stage, Player 1 must request the percentage of the pie they want to take, denoted by $a_1 \in [0, 100]$, with 100 giving the highest payoff (i.e., they receive the entire pie). However, at the second stage, Player 1 encounters a fairness calculator (Player 2). Player 2 decides whether or not to approve Player 1's request, denoted by $A_2=\{\text{accept}, \text{reject}\}$. The decisions are based on the following utilities:

\begin{equation}
\begin{split}
U_1[a_1] &= a_1 \times f_2[\text{accept} \mid a_1] \\
U_2[\text{accept}_{2} \mid a_1] &= 100 - a_1 \\
U_2[\text{reject}_{2} \mid  a_1] &= 50 \\
\end{split}
\end{equation}
where $U_n$ corresponds to Player $n$'s payoff, and $f_n$ the probability with which Player $n$ chooses the action. A player who has no look-ahead, i.e., one who assigns zero weight to future decisions (or assumes that their opponent has zero processing abilities), can be represented with $\beta \to \infty$ and $\gamma=0$. Such a player simply looks at the first stage and sees that it is in their best interest to request 100\% of the pie. However, this player fails to take into account the repercussions of their chosen action, as they did not consider the future decisions. They did not compute $f_2[a_2 \mid a_1]$, and thus assumed that $f_2[a_2 \mid a_1]$ is uniform and that their opponent would be indifferent to accepting or rejecting their request regardless of the value of $a_1$.

A perfectly rational player with unlimited computational resources, i.e., $\beta=\infty$ and $\gamma=1$, would request $49$ (assuming integer requests). They assign weight to the future of their actions, and can see that for any $a > 50$, the fairness calculator will deny their request, and they will be left with nothing (at $a=50$, the calculator will be indifferent to accepting or rejecting their request). This corresponds to the subgame perfect equilibrium, where the player performed backward induction. That is, the player examined the future until they reached the end of the game and then reasoned backwards to request the optimal choice.

A player with limited computational resources, i.e., $\beta < \infty$, requests the best action they can subject to their resource constraint. For example, they may only request $a=40$, as they are unable to complete the search for $a=49$. A player with no information processing abilities, i.e., $\beta=0$, cannot search for any optimal choices and therefore chooses based on the prior distribution, which we assumed to be uniform. Therefore, the player is equally likely to choose any $a \in A$.

An interesting question that arises is what if we assume that the "fairness" calculator may make errors, and that it is not necessarily defined by a step function that rejects all requests above $50$ and accepts all below. For example, there may be a range where $100$ will get rejected, but perhaps $75$ would not. This can be captured with $\beta\gamma < \infty$, where the calculator is assumed to make errors for low values, and for $\beta\gamma \to \infty$, it is assumed to be perfectly rational. If the fairness calculator is broken and is indifferent to accepting or rejecting values, this can be represented with $\gamma=0$, which gives $0$ processing ability to the calculator. This means $f_2[\text{accept}_2 \mid a_1] = f_2[\text{reject}_2 \mid a_1] = 0.5$, and therefore, a rational Player 1 should request $a_1=100$.

{This example shows the usefulness of the proposed approach, and how modifying $\beta$ and $\gamma$ can capture a variety of heterogeneous behaviours between the two players.}

\section{Results}\label{secExamples}

In this section, we perform out-of-sample comparisons across various canonical economic games, {including both sequential and simultaneous games}. We compare the proposed Quantal Hierarchy model against well-known approaches to capturing bounded rational reasoning, including QRE, level-$K$ and Cognitive Hierarchy, as well as the Nash equilibrium predicted solutions. {To assess the performance of each method, we fit the corresponding parameter values to experimental data and then evaluate the performance on hold-out data. For the Quantal Hierarchy method, these parameter values are $\beta$ and $\gamma$. For QRE, the parameter value is $\lambda$, which serves a similar purpose as $\beta$ in our approach, i.e., relaxing best response. For level-$k$, the parameter value is the steps of reasoning $k$. For Cognitive Hierarchy, the parameter value is $\tau$, corresponding to the Poisson distribution of level-$k$ thinkers. Further information on model fitting is given in \cref{secModelFit}.}

We show how the proposed approach convincingly captures human behaviour and generalises beyond the training examples, outperforming existing approaches on a wide range of games.

\subsection{Performance on Canonical Games}
For this work, we use various experimental data from canonical economic {sequential and simultaneous} games. Specifically, {for simultaneous games}, we analyse market entrance and beauty contest games, and {for sequential games}, centipede, and bargaining games. 

For market entrance games, we use the data of \cite{camerer2011behavioral}, originally presented in  \citep{sundali1995coordination}. For the beauty contest game, we use $p$-beauty contest results from \citep{bosch2002one}. For the Centipede games, we use the four and six-level data from \citep{mckelvey1992experimental}. For the sequential bargaining games, we use the Ultimatum game and two-stage game from \cite{binmore2002backward} (Game 1 and 3 in their paper). Further discussion on game specifics, utilities, and experimental analysis is given in \cref{secExamplesAppendix}.

\begin{figure}
    \centering
    \includegraphics[width=.75\textwidth]{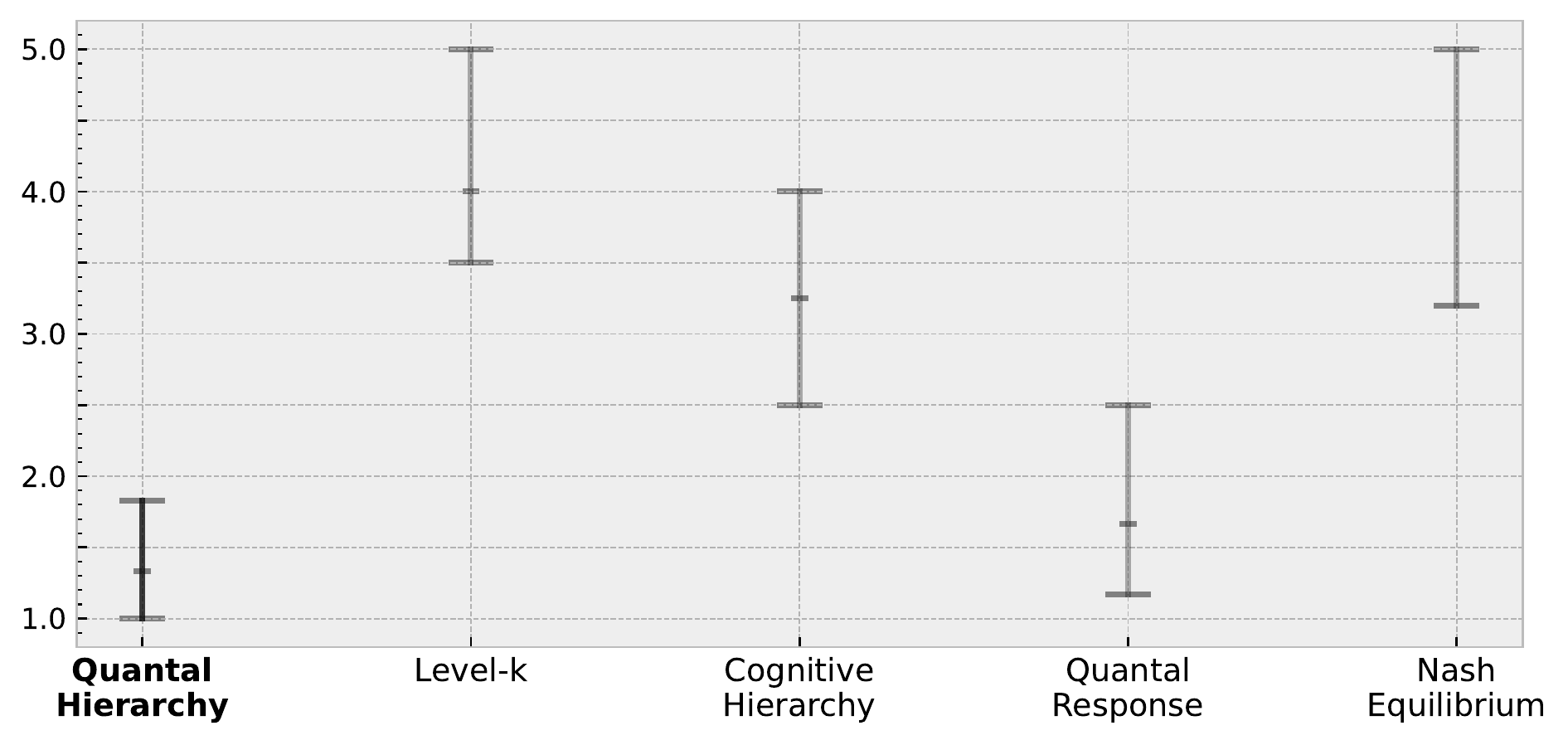}
    \caption{Overall rankings for out-of-sample errors across the various game classes trialled. The bars indicate the range of achieved ranks (worst achieved to best achieved ranking), with the middle bar indicating the median ranking. A lower ranking is better (with $1$=best).}
    \label{figRankings}
\end{figure}

For each game, we perform 5x2 repeated cross-fold validation \citep{dietterich1998approximate}, {analysing the out-of-sample performance. This analysis ensures that the inclusion of an additional parameter does not overfit to the original training data, and instead, ensures the approach generalises well to unseen data}.  We present the average RMSE on the unseen data and the resulting rankings \citep{demvsar2006statistical} of each method in \cref{tblResults}. The rankings account for the independence of the games, and the inability to compare errors directly across game classes. A visualisation of the resulting ranks in \cref{figRankings}. {By using these evaluation metrics, we are able to determine the effectiveness of the proposed method in comparison to existing approaches for predicting (out-of-sample) human behaviour on a range of canonical games.}

The proposed Quantal Hierarchy method consistently performs well across the various games trialled, resulting in the best (lowest) overall rank (\cref{tblResults}), as well as the most consistent (narrowest distribution of results, \cref{figRankings}), always performing in the top 2. These results validate the modelling assumption that it becomes more difficult to reason at deeper levels of reasoning, and thus, the reasoning process becomes more erroneous. This motivates the usage of the Quantal Hierarchy model for capturing human decision-making in a wide-range of settings.

In the following subsections, we analyse the game results in more detail.

\begin{table}[!htb]
\resizebox{\textwidth}{!}{
\begin{tabular}{rlccccc}
\toprule
\rowcolor[HTML]{C0C0C0} & \textbf{} & \textbf{\begin{tabular}[c]{@{}c@{}}Quantal\\ Hierarchy\end{tabular}} & Level-$k$ & \begin{tabular}[c]{@{}c@{}}Cognitive\\ Hierarchy\end{tabular} & \begin{tabular}[c]{@{}c@{}}Quantal Response\\ Equilibrium\end{tabular} & Nash \\ \midrule
\textbf{Market Entrance}
& Block 1 & 0.565 (2) & 1.503 (5) & 0.725 (3) & 0.564 (1) & 1.242 (4) \\
& Block 2 & 0.426 (1) & 2.115 (5) & 1.077 (4) & 0.442 (2) & 0.726 (3) \\
& Block 3 & 0.387 (1) & 2.230 (5) & 1.283 (4) & 0.406 (2) & 0.548 (3) \\
& Block 4 & 0.493 (2) & 2.344 (5) & 1.365 (4) & 0.431 (1) & 0.559 (3) \\
& Block 5 & 0.489 (1) & 2.513 (5) & 1.387 (4) & 0.519 (2) & 0.574 (3) \\
\rowcolor[HTML]{EFEFEF} 
\textbf{} & Average Rank & 1.4 & 5 & 3.8 & 1.6 & 3.2\\ \midrule
\textbf{Beauty Contest} & Lab & 0.020 (1) & 0.191 (4) & 0.175 (3) & 0.027 (2) & 0.194 (5) \\
& Classroom & 0.042 (2) & 0.188 (4) & 0.162 (3) & 0.019 (1) & 0.190 (5) \\
& Take Home & 0.056 (2) & 0.188 (4) & 0.162 (3) & 0.020 (1) & 0.192 (5) \\
& Internet & 0.053 (2) & 0.180 (4) & 0.149 (3) & 0.024 (1) & 0.181 (5) \\
& Newspaper & 0.061 (2) & 0.186 (4) & 0.149 (3) & 0.024 (1) & 0.187 (5) \\
& Theorists & 0.071 (2) & 0.171 (4) & 0.135 (3) & 0.040 (1) & 0.172 (5) \\
\rowcolor[HTML]{EFEFEF} 
 & Average Rank & 1.83 & 4 & 3 & 1.17 & 5\\ \midrule
\textbf{Centipede} & 4-level Centipede  & 0.469 (1) & 1.774 (4) & 0.611 (3) & 0.606 (2) & 3.715 (5) \\
\textbf{} & 6-level Centipede  & 0.350 (1) & 1.950 (4) & 0.439 (2) & 1.120 (3) & 2.837 (5) \\
\rowcolor[HTML]{EFEFEF} 
\textbf{} & Average Rank & 1 & 4 & 2.5 & 2.5 & 5 \\ \midrule
\textbf{Bargaining} & Ultimatum  & & & & & \\
\textbf{} & - (10, 10)  & 0.051 (1.5) & 0.098 (3.5) & 0.098 (3.5) & 0.051 (1.5) & 0.197 (5) \\
\textbf{} & - (10, 60) & 0.030 (1) & 0.093 (3.5) & 0.093 (3.5) & 0.057 (2) & 0.192 (5) \\
\textbf{} & - (70, 10) & 0.048 (2) & 0.090 (3.5) & 0.090 (3.5) & 0.047 (1) & 0.187 (5) \\
\textbf{} & Two-stage Bargaining & & & & & \\
& -D=0.9 & 0.040 (1) & 0.096 (3.5) & 0.096 (3.5) & 0.084 (2) & 0.198 (5) \\
& -D=0.8 & 0.054 (1) & 0.095 (3.5) & 0.095 (3.5) & 0.076 (2) & 0.198 (5) \\
& -D=0.7 & 0.048 (1) & 0.099 (3.5) & 0.099 (3.5) & 0.075 (2) & 0.197 (5) \\
& -D=0.6 & 0.067 (1) & 0.128 (3.5) & 0.128 (3.5) & 0.095 (2) & 0.197 (5) \\
& -D=0.5 & 0.037 (1) & 0.111 (3.5) & 0.111 (3.5) & 0.054 (2) & 0.190 (5) \\
& -D=0.4 & 0.030 (1) & 0.105 (3.5) & 0.105 (3.5) & 0.039 (2) & 0.191 (5) \\
& -D=0.3 & 0.024 (1.5) & 0.081 (3.5) & 0.081 (3.5) & 0.024 (1.5) & 0.192 (5) \\
& -D=0.2 & 0.052 (2) & 0.117 (3.5) & 0.117 (3.5) & 0.050 (1) & 0.196 (5) \\
\rowcolor[HTML]{EFEFEF} 
& Average Rank & 1.27 & 3.5  & 3.5  & 1.73 & 5  \\
\midrule
\rowcolor[HTML]{C0C0C0} 
\textbf{Overall} & Rank  & 1.37 & 4.12 & 3.2 & 1.75 & 4.55 \\ \bottomrule 
\end{tabular}
}
\caption{Average out-of-sample (5x2-fold cross-validation) error. Resulting ranks are indicated in brackets. In both cases, lower is better. Tied values receive the average rank between the ranks which would have been achieved had there been no ties. The overall rank is determined as the mean rank of the average rank across the game classes, meaning each game has an equal weighting in the overall rank, and the number of experiments for a game class does not affect this overall weight.
}
\label{tblResults}
\end{table}

\begin{table}[!htb]
\resizebox{\textwidth}{!}{
\begin{tabular}{rlccccc}
\toprule
\rowcolor[HTML]{C0C0C0} & \textbf{} & \textbf{\begin{tabular}[c]{@{}c@{}}Quantal\\ Hierarchy\\$\beta$, $\gamma$\end{tabular}} & \begin{tabular}[c]{@{}c@{}}Level-$k$\\\\$k$\end{tabular} & \begin{tabular}[c]{@{}c@{}}Cognitive\\ Hierarchy\\$\tau$\end{tabular} & \begin{tabular}[c]{@{}c@{}}Quantal Response\\ Equilibrium\\$\lambda$\end{tabular} & Nash \\ \midrule
\textbf{Market Entrance}
& Block 1 & 0.43, 0.24 & 0.0 & 0.72 & 10.39 & \\
& Block 2 & 0.67, 0.18 & 0.0 & 0.68 & 34.56 & \\
& Block 3 & 0.62, 0.28 & 0.0 & 0.65 & 59.44 & \\
& Block 4 & 0.32, 0.54 & 0.0 & 0.69 & 83.51 & \\
& Block 5 & 0.75, 0.19 & 0.2 & 0.72 & 80.8 & \\ \midrule
\textbf{Beauty Contest} & Lab & 0.08, 0.76 & 1.2 & 5.45 & 1.21 & \\
& Classroom & 0.1, 0.69 & 1.7 & 5.52 & 1.9 & \\
& Take Home & 0.06, 0.79 & 2.4 & 5.36 & 2.2 & \\
& Internet & 0.07, 0.72 & 3.3 & 5.67 & 2.31 & \\
& Newspaper & 0.08, 0.64 & 6.3 & 5.76 & 2.8 & \\
& Theorists & 0.05, 0.67 & 5.6 & 5.84 & 3.02  & \\\midrule
\textbf{Centipede} & 4-level Centipede  & 12.43,  0.22 & 0.0 & 1.82 & 2.09 & \\
\textbf{} & 6-level Centipede  & 19.1, 0.14 & 0.3 & 2.3 & 1.09 & \\ \midrule
\textbf{Bargaining} & Ultimatum &  & & & & \\
\textbf{} & - (10, 10)  & 0.08, 0.92 & 0.0 & 4.36 & 0.08 & \\
\textbf{} & - (10, 60)  & 0.2, 0.32 & 0.0 & 4.84 & 0.09 & \\
\textbf{} & - (70, 10) & 0.06, 0.88 & 0.0 & 3.68 & 0.06 & \\
\textbf{} & Two-stage Bargaining & & & & & \\
& -D=0.9 & 0.24, 0.13 & 0.0 & 4.07 & 0.04 & \\
& -D=0.8 & 0.2, 0.22 & 0.0 & 4.29 & 0.05 & \\
& -D=0.7 & 0.22, 0.28 & 0.0 & 3.73 & 0.06 & \\
& -D=0.6 & 0.52, 0.2 & 0.0 & 3.89 & 0.08 & \\
& -D=0.5 & 0.2, 0.36 & 0.0 & 2.97 & 0.11 & \\
& -D=0.4 & 0.19, 0.38 & 0.0 & 3.78 & 0.1 & \\
& -D=0.3 & 0.13, 0.49 & 0.0 & 3.69 & 0.08 & \\
& -D=0.2 & 0.17, 0.52 & 0.0 & 8.02 & 0.11 & \\
 \bottomrule 
\end{tabular}
}
\caption{Average fitted parameter values {for each approach. Explanation of the parameters and the fitting procedure is given in \cref{secModelFit}. }}
\label{tblParams}
\end{table}

\subsection{Simultaneous Games}
\paragraph{Market Entrance}

\begin{figure}[htb]
\centering
\begin{subfigure}{0.18\textwidth}
  \includegraphics[width=\linewidth]{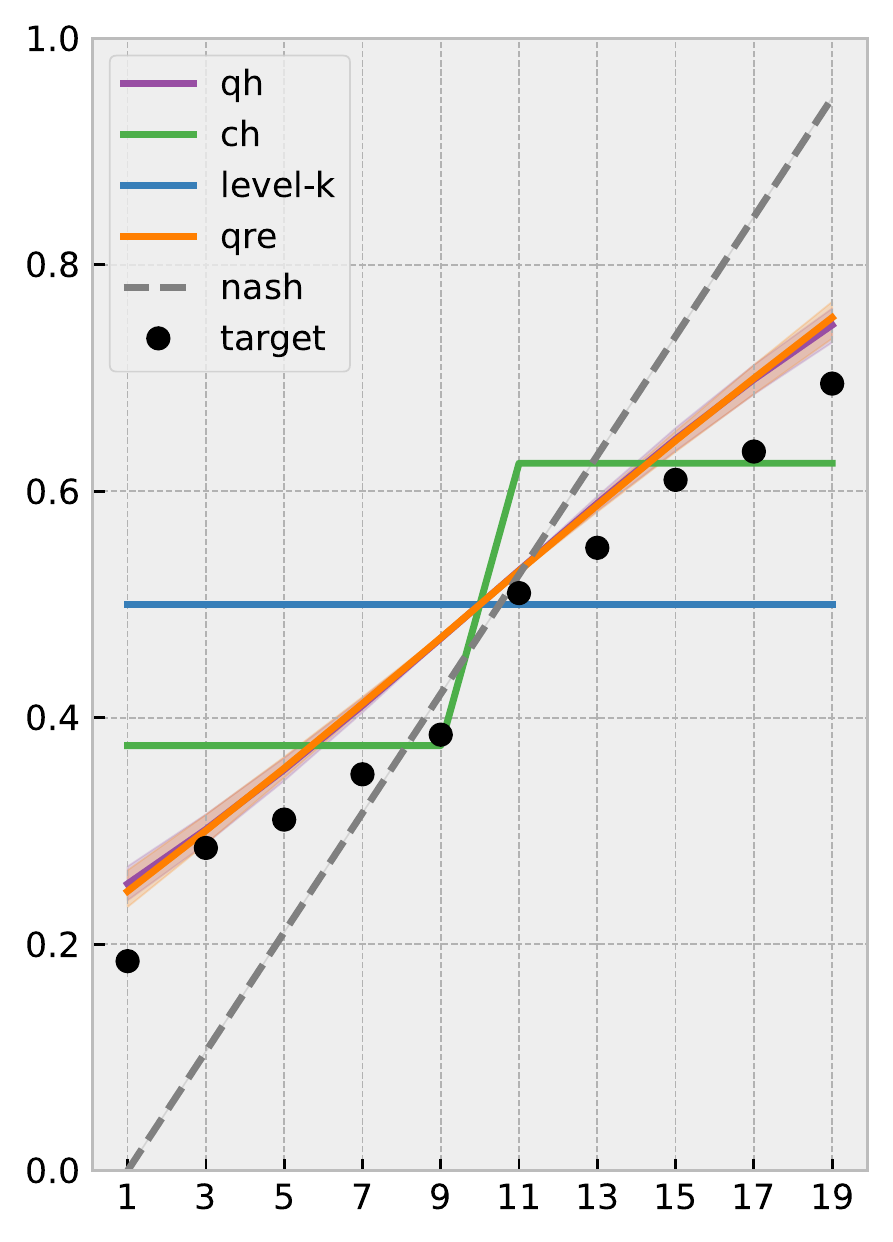}
  \caption{Block 1}\label{figMarketEntranceBeginning}
\end{subfigure}\hfil
\begin{subfigure}{0.18\textwidth}
  \includegraphics[width=\linewidth]{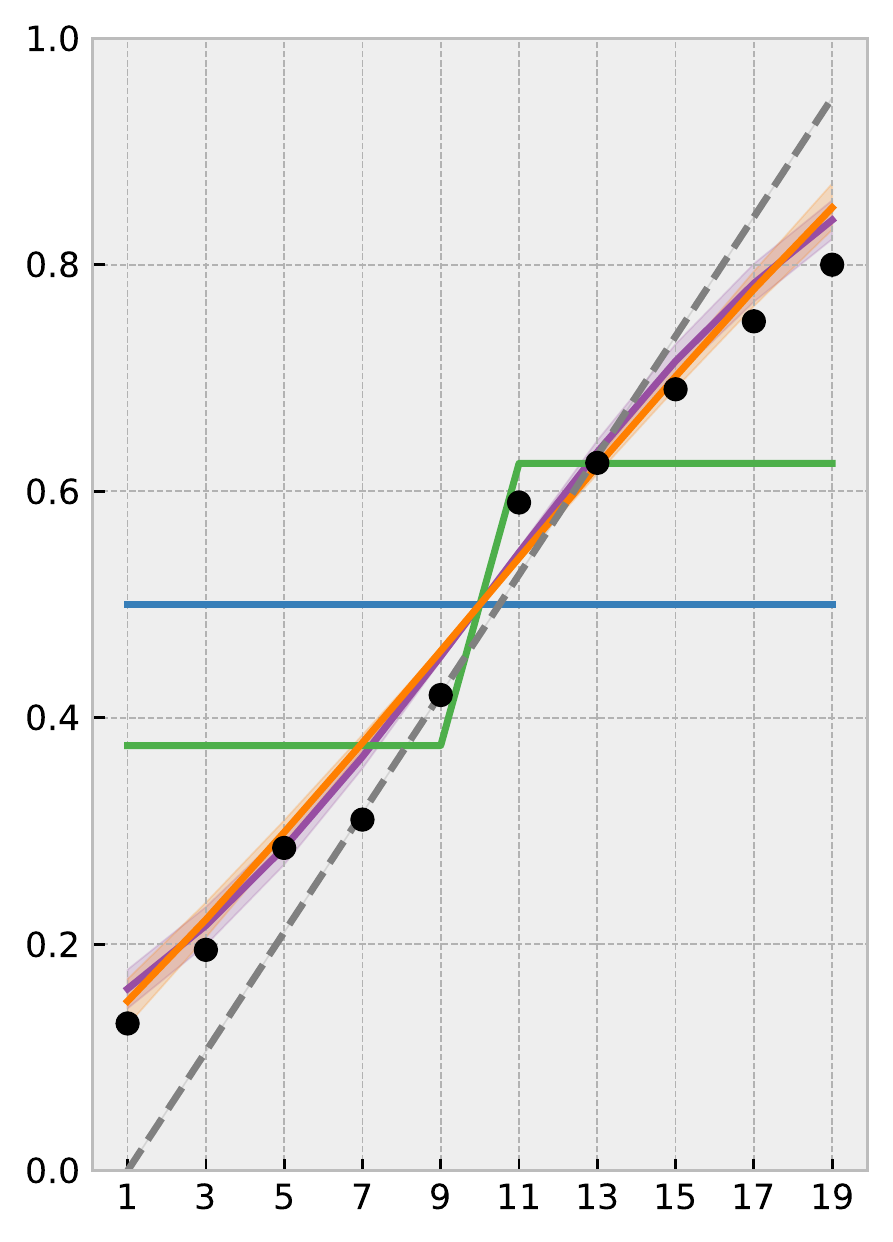}
  \caption{Block 2}
\end{subfigure}\hfil
\begin{subfigure}{0.18\textwidth}
  \includegraphics[width=\linewidth]{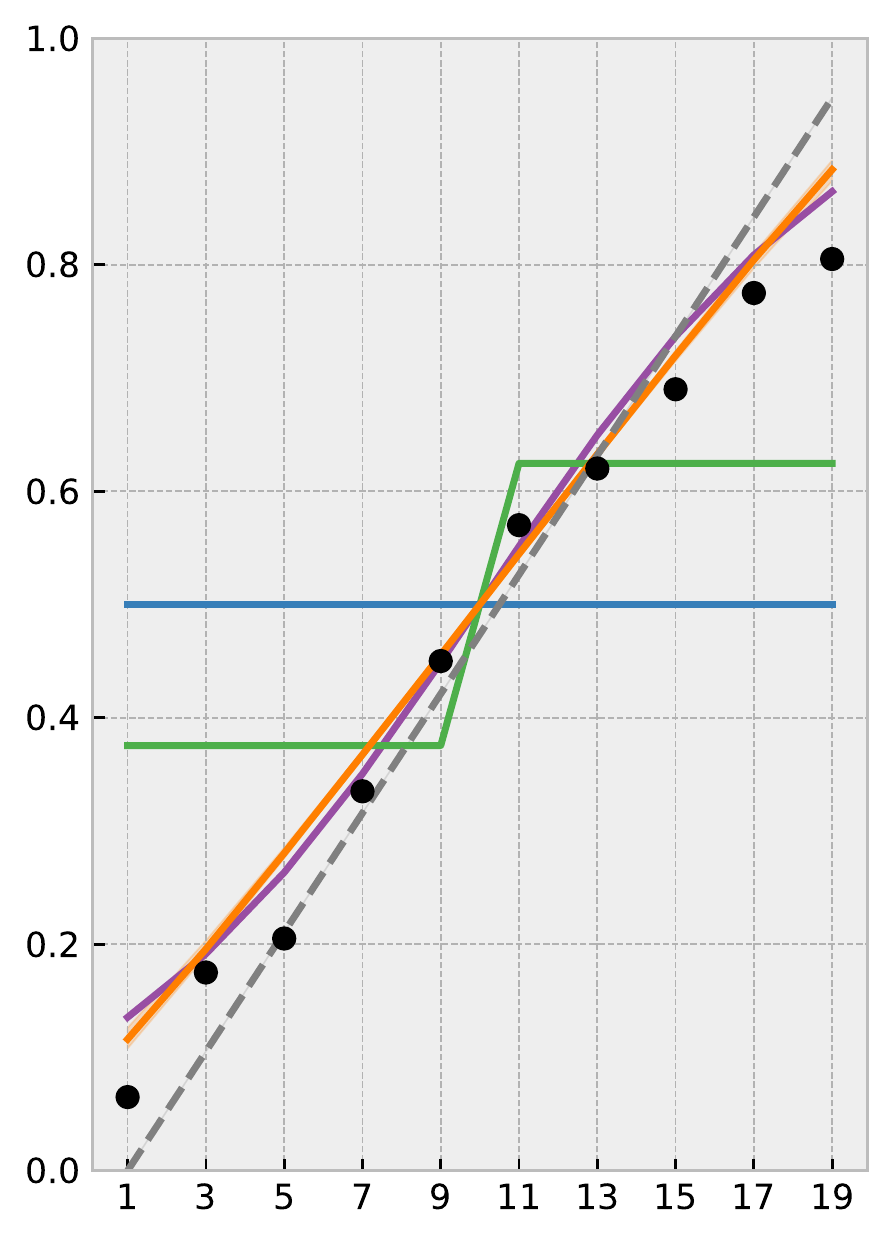}
  \caption{Block 3}
\end{subfigure}\hfil
\begin{subfigure}{0.18\textwidth}
  \includegraphics[width=\linewidth]{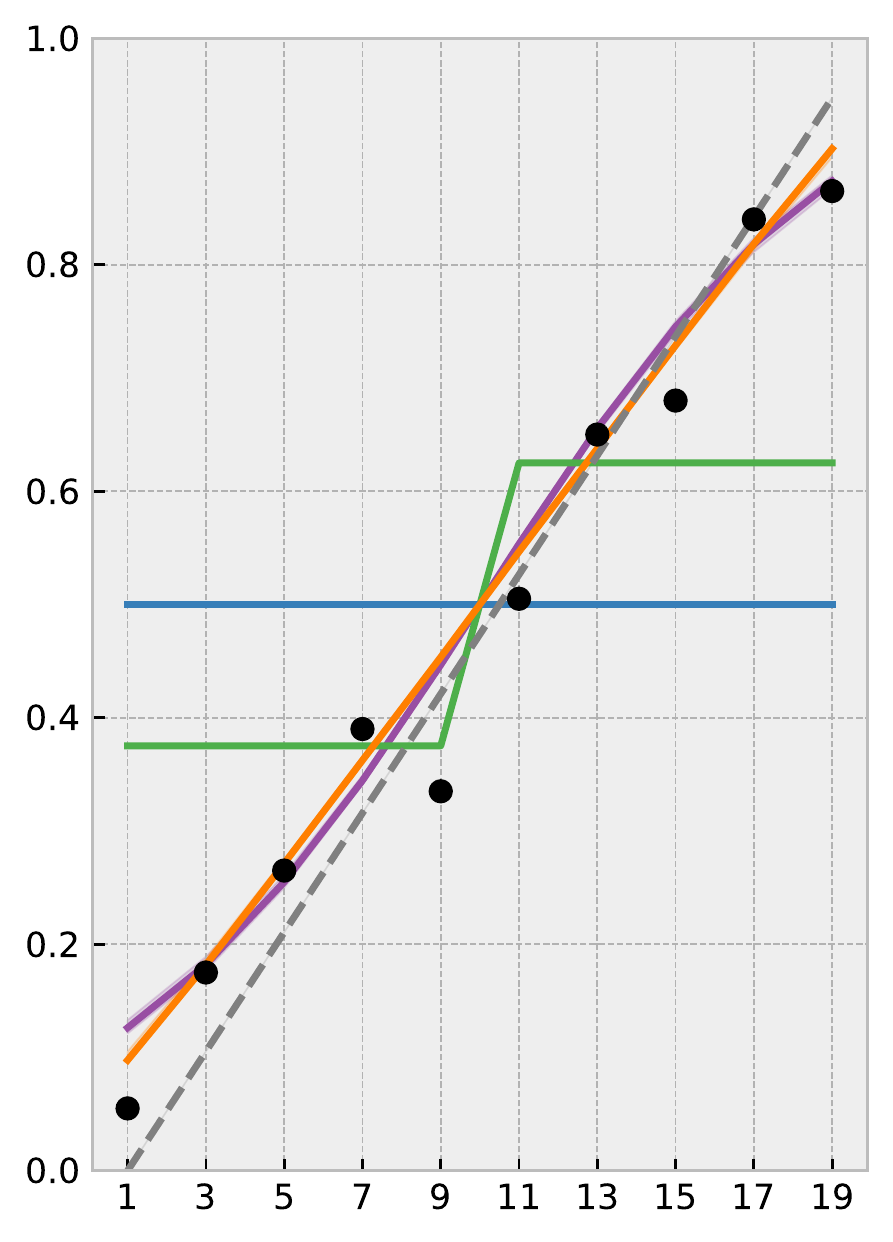}  \caption{Block 4}
\end{subfigure}\hfil
\begin{subfigure}{0.18\textwidth}
  \includegraphics[width=\linewidth]{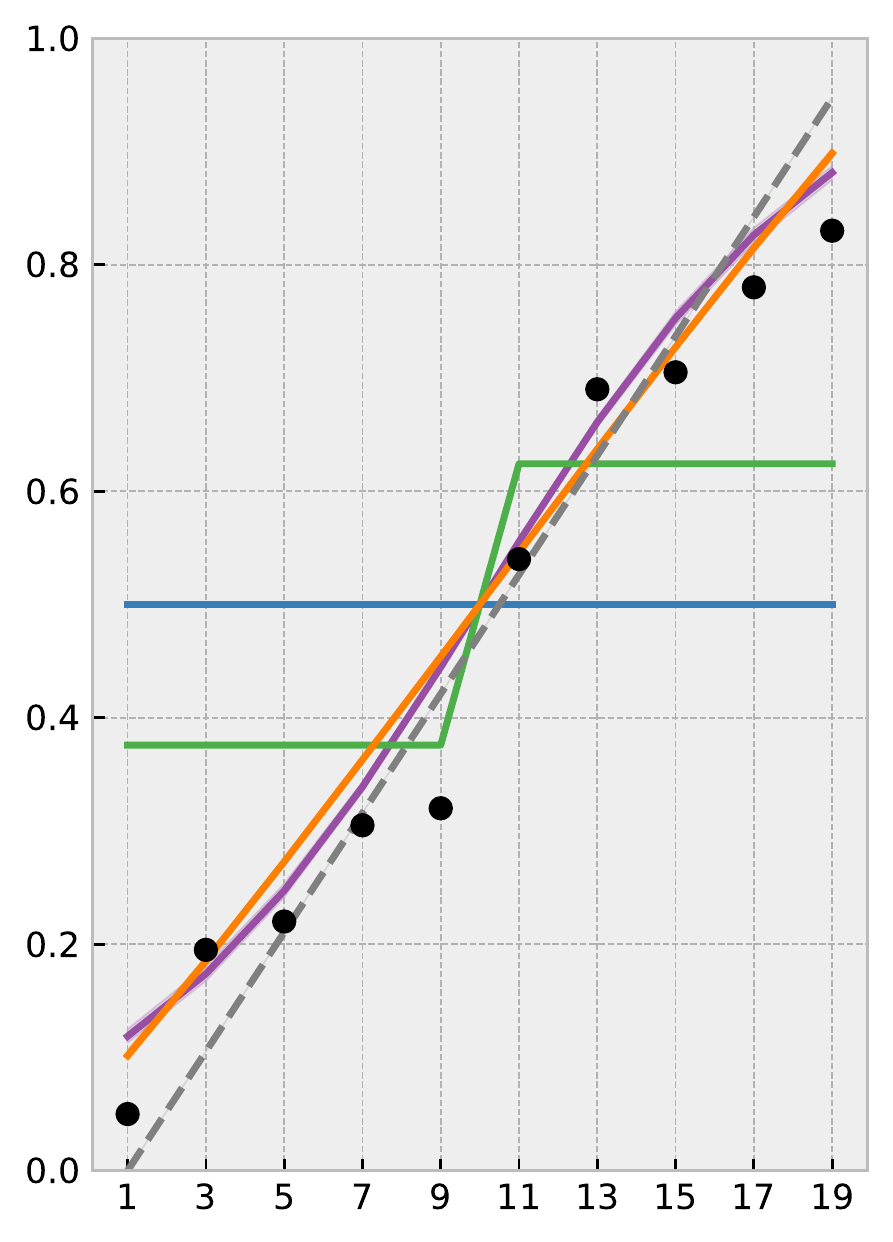}
  \caption{Block 5}\label{figMarketEntranceEnd}
\end{subfigure}
\caption{Market Entrance Game. The darker lines indicate the mean result from the 5x2 cross-validation. The shaded regions indicate $\pm$ one standard deviation. The out-of-sample data are shown as the black circles. The proposed Quantal Hierarchy model is the purple line. Quantal Response Equilibrium is the orange line, level-$k$ is the blue line, and Cognitive Hierarchy is the green line. The Nash equilibrium solution is indicated as the diagonal dashed grey line.
}\label{figMarketEntrance}
\end{figure}

In the market entrance game, players must simultaneously decide whether to enter or stay out of a market, where the payoff depends on the decisions of the other players and market capacity $c$ (see \cref{secAppendixMarket}). Experimental data show that player behaviour in market games is inconsistent with either mixed or pure Nash equilibria, although, with repeated play, players begin to approach the mixed strategy equilibrium \citep{duffy2005learning}.

These deviations from equilibrium are captured well by the proposed Quantal Hierarchy model (\cref{figMarketEntrance}). We see that in the beginning (before learning, \cref{figMarketEntranceBeginning}), the players overestimate for low $c$ and underestimate for high $c$. Towards the final rounds (after learning, e.g. \cref{figMarketEntranceEnd}), the behaviour approaches equilibrium, and the proposed QH model approximates this well using an increase in processing resources $\beta$ and/or $\gamma$ (see \cref{tblParams}) to capture this player ``learning". These changes highlight an important property of the QH model. If a player is learning, i.e., becoming closer to rational, this should correspond to an increase in $\beta$ (and/or an increase in $\gamma$).

The level-$k$ model fails to capture the overall trend, and in fact, is best fitted with $k=0$, performing worse than the mixed strategy equilibrium (and all other alternatives). The reason for this is simple. Level-$k$ ($k\geq1$) implies a step function, where for $c > T$ where $T$ is some threshold, the player enters with certainty, and for $c \leq T$, the player stays out with certainty. The distance to the experimental data from this step function is greater than the uniform case ($k=0$), so the uniform case is chosen. The cognitive hierarchy model improves upon level-$k$, by fitting a distribution of $k$ thinkers, able to ``smooth" out this step function, with the line shown in \cref{figMarketEntrance}. While this captures the qualitative trend (over entry for low $c$, under entry for high $c$, near equilibrium for mid $c$), quantitatively, the approach is not as strong as QRE, Quantal Hierarchy, or even the mixed-strategy equilibrium in most cases.

The QRE model is also a good fit here, however, due to the representation is constrained to linear lines. In contrast, the QH representation can capture such ``S" shape curves, better approximating the experimental data in 3 out of the 5 blocks. QRE and QH significantly outperform the approaches which just relax mutual consistency (level-$k$ and CH), motivating the relaxation of best-response in addition to mutual consistency.

\paragraph{$p$-Beauty contest}

\begin{figure}[htb]
\centering
\begin{subfigure}{0.3\textwidth}
  \includegraphics[width=\linewidth]{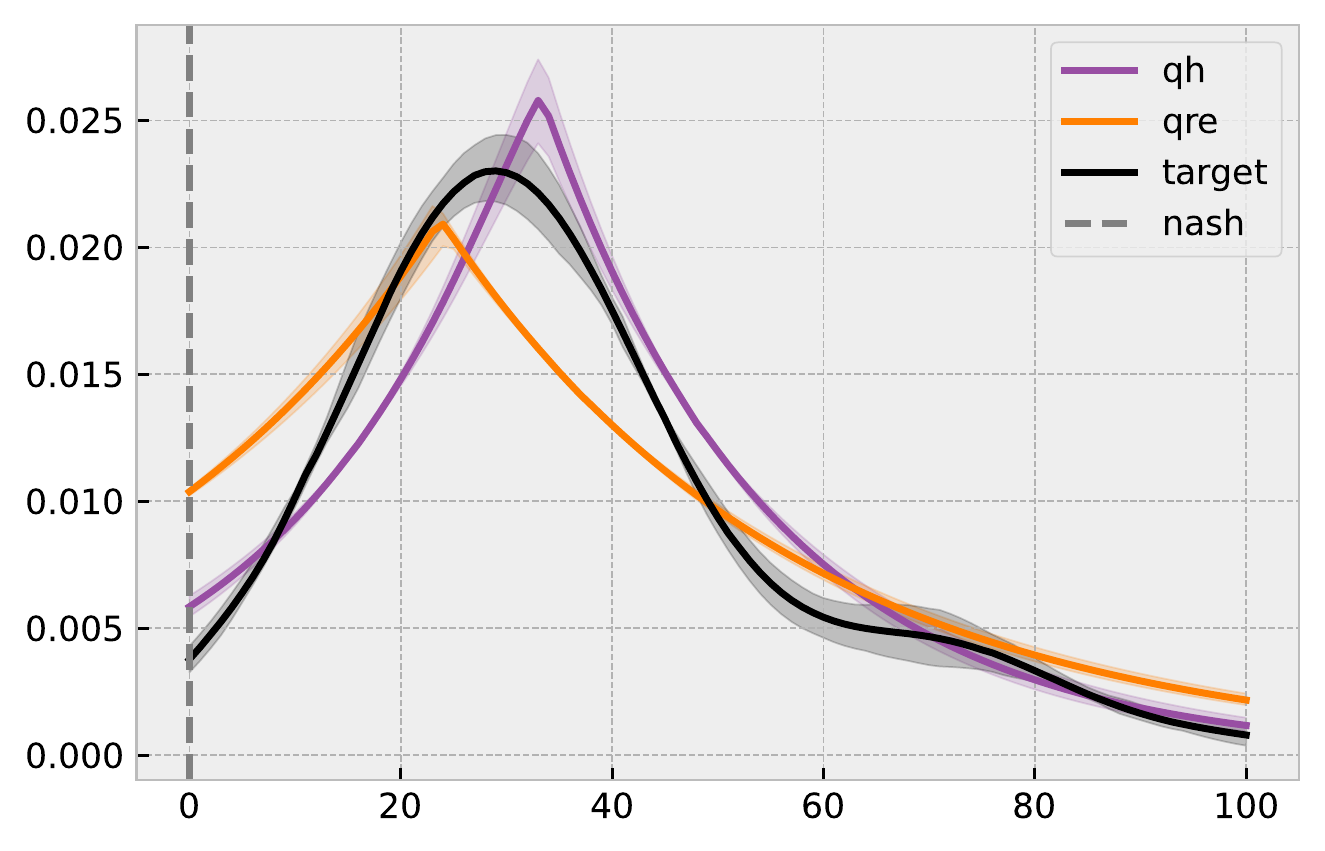}
  \caption{Lab}
\end{subfigure}\hfil
\begin{subfigure}{0.3\textwidth}
  \includegraphics[width=\linewidth]{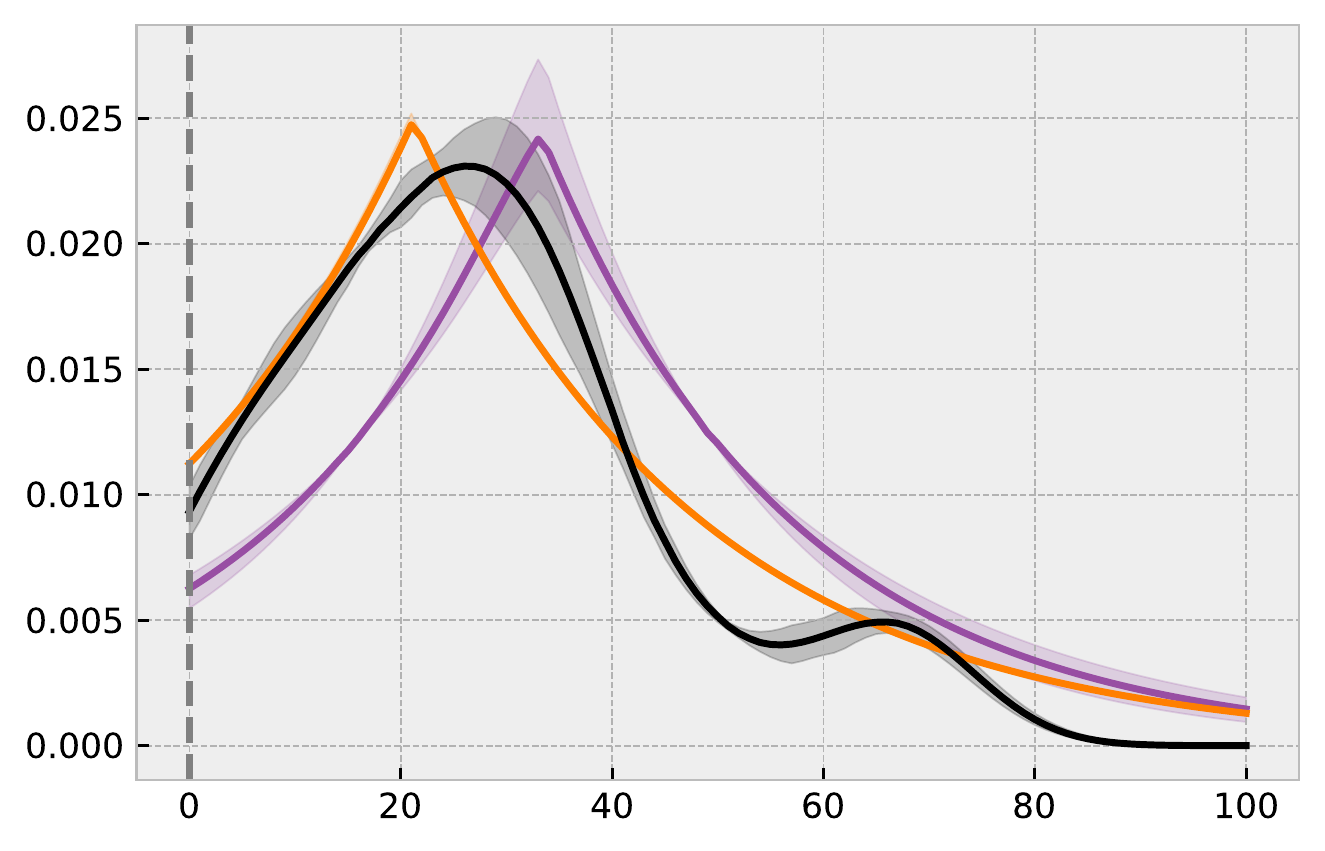}
  \caption{Classroom}
\end{subfigure}\hfil
\begin{subfigure}{0.3\textwidth}
  \includegraphics[width=\linewidth]{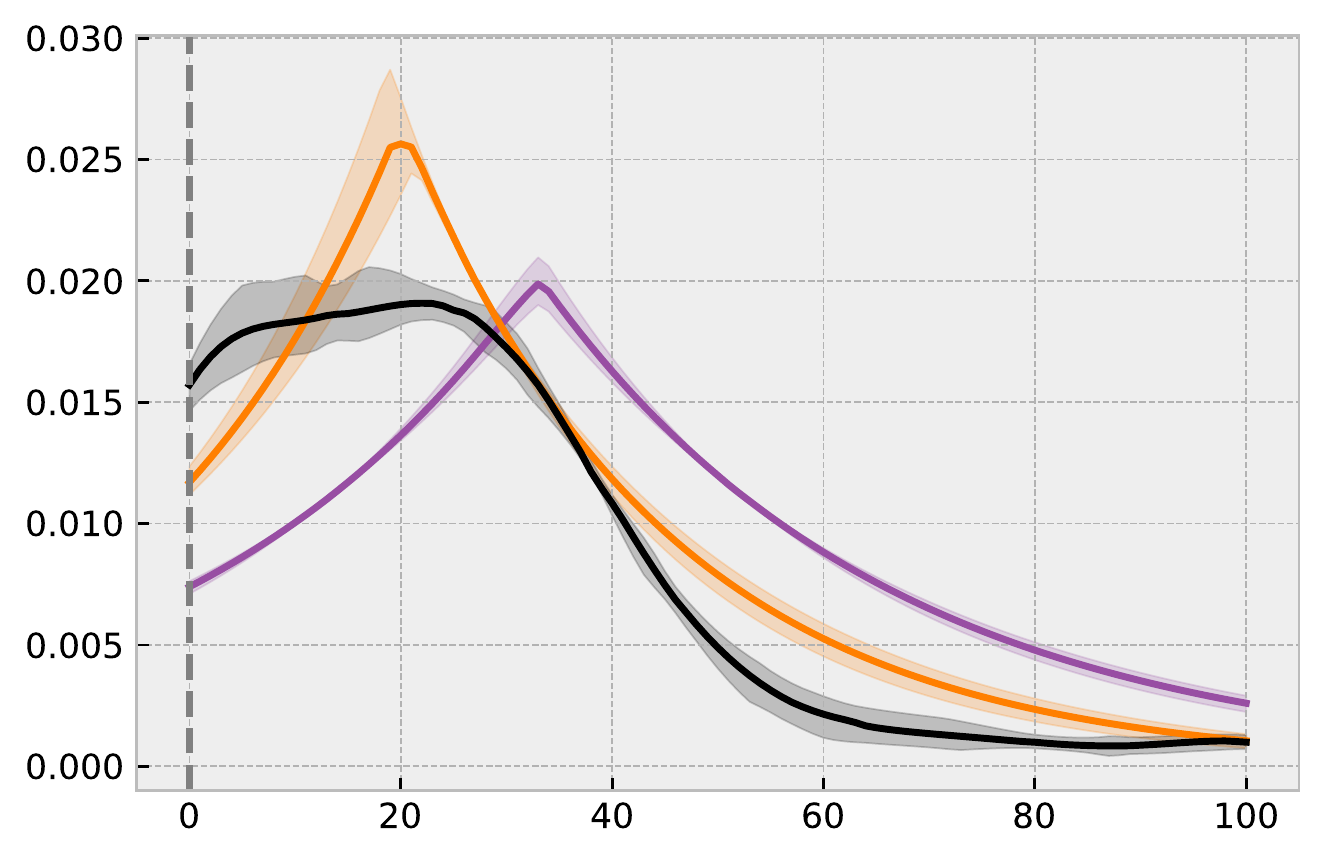}
  \caption{Internet}
\end{subfigure}\hfil
\begin{subfigure}{0.3\textwidth}
  \includegraphics[width=\linewidth]{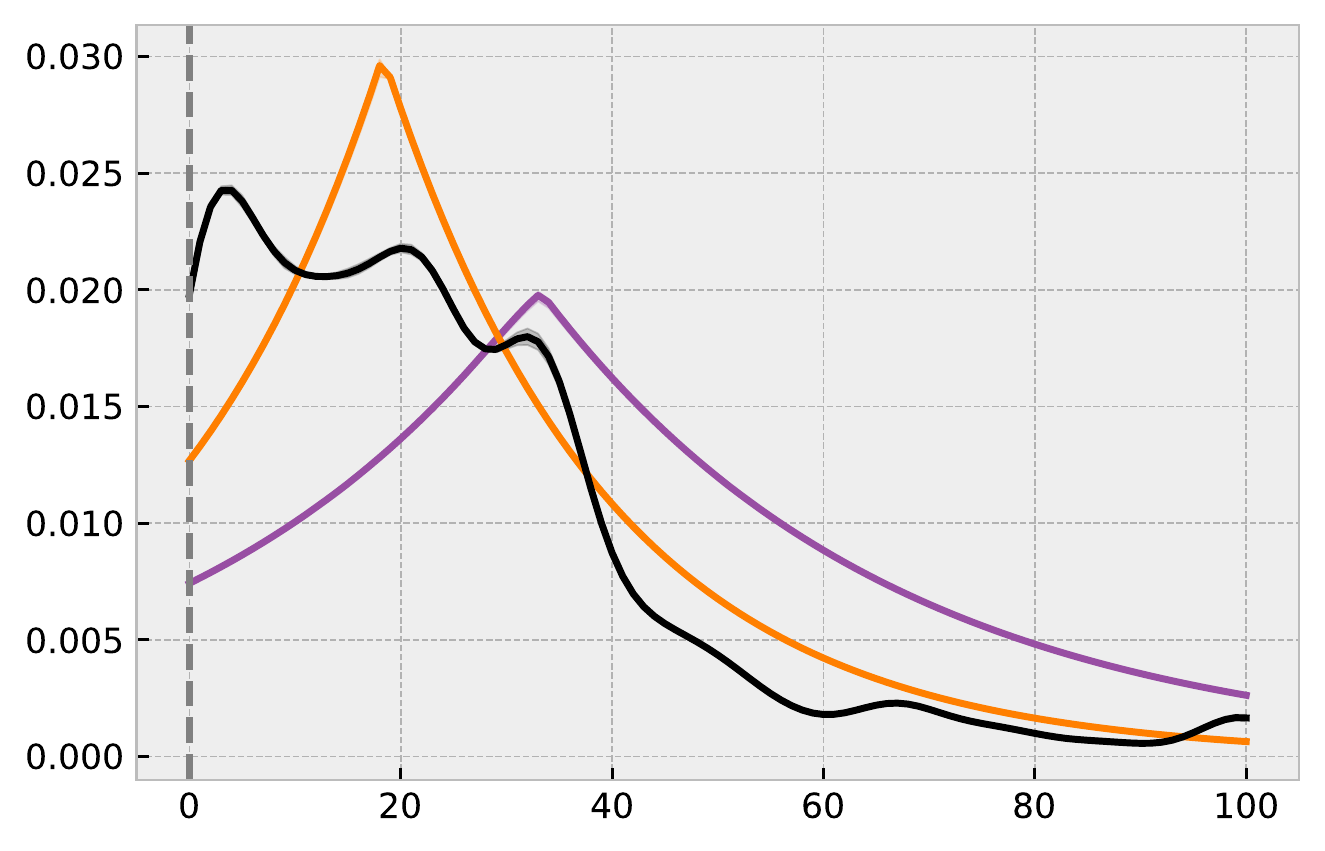}
  \caption{Newspaper}
\end{subfigure}\hfil
\begin{subfigure}{0.3\textwidth}
  \includegraphics[width=\linewidth]{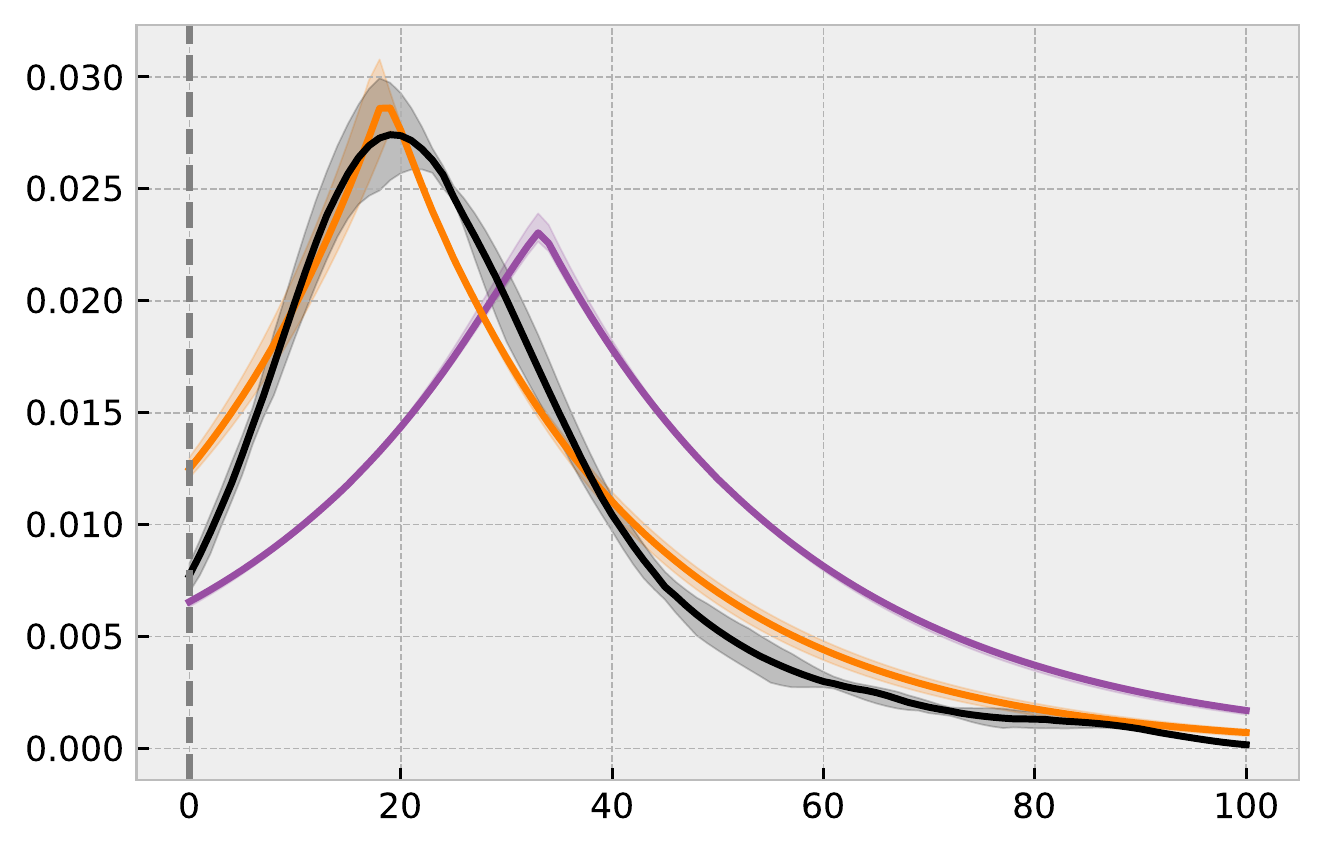}
  \caption{Take Home}
\end{subfigure}\hfil
\begin{subfigure}{0.3\textwidth}
  \includegraphics[width=\linewidth]{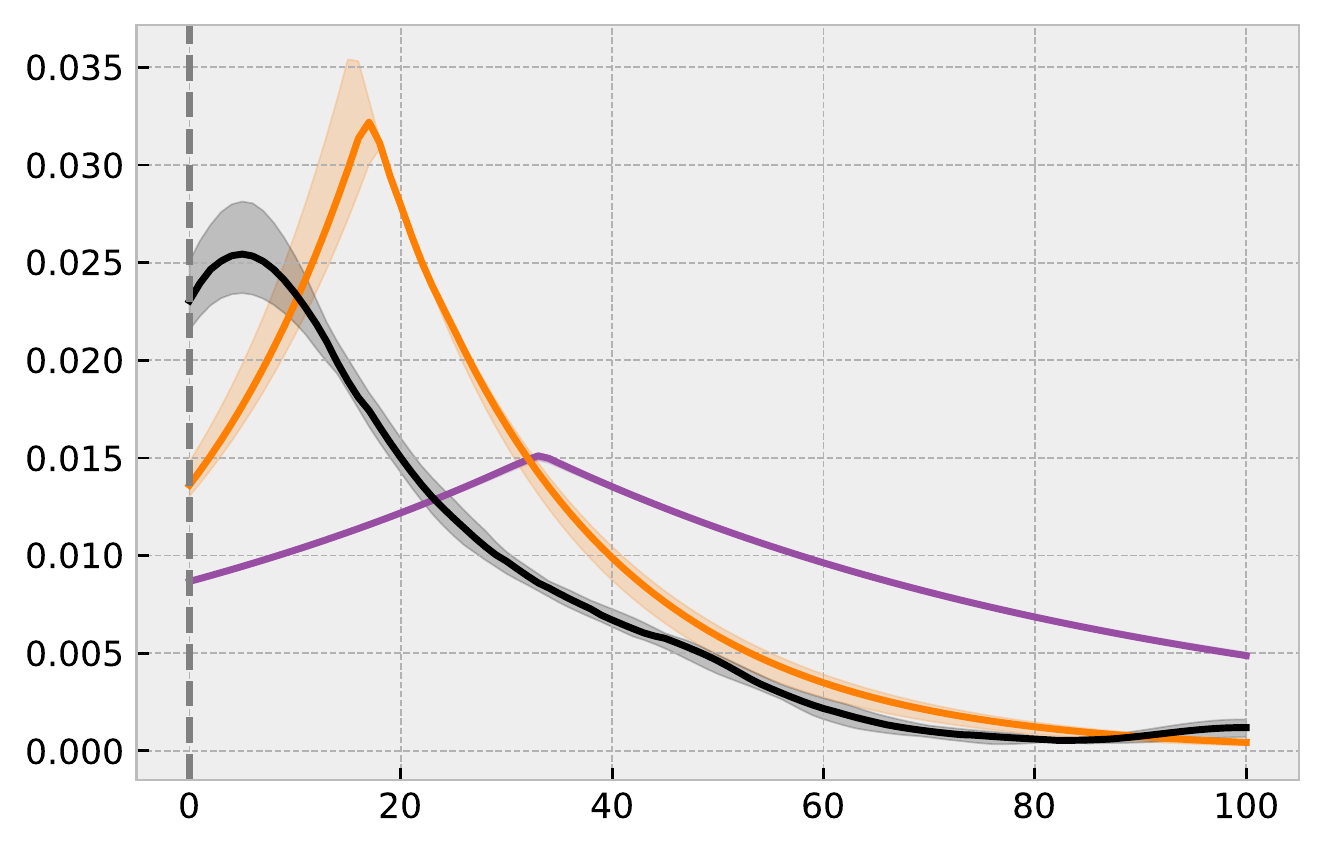}
  \caption{Theorists}
\end{subfigure}\hfil
\caption{Beauty contest games. The darker lines indicate the mean result from the 5x2 cross-validation. The shaded regions indicate $\pm$ one standard deviation. The out-of-sample data are shown as the black line. The proposed Quantal Hierarchy model is the purple line. Quantal Response Equilibrium is the orange line. The level-$k$ and Cognitive Hierarchy plots are shown in \cref{figBeautyExtended} due to the large difference in scales, distorting the figure. The Nash equilibrium solution is indicated as the diagonal dashed grey line.
}\label{figBeauty}
\end{figure}

In the $p$-beauty contest \citep{moulin1986game}, players must try and guess $p$ times the average guess (in the range $[0,100]$) of other competitors (see \cref{secAppendixBeauty}). The Nash equilibrium is for all players to guess $0$, however, experimentally, we see large deviations from this behaviour. 

Analysing the experimental results (\cref{figBeauty}), we see very strong performance for the QH model when modelling the less experienced players, e.g. in the Lab experiments. The QH model fits the data well, capturing the overall distribution and achieving the lowest error rate. However, for the other experiments composed of more experienced players or players with more time (take home, newspaper), we see the distribution is better approximated by QRE.

\begin{figure}[ht]
    \centering
    \includegraphics[width=.5\textwidth]{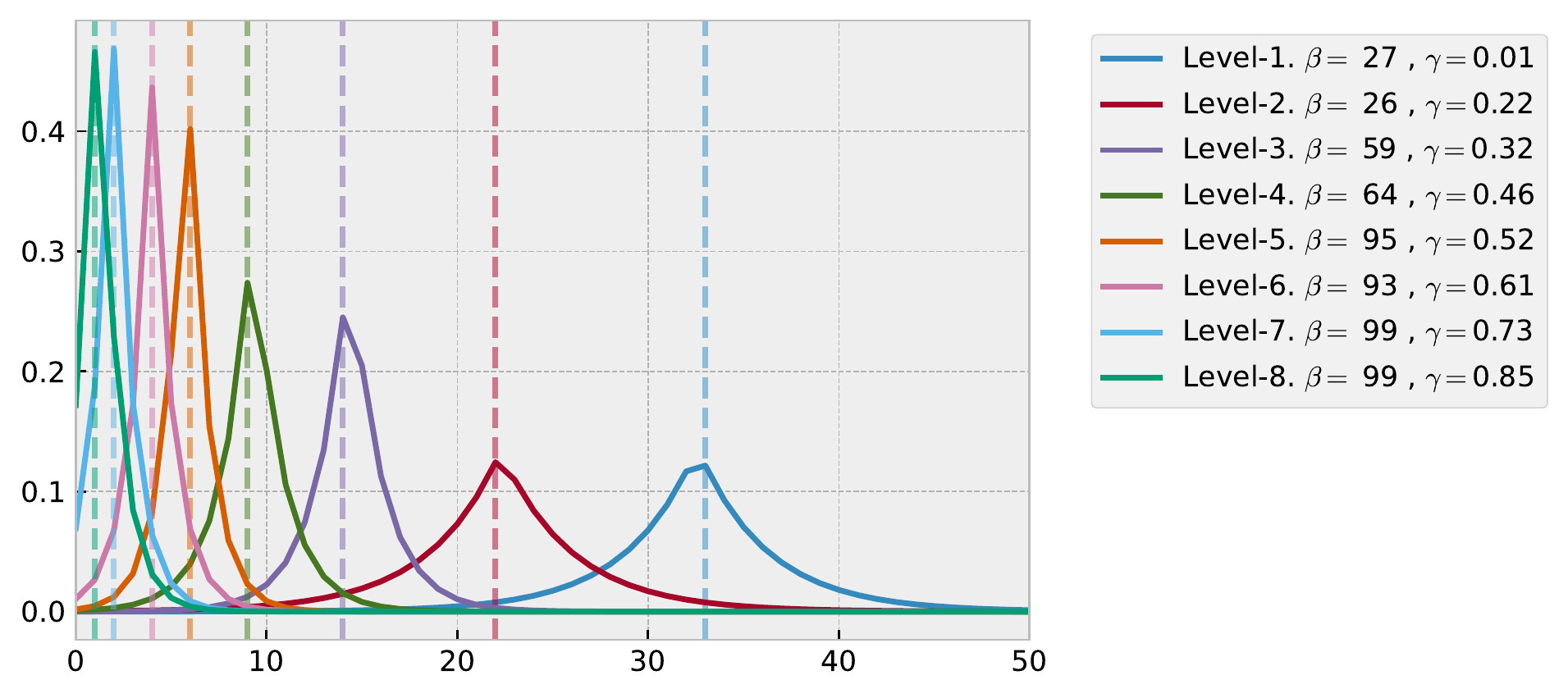}
    \caption{Example comparing the decisions of level-$k$ (dashed vertical lines) to the proposed QH model (solid lines) in the $p$-beauty contest for various settings.
    }
    \label{figKLevelApproximation}
\end{figure}

The reason for this performance is because under the proposed approach, as a player becomes more rational, the distribution of choices narrows in to the optimal choice (see \cref{figKLevelApproximation}). However, under these experimental settings, even in the theorist case, there is bounded rational (and in fact, anti-rational) behaviour. These deviations are captured well by QRE. However, it is difficult for the QH model to capture the wide distribution of choices, as well as the bulk probability mass around the optimal case. For example, in \cref{figKLevelApproximation} we show the proposed approach approximating level-$k$. As $k$ increases, the distribution narrows. Here, this narrowing of the distribution makes it difficult to capture the entire prediction range for the more advanced subjects, due to the fact there are many sub-rational choices mixed in. As a result, we see similar fitted models for each case, despite the fact that the theorists clearly have a higher level of reasoning. If, instead, we tried to approximate the average player for each case, we could capture this increase in reasoning effort, and it would reflect an increase in $\beta$ and/or $\gamma$ as expected.

Nevertheless, in all cases, the model still significantly outperforms level-$k$, Cognitive Hierarchy, and the mixed strategy equilibrium. The level-$k$ model predicts some of the representative spikes in the experimental data (e.g., with $k=1$ guesses of $33$, $k=2$ of $22$, etc.). However, we can see that the players do not necessarily choose according to level-$k$, and may make errors around the best response suggested by level-$k$ reasoning. Level-$k$ thinking presupposes that players will predict a multiple of $p$, i.e., with $p=\frac{2}{3}$, we get $p \times 50, p^2 \times 50, \dots,  p^k \times 50$, as the players at each level are best responding to lower-level players. In the proposed QH model, level-$k$ reasoning can be recovered if $\beta_t=\infty$ for $t \leq k$ and $\beta_t=0$ for $t > k$. However, with $\beta_t < \infty$, the proposed QH model produces a distribution around these best-responding values anticipating potential errors in player reasoning, with these errors growing throughout the chain of reasoning.

The cognitive hierarchy model can improve upon the level-$k$ approach here by weighting the ``spikes" of the level-$k$ model differently, however, this still fails to capture the underlying distribution. A large reason for this is that certain predictions in the $p$-beauty contest are considered irrational, for example, any prediction over 67. However, we can see experimentally that such predictions occur, for example, in \cref{figBeauty}. If a player believes that other players would choose the maximal offer of $100$, then the player should choose $\frac{2}{3}100=~67$. Level-$k$ or cognitive hierarchy models cannot capture such irrational behaviour where players choose $>67$. That is, there is no distribution of level-$k$ thinkers that would predict $100$. However, this feature can be captured directly under the proposed QH and QRE models due to the errors in play, again motivating the usefulness of relaxing best response in addition to mutual consistency.

In summary, we see that under the lab experiment, the QH approach is the best fit. However, QRE is a better fit in other cases of the beauty contest game.

\subsubsection{Sequential Games}
\paragraph{Centipede Games}

In the centipede game (see \cref{secAppendixCentipede}), ``two players alternately get a chance to take the larger portion of a continually escalating pile of money. As soon as one person takes, the game ends with that player getting the larger portion of the pile, and the other player getting the smaller portion" \citep{mckelvey1992experimental}.

\begin{figure*}
    \centering
    \begin{subfigure}[t]{0.4\textwidth}
        \centering
        \includegraphics[width=\textwidth]{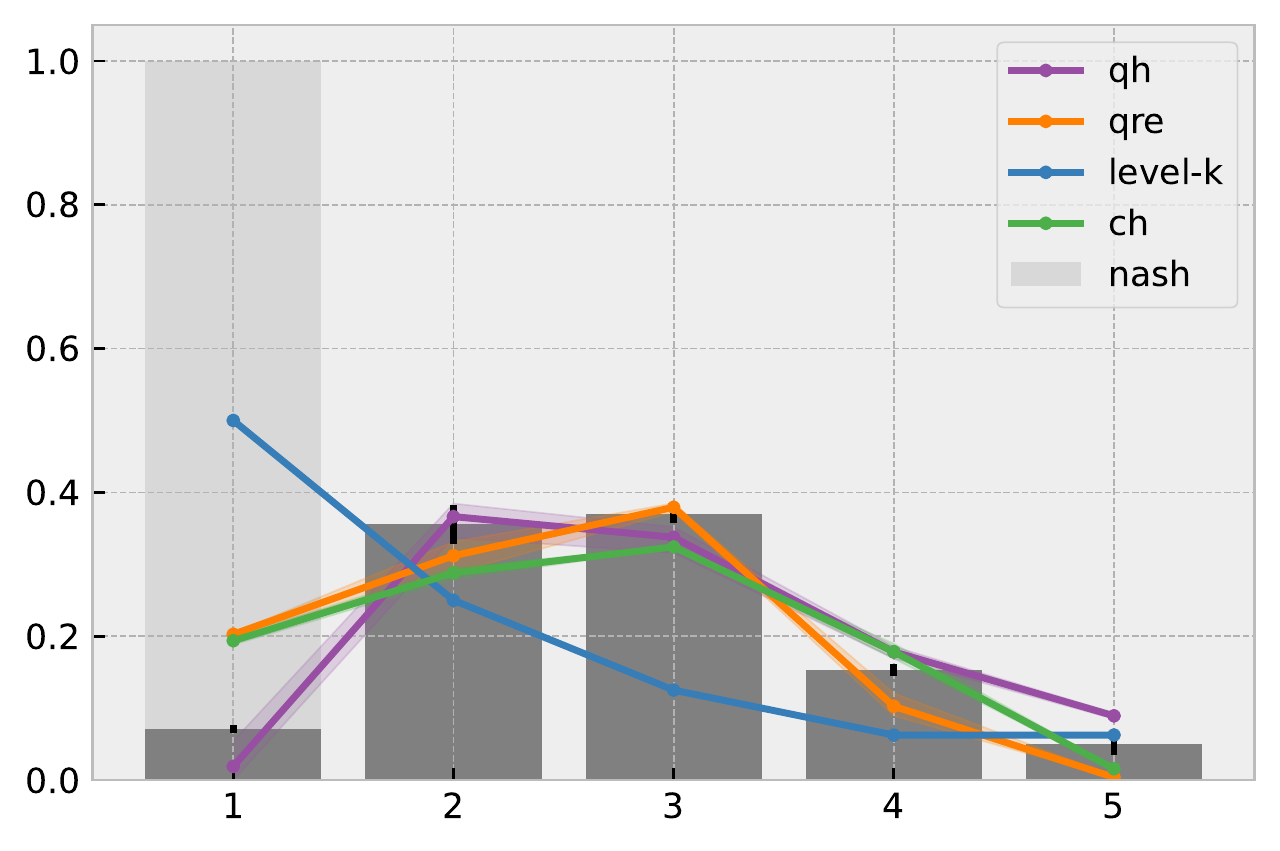}
        \caption{Four Move.}\label{figCentipede4}
    \end{subfigure}
    \begin{subfigure}[t]{0.4\textwidth}
        \centering
        \includegraphics[width=\textwidth]{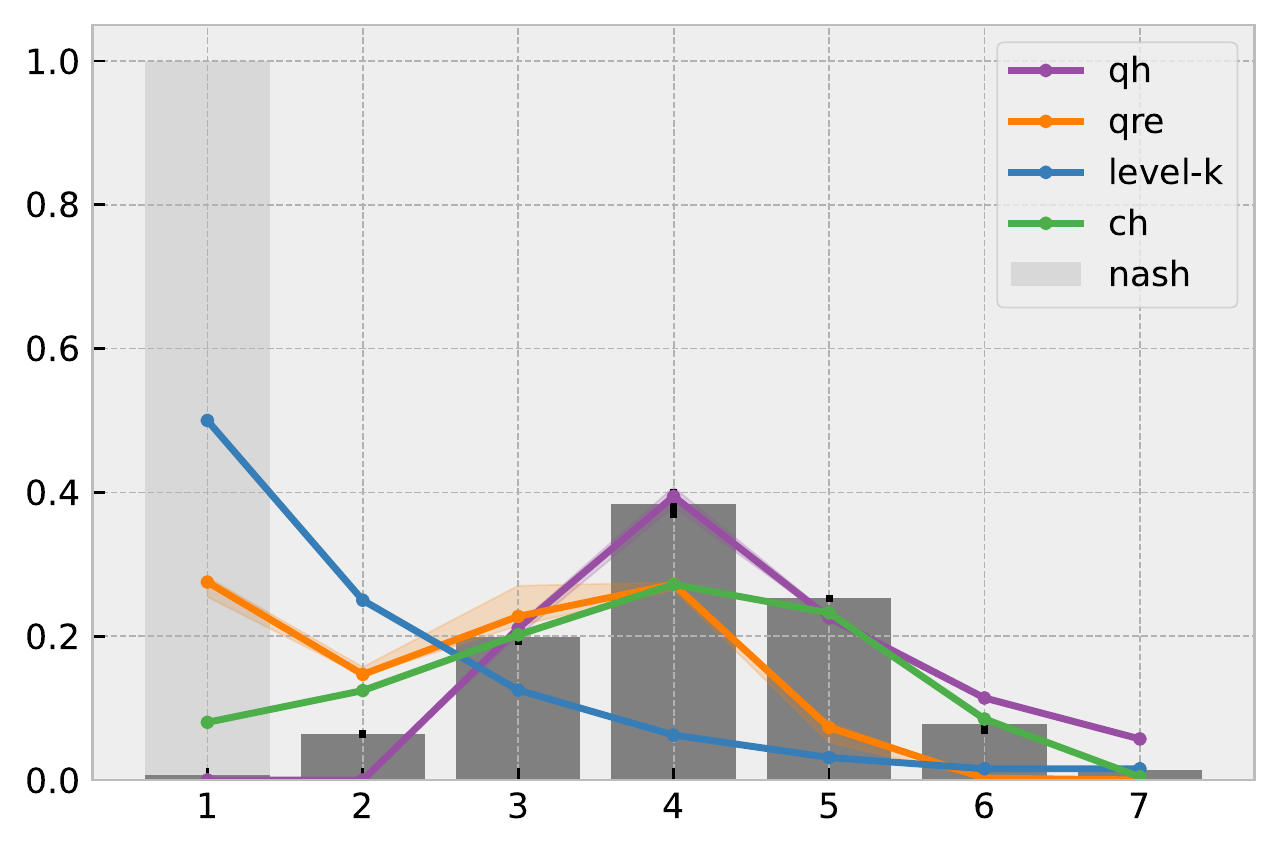}
        \caption{Six Move.}\label{figCentipede6}
    \end{subfigure}
    \caption{Four and Six-level Centipede Games.
    The darker lines indicate the mean result from the 5x2 cross-validation. The shaded regions indicate $\pm$ one standard deviation. The out-of-sample data are shown in the dark grey bars. The proposed Quantal Hierarchy model is the purple line. Quantal Response Equilibrium is the orange line, level-$k$ is the blue line, and Cognitive Hierarchy is the green line. The Nash equilibrium solution is indicated as the light grey bar at the first move.
    }\label{figCentipede}
\end{figure*}

The subgame perfect equilibrium of the centipede game is for each player to immediately take the pot without proceeding to any further rounds, however, we see this is not the case experimentally, where players behave far from the subgame perfect equilibrium (\cref{figCentipedeExperimental}). In general, many players take towards the middle of the game. The proposed QH model can capture this trend well, with QRE and Cognitive Hierarchy generally over-weighting the earlier nodes and under-weighting the later modes (\cref{figCentipede}).

The Quantal Hierarchy model provides the best fit for both the four and six-level centipede games, capturing realistic beliefs. When modelling this reasoning process, the player believes they are reasoning at a higher level than their opponent, but in addition, it is \textit{as if} the player overestimates how noisy their own play will be when faced with a decision at later nodes. This overestimation is because once actually faced with the decision, there will be a smaller game tree for the player to consider. This reasoning process was shown to approximate the experimental results well, motivating the discounting of information processing resources for capturing future beliefs. When comparing the resulting parameters ($\beta$ and $\gamma$) from the four and six-level variants (\cref{tblParams}), we note that the six-level variant results in additional information processing costs for the player (larger $\beta$ and $\gamma$). The additional processing costs result from the longer chain of reasoning, requiring higher processing resources. 

The significantly improved performance over quantal response equilibrium on both games motivates the usefulness of relaxing mutual consistency in addition to best response. By relaxing mutual consistency, we captured the perceived ``lapse" in reasoning when considering the full extensive form game tree by reducing the information processing resources the further the player tries to reason through the tree. 

\paragraph{Bargaining Games}

In bargaining games, players alternately bargain over how to divide a sum (see \cref{secAppendixBargaining}). We examine two types, single-stage (Ultimatum) and two-stage bargaining games. These are extensions of the example game considered in \cref{secExample}.

\begin{figure}[htb]
\centering
\begin{subfigure}{0.3\textwidth}
  \includegraphics[width=\linewidth]{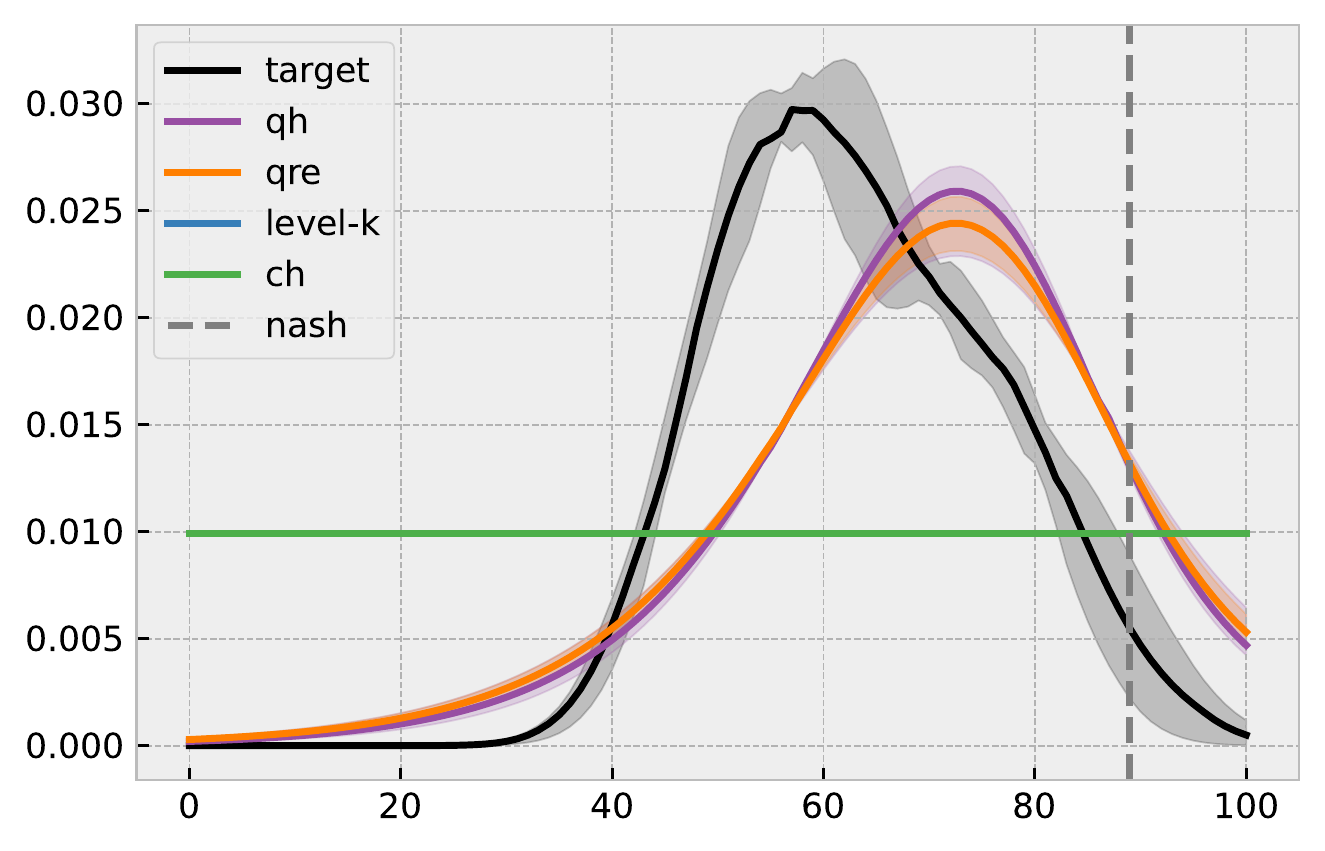}
  \caption{(10,10)}
  \label{figRequested10}
\end{subfigure}\hfil
\begin{subfigure}{0.3\textwidth}
  \includegraphics[width=\linewidth]{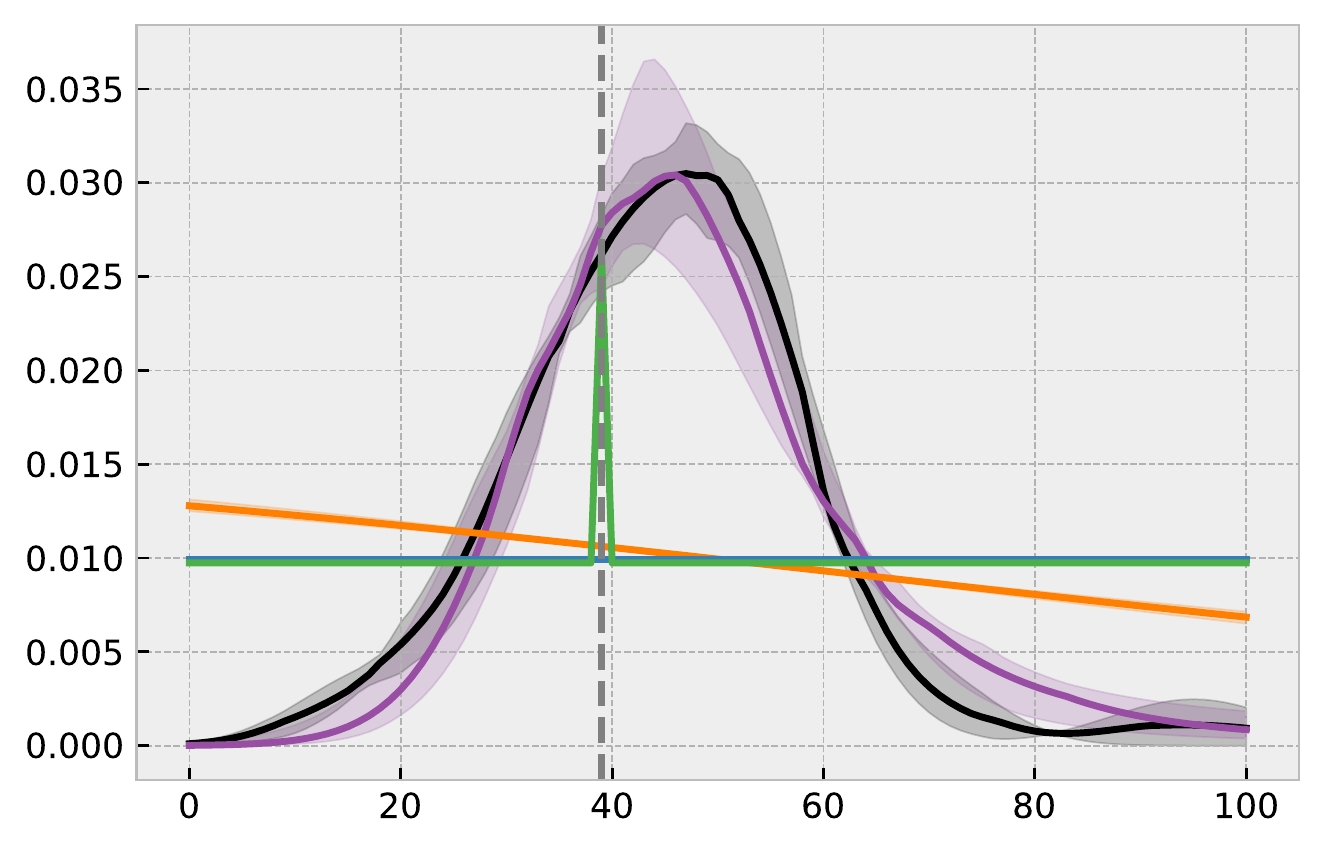}
  \caption{(10,60)}
  \label{fig:3}
\end{subfigure}\hfil
\begin{subfigure}{0.3\textwidth}
  \includegraphics[width=\linewidth]{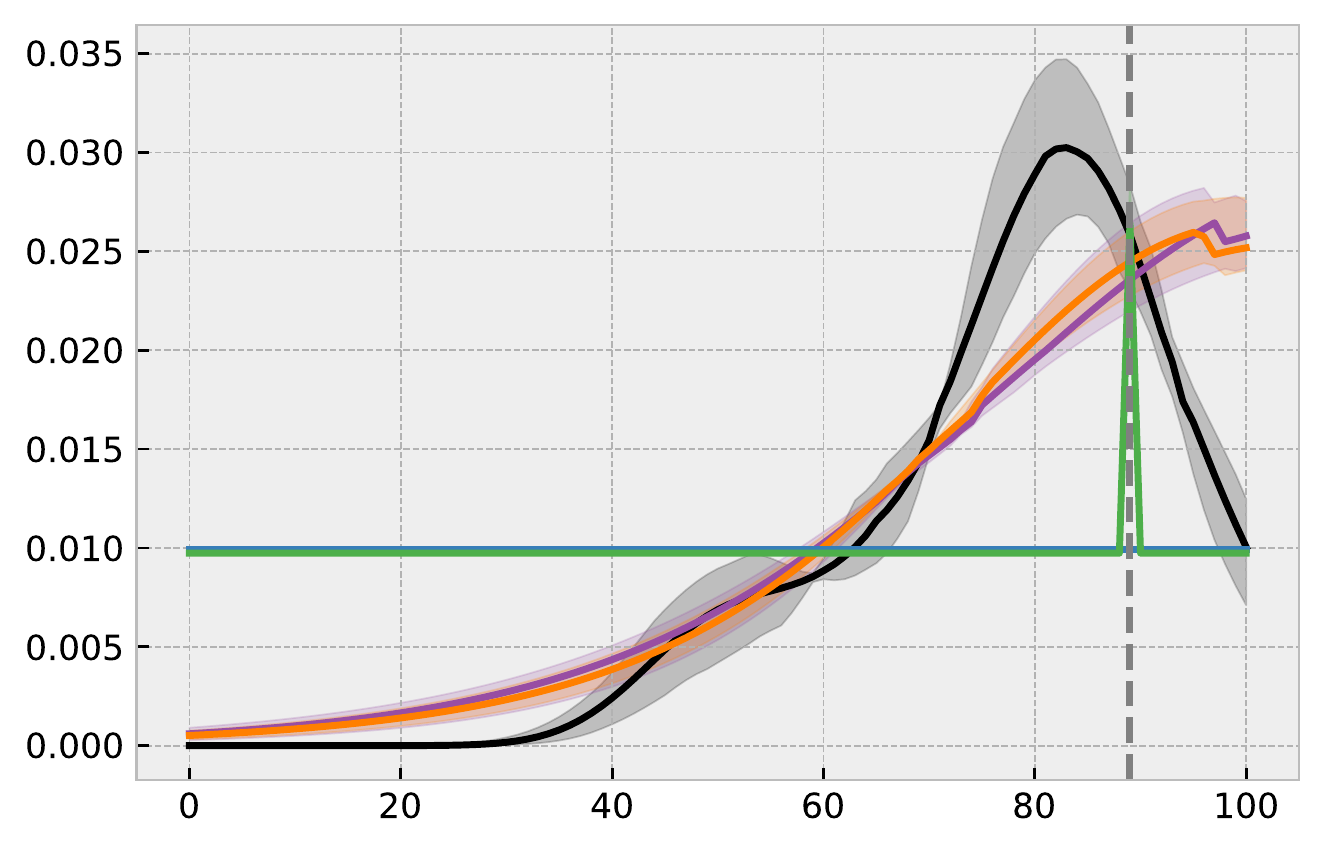}
  \caption{(70,10)}
  \label{figRequested70}
\end{subfigure}\hfil
\caption{Ultimatum game. The darker lines indicate the mean result from the 5x2 cross-validation. The shaded regions indicate $\pm$ one standard deviation. The out-of-sample data are shown as the black line. The proposed Quantal Hierarchy model is the purple line. Quantal Response Equilibrium is the orange line, level-$k$ is the blue line, and Cognitive Hierarchy is the green line. The Nash equilibrium solution is indicated as the vertical dashed grey line.
}\label{figUltimatum}
\end{figure}

\textbf{Ultimatum}
The results are presented in \cref{figUltimatum} for the ultimatum game. We see the QH model explains important features of the observed experimental behaviour. For example, with higher $V_1$, Player 1 is likely to make a larger initial request (\cref{figRequested70} vs \cref{figRequested10}). Perfect rationality does not capture this (with the Nash equilibrium remaining unchanged), whereas the QH model suggests higher initial requests (due to higher $V_1$ if the request is rejected). The QRE model and the QH model behave similarly here. The reason for this similar behaviour is because the ultimatum game is nearly a single-stage decision, meaning the QH model ``collapses" to QRE. This is also confirmed in the fitted parameters \cref{tblParams}, with the $(10,10)$ and $(70,10)$ having almost identical values for $\beta$ and $\lambda$, and QH having relatively high values of $\gamma$, meaning the differences between Player 1 and Player 2 processing resources are small. However, something interesting happens in the $(10,60)$ case, where QRE cannot capture the distribution. For example, if we look to the right of the rational rejection region for Player 2 (\cref{figUltimatumStep}), we do not see any acceptances in (10,10) or (70,10). Whereas, if we look in the rational rejection region of (10,60), we see several acceptances. This behaviour is irrational, because if the player rejected the request, they would have received a higher payoff. In contrast, Player 1's initial requests are relatively rational, with the peak occurring around the rational request of $40$. This mismatch in player rationality is captured under the proposed model with a small $\gamma$, i.e., a large discount in processing resources. This mismatch in rationality cannot be captured with the standard QRE, which assumes a fixed $\beta$ for both players. These results motivate the discounting of player resources, which can capture heterogeneous information processing resources between the two players. Alternate forms of QRE, such as Heterogenous QRE have also been proposed to deal with such dilemmas \citep{rogers2009heterogeneous}, however, this is captured natively by the QH model.

The level-$k$ model fails to capture any of the trends, with the uniform level-$0$ case being the best fit. The cognitive hierarchy model predicts a representative spike at the rational capacity in the $(70,10)$ and $(10,60)$ cases, however, it is still clearly outperformed by both QRE and QH.

These results confirm the usefulness of QH. When both players behave with similar levels of rationality, this can be captured with $\gamma \to 1$, and the model acts the same as QRE. However, when there is a mismatch in player rationality, this heterogeneity can be captured directly with $\gamma < 1$, which became most pronounced in the $(10,60)$ case.


\begin{figure}[htb]
\centering
\begin{subfigure}{0.24\textwidth}
  \includegraphics[width=\linewidth]{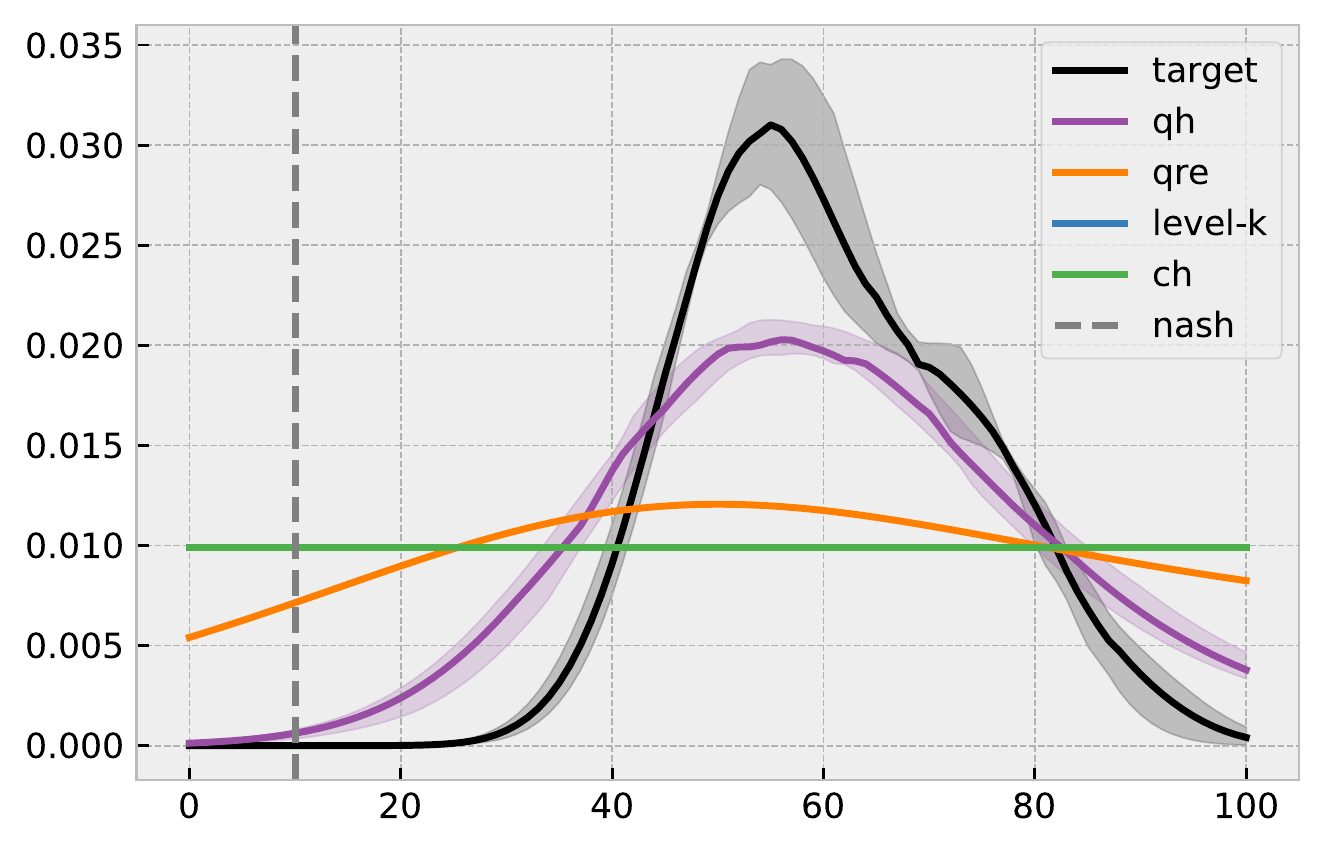}
  \caption{D=0.9}
\end{subfigure}\hfil
\begin{subfigure}{0.24\textwidth}
  \includegraphics[width=\linewidth]{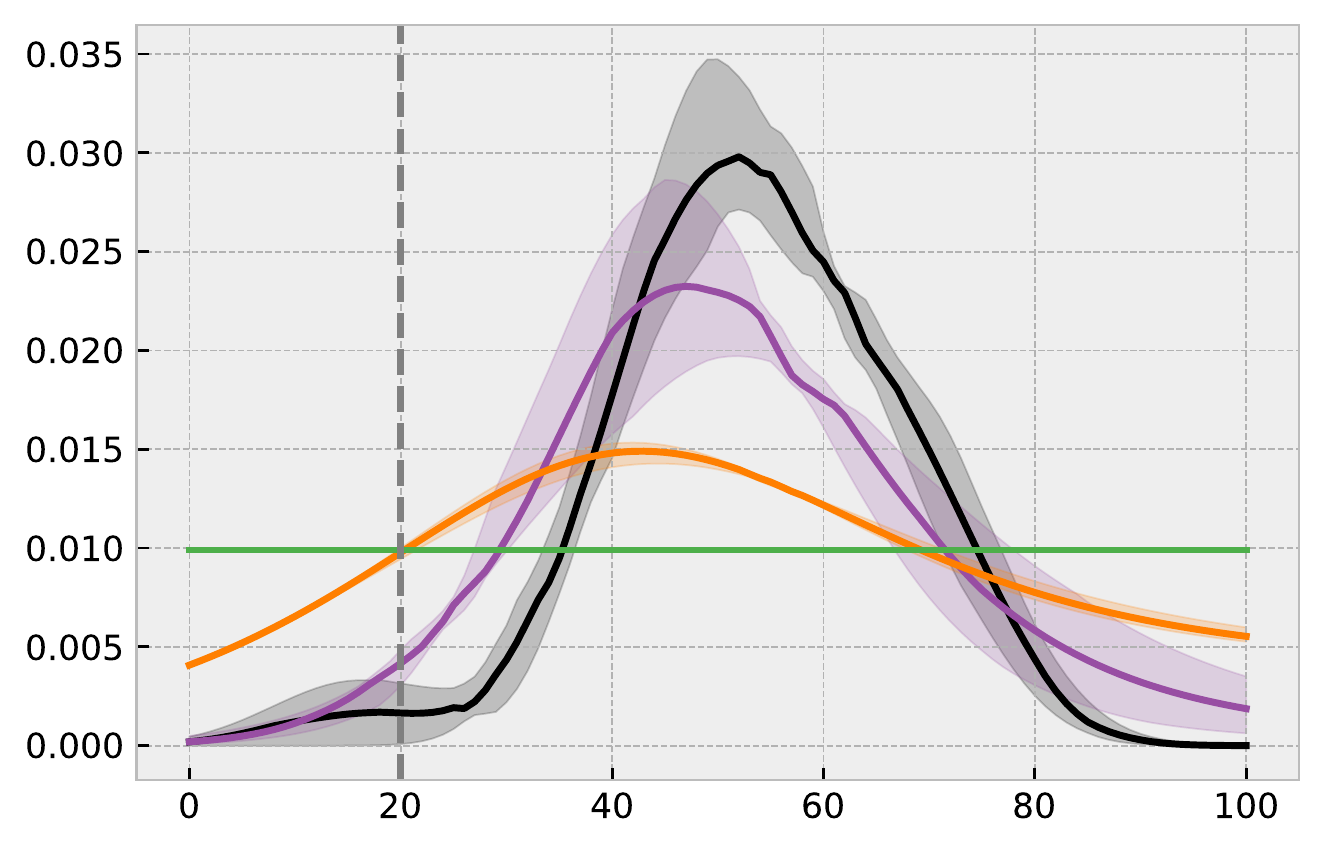}
  \caption{D=0.8}
\end{subfigure}\hfil
\begin{subfigure}{0.24\textwidth}
  \includegraphics[width=\linewidth]{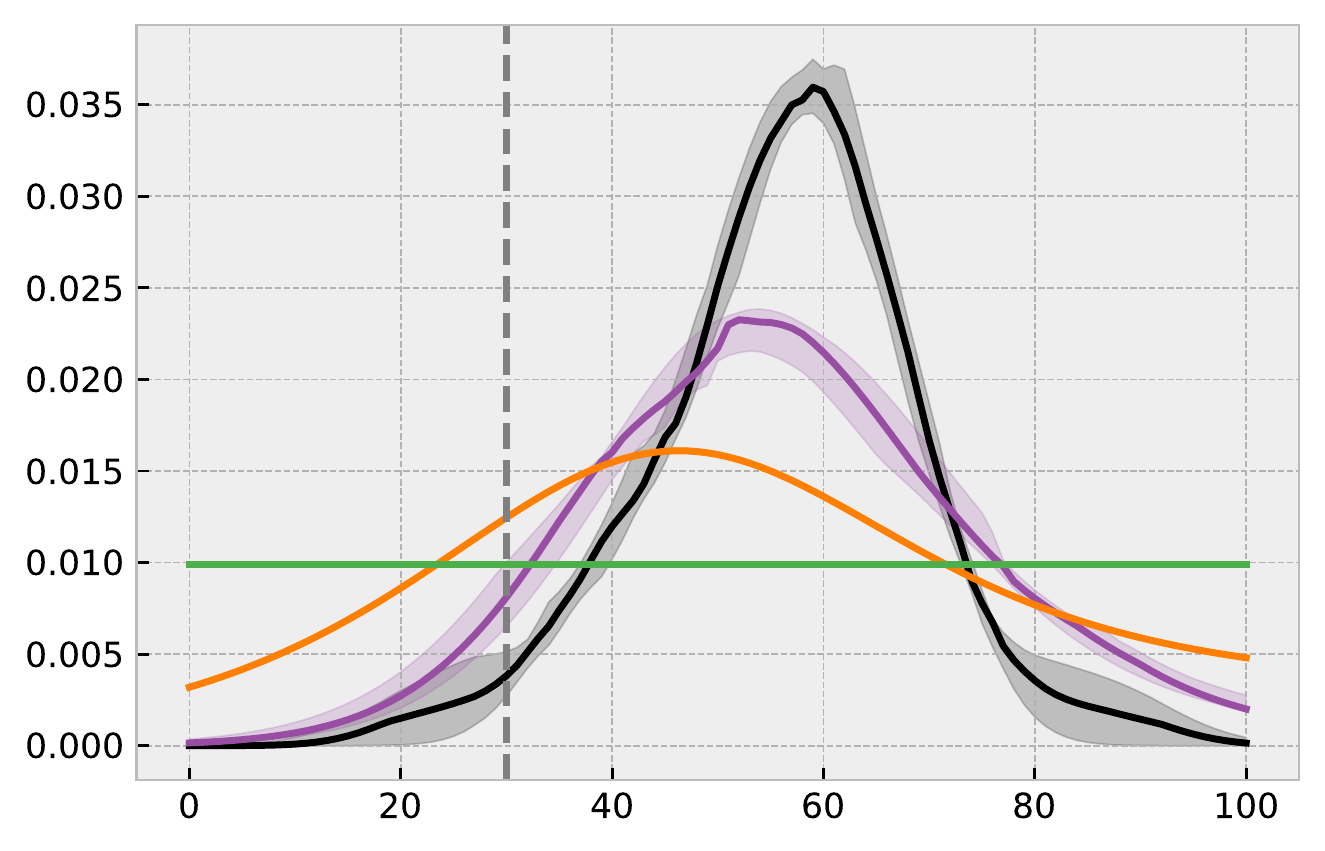}
  \caption{D=0.7}
\end{subfigure}\hfil
\begin{subfigure}{0.24\textwidth}
  \includegraphics[width=\linewidth]{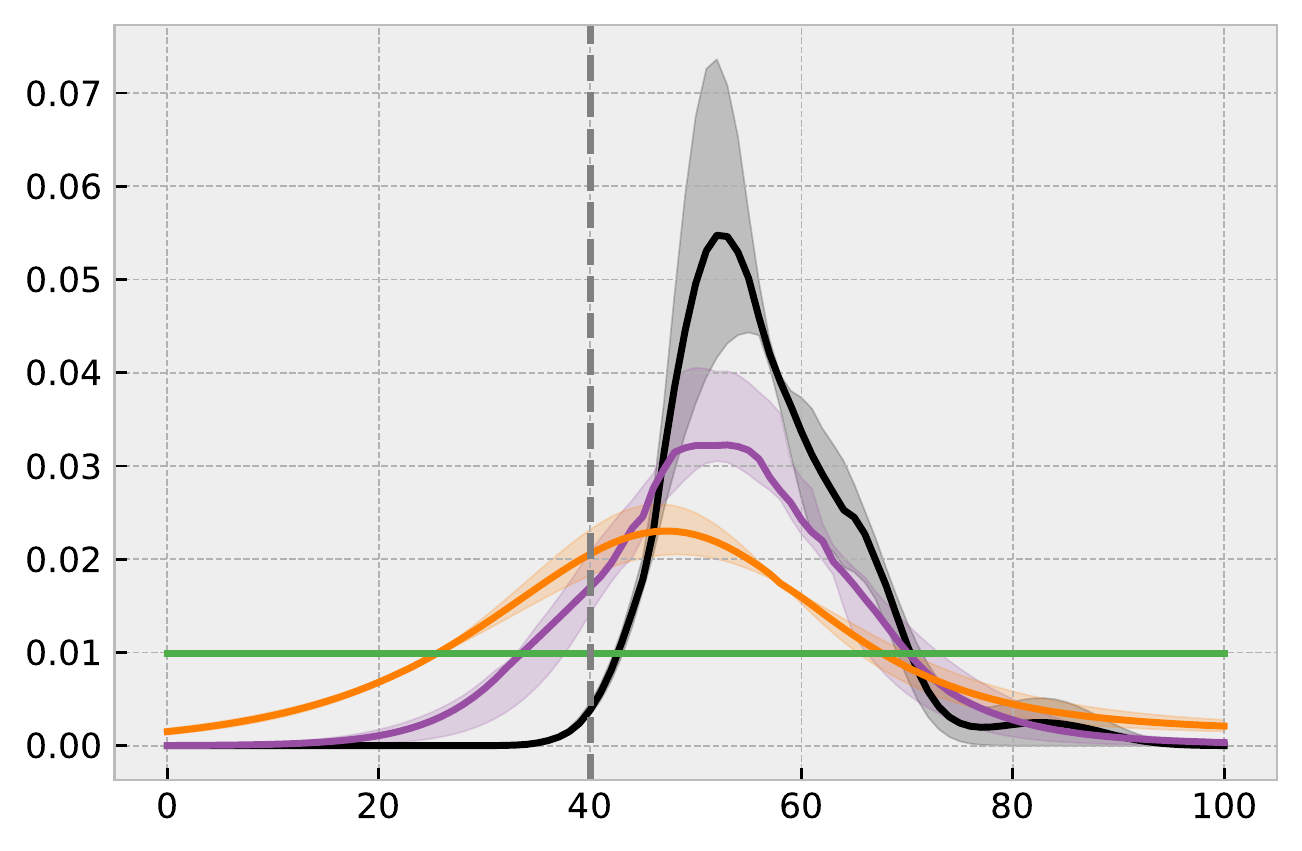}
  \caption{D=0.6}
\end{subfigure}\hfil
\begin{subfigure}{0.24\textwidth}
  \includegraphics[width=\linewidth]{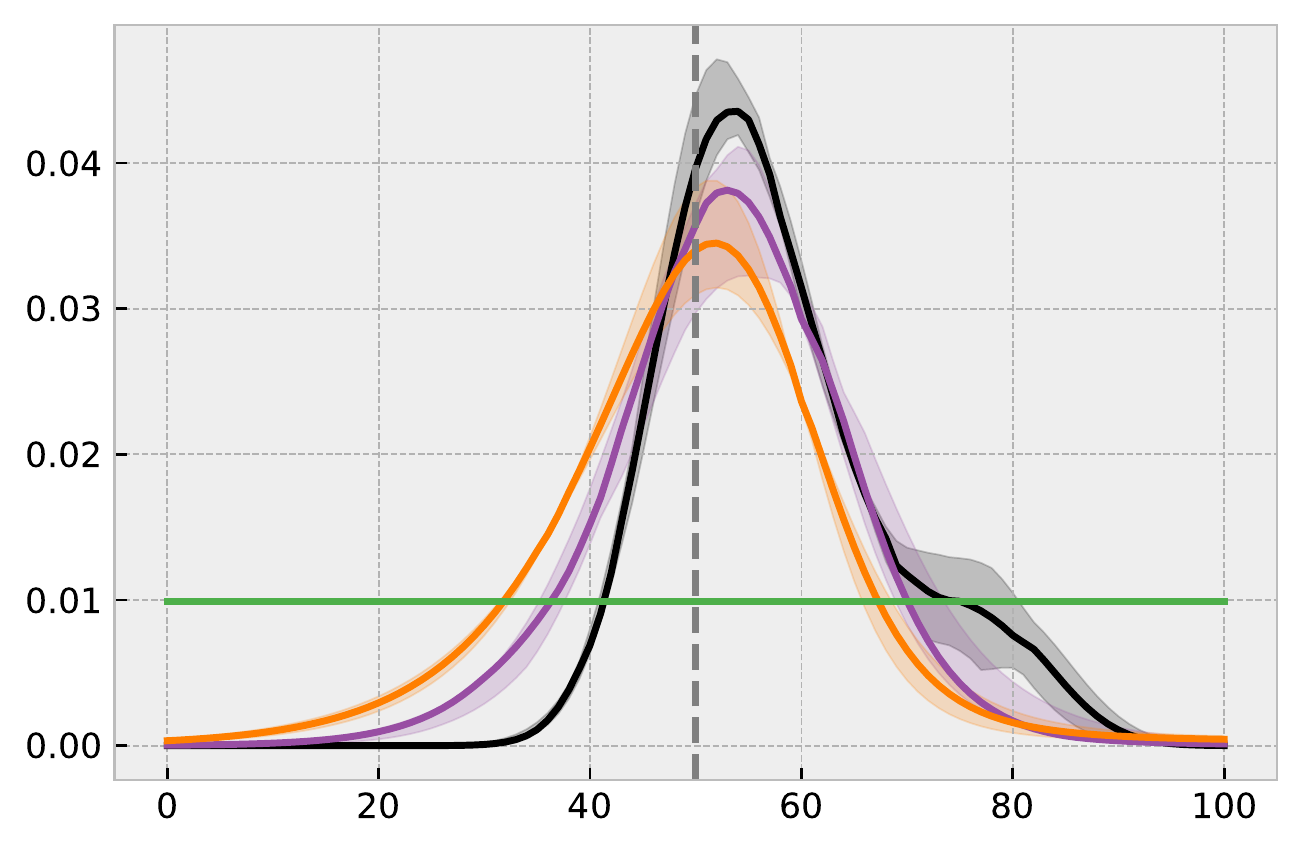}
  \caption{D=0.5}
\end{subfigure}\hfil
\begin{subfigure}{0.24\textwidth}
  \includegraphics[width=\linewidth]{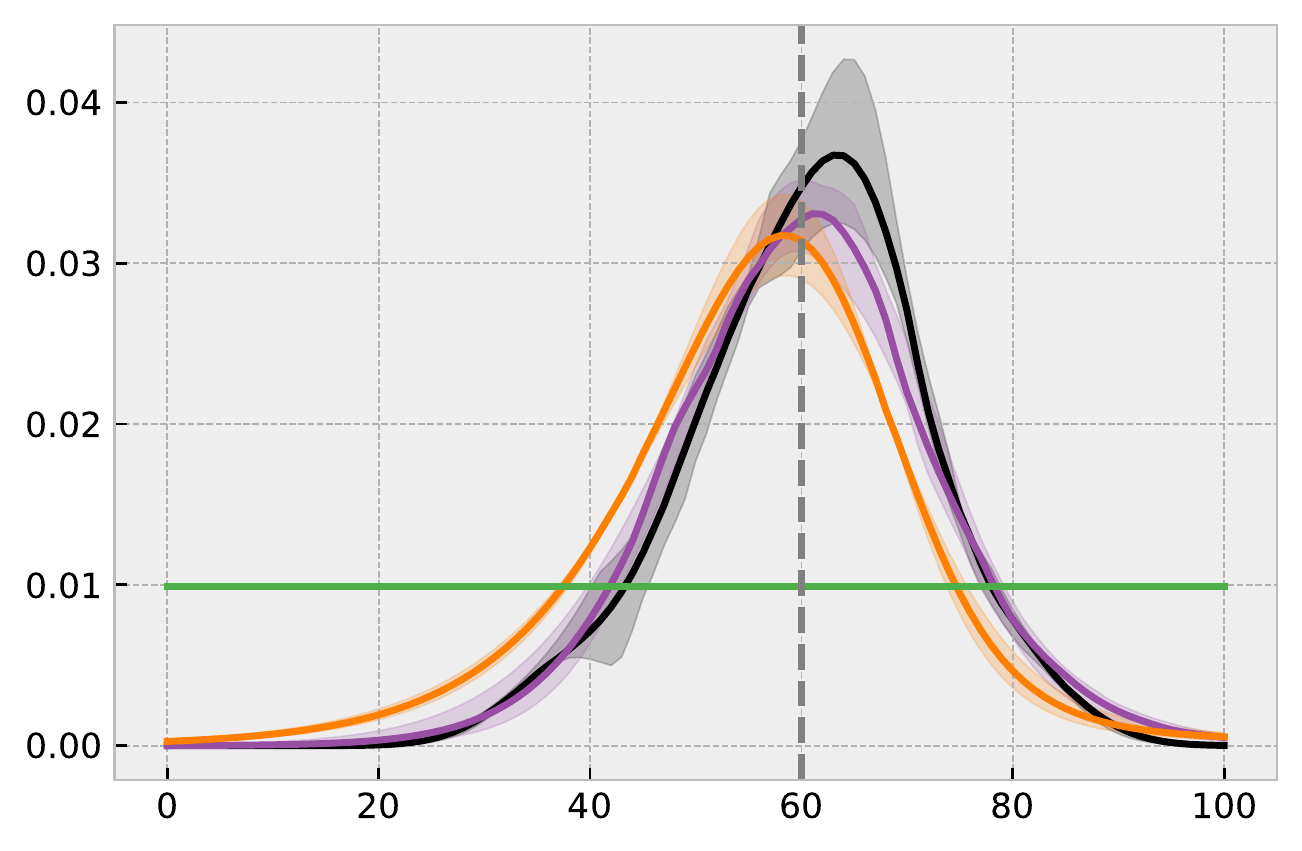}
  \caption{D=0.4}
\end{subfigure}\hfil
\begin{subfigure}{0.24\textwidth}
  \includegraphics[width=\linewidth]{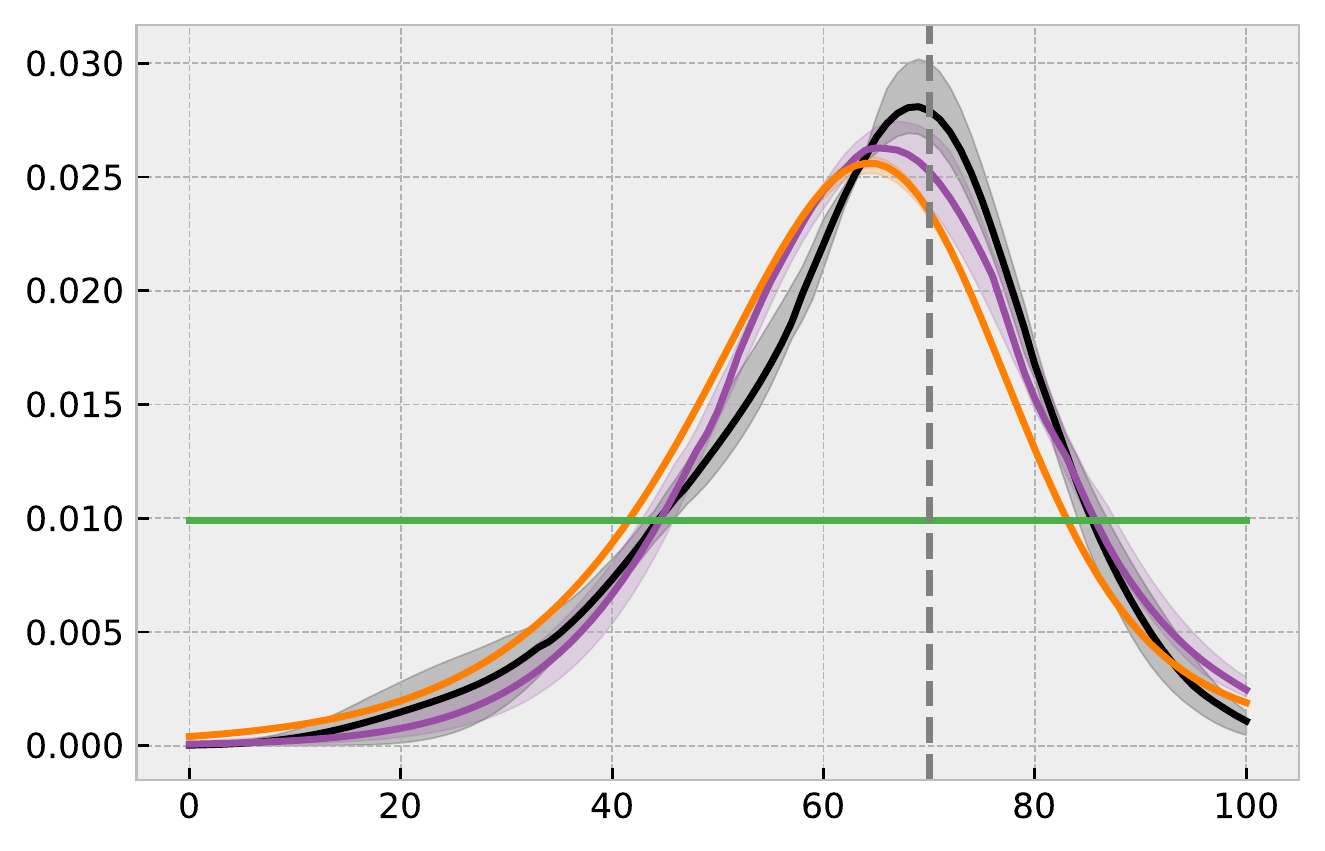}
  \caption{D=0.3}
\end{subfigure}\hfil
\begin{subfigure}{0.24\textwidth}
  \includegraphics[width=\linewidth]{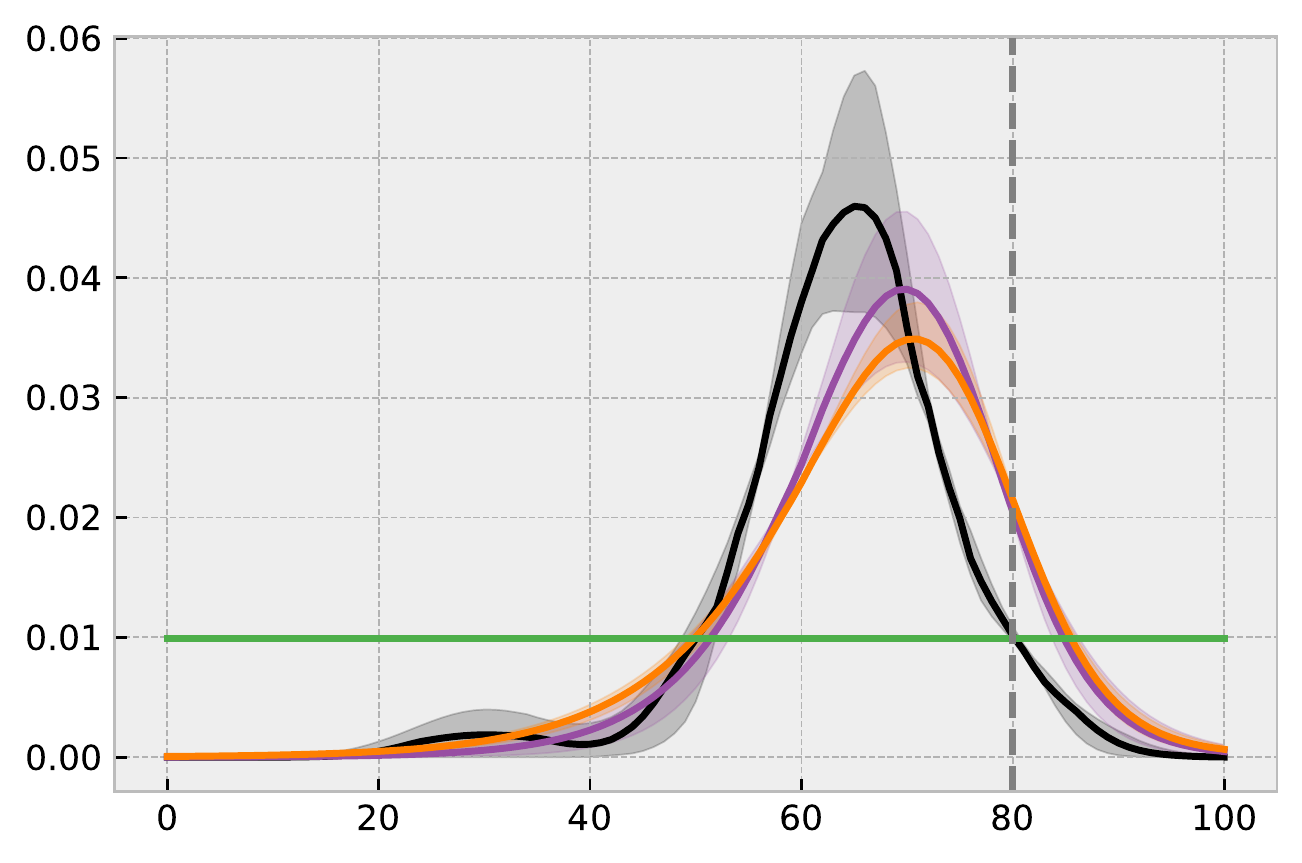}
  \caption{D=0.2}
\end{subfigure}\hfil
\caption{Two-stage bargaining game. The darker lines indicate the mean result from the 5x2 cross-validation. The shaded regions indicate $\pm$ one standard deviation. The out-of-sample data are shown as the black line. The proposed Quantal Hierarchy model is the purple line. Quantal Response Equilibrium is the orange line, level-$k$ is the blue line, and Cognitive Hierarchy is the green line. The Nash equilibrium solution is indicated as the vertical dashed grey line.}\label{figTwoStageBargain}
\end{figure}

\textbf{Two-stage} While we saw similar behaviour between QRE and QH in the ultimatum game, under the two-stage game, the differences between the approaches become more pronounced due to the longer game tree. Under such conditions, the usefulness of discounting future paths (and relaxing mutual consistency) becomes more noticeable. The QH model convincingly outperforms QRE across the experimental results with the small disagreement penalties \footnote{These are referred to as ``discount" rates in \cite{binmore2002backward}. We have used the term disagreement penalties to avoid confusion with the information processing ``discount" parameter $\gamma$.} (i.e., $D>0.5$), and still generally outperforms QRE for the larger disagreement penalties (i.e., $D \leq 0.5$), although the two methods become closer.

With the larger disagreement penalties (i.e. smaller $D$), the experimental data are closer to the perfectly rational case, as indicated with the peaks corresponding roughly to the rational request in \cref{figTwoStageBargain}. This distribution around the rational request is precisely the premise QRE is founded on, so QRE achieves adequate performance. However, with the smaller disagreement penalties (larger $D$, top row of \cref{figTwoStageBargain}), the distribution is not centred around the rational request, meaning QRE struggles to capture such phenomena. In contrast, the QH approach is robust to this shift due to the relaxation of mutual consistency, and is able to capture the varying distributions regardless of whether they are approximating the best-response case.

{
\subsection{Results Summary}

The Quantal Hierarchy method consistently performed well out-of-sample in all games, ranking the best overall and achieving either the first or second position in every game. The results analysis was categorised into two game types: sequential and simultaneous games, where reasoning is represented as an extensive-form game tree with depleting information-processing resources. Although the representation worked well in both game types, it showed more improvement over alternative methods in sequential games. This improvement in sequential games can be attributed to the discount parameter that captures the heterogeneity of players, allowing for different information processing resources between the players at each stage, relaxing mutual consistency, which was crucial in bargaining games.

On the other hand, in simultaneous games, the approach aims to fit a representative distribution of the entire group, but it can struggle to capture the entire distribution of players, particularly when they exhibit widely varying levels of rationality, as in certain versions of the beauty-contest game.  This highlights a potential limitation of the approach when attempting to capture multimodal distributions with varying levels of rationality, such as a bi-modal distribution with beginners and experts. To address this limitation, multiple versions of the model may need to be fitted, such as one for beginners and one for experts, as modelling more rational play narrows the distribution to the rational prediction and modelling less rational play widens the distribution to account for larger errors, as demonstrated in \cref{figKLevelApproximation}. However, despite this potential limitation, the method still performed exceptionally well overall.

}

\section{Discussion and Conclusions}\label{secConclusions}

The assumption of perfect rationality amongst players is violated in numerous experimental settings, particularly in non-repeated games. In this work, we utilised experimental datasets for several games, showing that the equilibrium behaviour is often a poor predictor of the observed actions. The proposed Quantal Hierarchy model offers a concise alternative representation, relaxing some traditional game-theoretic assumptions underlying rationality. The model is a good fit for experimentally observed behaviour on a range of canonical economic games, outperforming existing bounded rationality approaches on out-of-sample validation.


In the QH model, we represent higher-order reasoning as pseudo-sequential decision-making. At each level, players may reason erroneously, and this error grows the deeper one reasons (i.e., it becomes more difficult to reason about reasoning). The magnitude of the errors is governed by $\beta$, with $\beta=0$, players do not perform any reasoning, and with $\beta \to \infty$ players reason perfectly. Parameter $\beta$, therefore, relaxes the best response assumption of players at each level of reasoning. 

Decreasing $\beta$ at each level of reasoning was shown to work well on a wide variety of games, reinforcing the assumption that players cognitive abilities decrease throughout the depth of reasoning. This reduction in player cognition is captured with $\gamma$, introducing an implicit hierarchy of players, relaxing the mutual consistency assumption. Representing this hierarchy of players as extensive-form game trees allowed for an information-theoretic representation, where lower-level players are assumed to make more significant playing errors (constrained by lower information processing resources). With a single-step decision, this recovers the Quantal Response Equilibrium model. With multi-stage decisions, we recover an approximation of a generalised level-$k$ formulation, where at each step, players are assumed to have higher resources and reasoning ability than players below themselves, but may still play erroneously.

Similar to QRE, the resource parameter $\beta$  is problem dependent, and depends on the payoff magnitude \citep{MCKELVEY2000523}. This opens an area of research analysing whether a normalised $\beta$ can be used to measure problem difficulty, or whether some relationship holds between the the experimentally fitted $\beta$ and the $\beta$ which corresponds to the Nash solution. For example, a question arises if a normalised $\beta$ can provide insights across games, and if so, can this average distance to the Nash solution be generally useful across games. A similar consideration is given to whether such payoff perturbations in QRE can be related across different games \citep{haile2008empirical}, and whether the boundedness parameter can be endogenised \citep{FRIEDMAN2020620}. 

There is a clear relationship between the decision-making components proposed in this work and the decision-making in multi-agent systems, such as agent-based models (ABMs) and multi-agent reinforcement learning (RL) approaches. For example, \cite{wen2019modelling} outline a novel framework for hierarchical reasoning RL agents, which allows agents to best respond to other less sophisticated agents based upon level-$k$ type models. Likewise, \cite{latek2009bounded} propose a recursion based bounded rationality approach for ABMs. Replacing the agents in these multi-agent approaches with the informationally constrained agents presented in our work provides a distinct area of future research, where we could examine the resulting dynamics and out-of-equilibrium behaviour from heterogeneous QH agents. 

In summary, we proposed an information-theoretic model for capturing higher-order reasoning for boundedly rational players. Bounded rationality is achieved in the model by the relaxation of two central assumptions underlying rationality, namely, mutual consistency between players and best response decisions. {Through relaxing these assumptions, we showed how the predictions from the proposed Quantal Hierarchy model align well with the experimentally observed human behaviour in a variety of canonical economic games.}

\bibliographystyle{apalike}
\bibliography{bib}

\clearpage

\appendix

\section{Model Fitting}\label{secModelFit}

Each model is fit to the training portion of the data (from 5x2 cross-fold validation), using Bayesian hyperparameter optimisation \citep{bergstra2013making}. Every model is given 1000 evaluations for a fair comparison. With level-$k$, due to the integer parameter, rather than Bayesian optimisation, we instead perform an exhaustive search for $k \in [0,1,\dots,100]$, noting that this is an extensive range of $k$, easily capturing standard $k$'s reported in the literature. The fitted parameter values with the lowest mean squared error between the predictions and the training values are selected. The out-of-sample (testing) portion is never seen by the optimisation process and is only used for evaluation after the parameter optimisation has been complete.

For Quantal Response Equilibrium and Quantal Hierarchy, we sample from the range $0 \leq \beta < 100$. While $\beta$ is unbounded, we find this upper bound to be more than enough with no fitted values coming close to this upper threshold. For $\gamma$, this is bounded $0\leq \gamma \leq 1$. For Cognitive Hierarchy, we sample from the range $0 \leq \tau < 10$, which despite $\tau$ being unbounded, again provides a more than sufficient range for the experimental data, and covers common $\tau$'s reported in literature \citep{camerer2010behavioural,camerer2004cognitive}.

For the beauty contest games, as well as the bargaining games, due to the large action space ($a \in [0,\dots,100]$), rather than using the raw data directly, a fitted Gaussian kernel density estimate of the training and testing data is used to account for the large action space and the relatively small number of observations. Scott's rule is used to determine the bandwidth automatically \citep{scott2015multivariate}, and we validate the robustness of this rule choice in \cref{appendixDensityEstimate}. The same kernel density estimates are used across all methods to ensure fair comparisons. For the remaining game classes, the action space is sufficiently well sampled from the observations, so no density approximation is required. 

\section{Special Cases}
\subsection{Backwards Induction}\label{secBackwardsInduction}
Backwards induction can be recovered as a limiting case of the proposed model.  We can see this as follows from \cref{eqDiscountDecision} (and the expansion process from \cref{eqExpansion}), noting that softmax $e^{\beta U[a]} / Z$ converges to $\argmax$ with $\beta \to \infty$:

\begin{equation}\label{eqBellmanEquivalence}
\begin{split}
  f[a_k \mid a_{<k}] &=
    \frac{1}{Z_k} \times \underbrace{Z_{k+1} ^ {1 / \gamma}}_{\text{Future Contribution}} \times \underbrace{e ^{\beta\gamma^{k} U[a_k \mid a_{<k}]}}_{\text{Current Payoff}} \\
    &= 
    \argmax_{a_k\in A_k} \left( \underbrace{U[a_k \mid a_{<k}]}_{\text{Current Payoff}} +  \underbrace{V_{k+1}}_{\text{Future Contribution}} \right)\\
\end{split}
\end{equation}

where $V_{k+1}$ is derived recursively based on choosing $a_k$. Backward induction assumes that all future decisions will be rational, i.e., at each stage, players choose rationally. 

\section{Canonical Games and Experimental Data}\label{secExamplesAppendix}

\subsection{Market Entrance/El Farol Bar Problem}\label{secAppendixMarket}

The market entrance game was outlined in the context of cognitive hierarchies in \cite{camerer2004cognitive}, and has also been considered in prior studies, e.g., \cite{rapoport1998equilibrium}. This game is fundamentally similar to the El Farol bar problem of \cite{arthur1994inductive}, and minority games of \cite{challet2013minority}. A player will profit (enjoy) in the market (bar) if less than $d \times N, d \in [0,1]$ players also enter the same market (bar).

For this work, we use the experimental data from \cite{camerer2011behavioral}, specifically the results originally presented in \cite{sundali1995coordination}. The payoff for staying out is fixed 
\begin{equation}
    U[\text{stay out}_k] = 1
\end{equation}
However, the payoffs for entering are dependent on the total demand from the other (lower-level) players and a preferential capacity $c= d \times N$:
\begin{equation}
    \begin{split}
        U[\text{enter}_k] &=  1+2(c-f[\text{enter}_{k+1}])
    \end{split}
\end{equation}
There were $N=20$ subjects, and various $c$'s trialled $c \in [1,3,5\dots,19]$.

\subsubsection{Comparison Methods}

We can represent this pseudo-sequential structure \citep{camerer2004cognitive} as an extensive-form game, with each level of reasoning forming a new node in the game tree.

\paragraph{Level-$k$} Under this configuration, a level-$0$ player is assumed to randomise, i.e., enter or stay out with equal probability (the same level-$0$ configuration is used for the naive player in the QH model). A level-$1$ player exploits this and attends the bar if $d > 0.5$, or stay home with $d < 0.5$, at $d=0.5$ the player is indifferent and would attend with $50\%$ probability. Level-$2$ players then base their decision assuming other players are level-$1$, and enter only if the level-$1$ players underestimated the expected capacity. Likewise, level-$3$ players base their decision on reasoning about level-$2$. Level-$k$ behaviour necessitates step functions in the response, where players only enter at a capacity $c$ if they believe lower-level thinkers have over or under entered. 

\paragraph{Cognitive Hierarchy} Rather than assuming all players are at $k-1$, the cognitive hierarchy model fits a distribution to these $k$ players, and best responds according to this distribution of lower level thinkers. Following \cite{camerer2004cognitive}, we use the Poisson distribution.

\paragraph{Quantal Response Equilibrium} To derive the (mixed strategy) quantal response equilibrium, we use the logistic function of the differences in payoffs (between enter and stay out) as the distribution function, with numeric estimations for the fixed-point solution since no analytical solution exists, following \cite{goeree2016quantal} (Section 2.2.2 and Section 8) and \cite{goeree2005explanation}.

\begin{figure*}[ht]
\centering
    \begin{subfigure}[t]{0.18\textwidth}
        \centering
        \includegraphics[width=\textwidth]{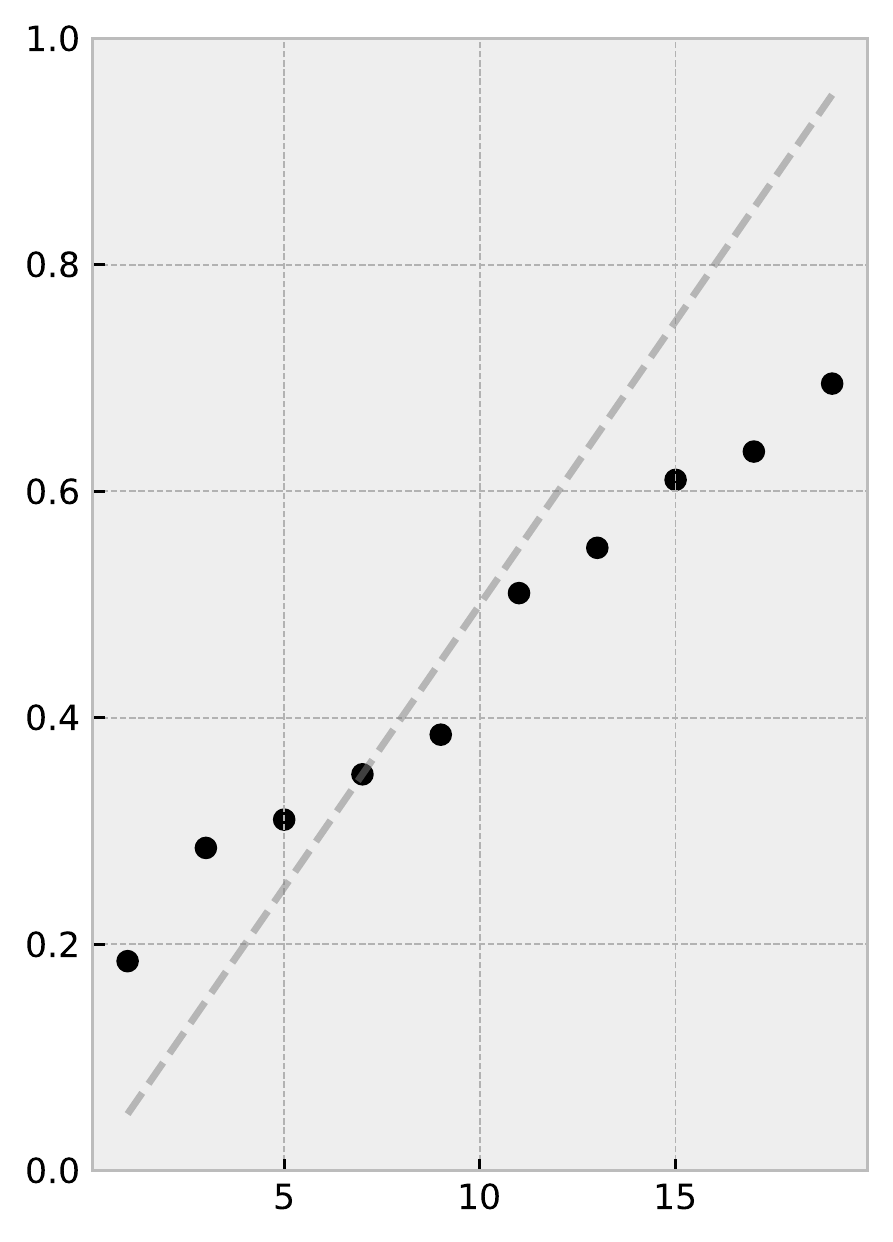}
        \caption{Block 1}
    \end{subfigure}
    \begin{subfigure}[t]{0.18\textwidth}
        \centering
        \includegraphics[width=\textwidth]{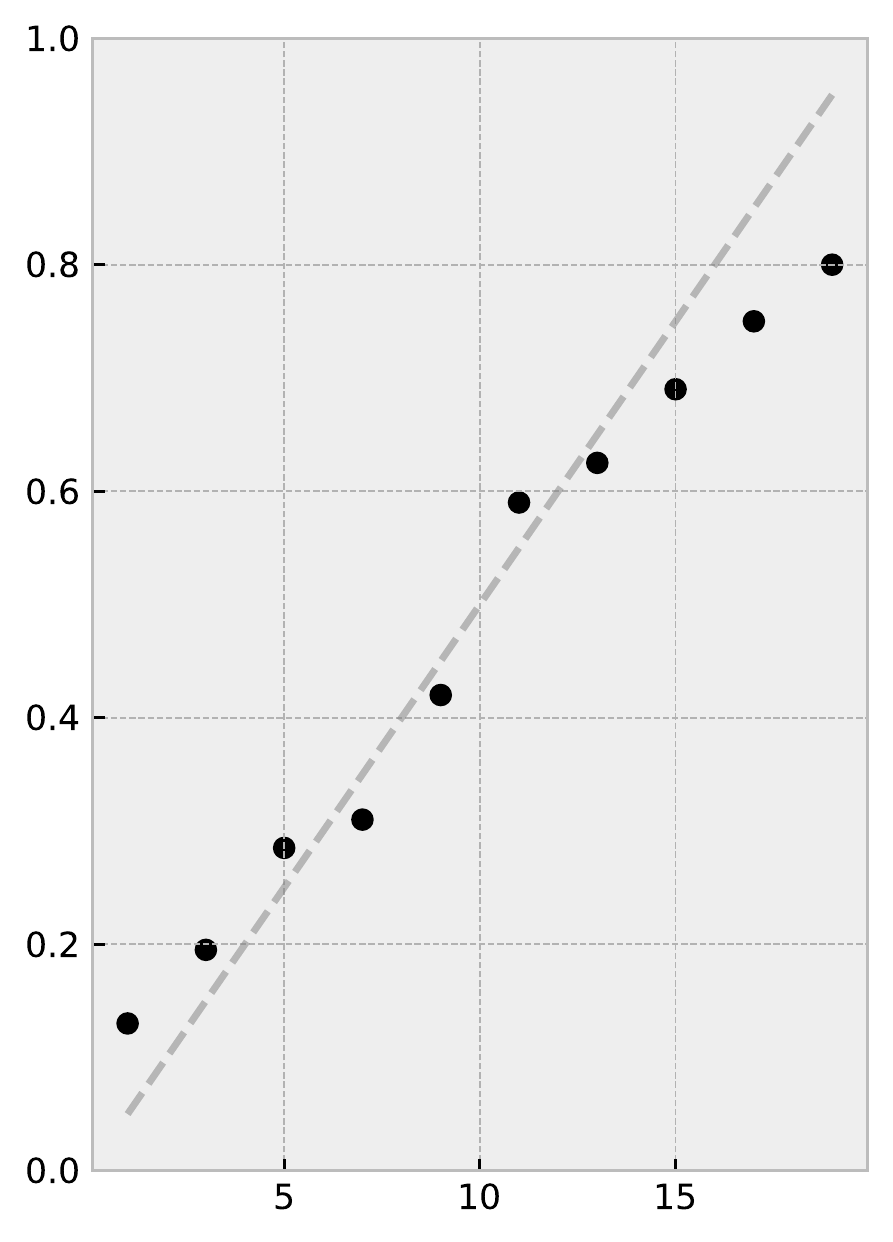}
        \caption{Block 2}
    \end{subfigure}
    \begin{subfigure}[t]{0.18\textwidth}
        \centering
        \includegraphics[width=\textwidth]{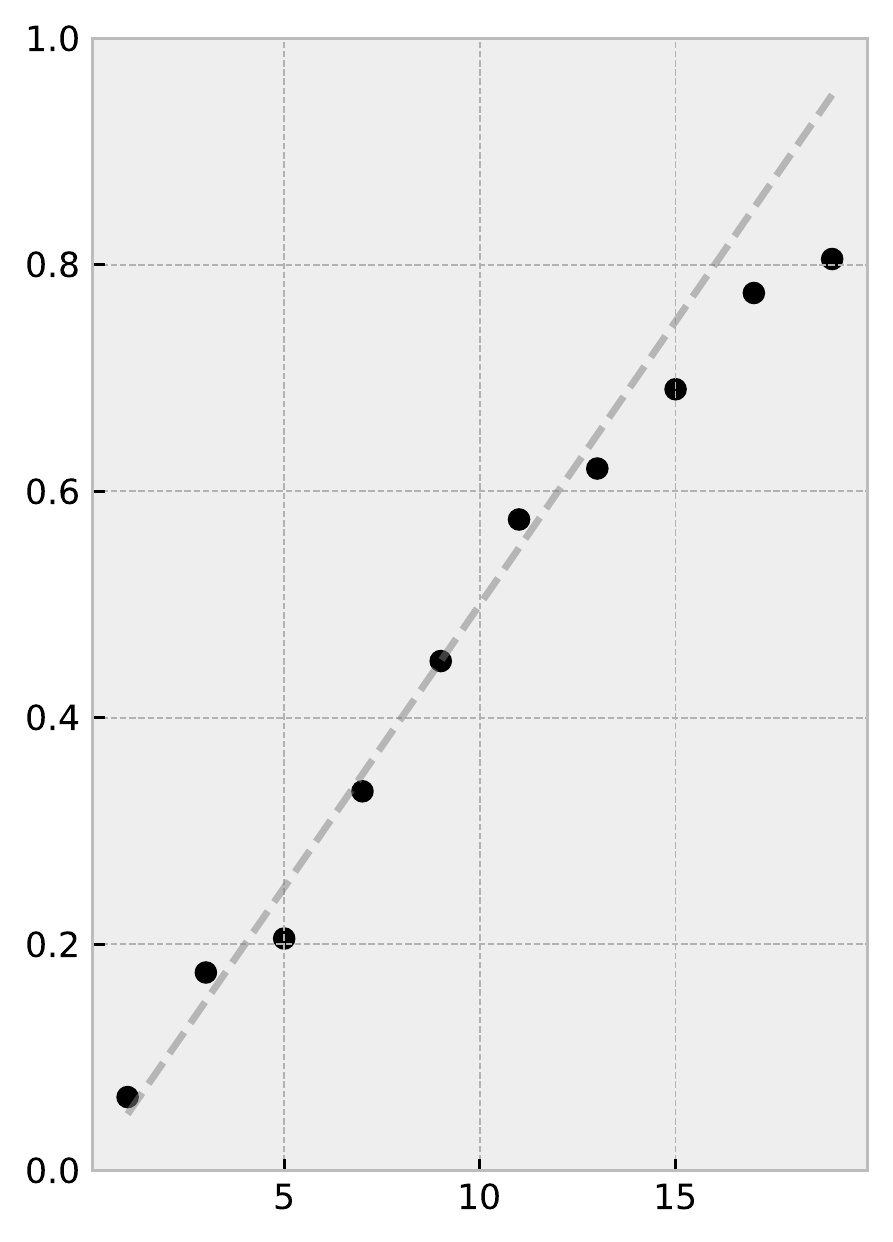}
        \caption{Block 3}
    \end{subfigure}
        \begin{subfigure}[t]{0.18\textwidth}
        \centering
        \includegraphics[width=\textwidth]{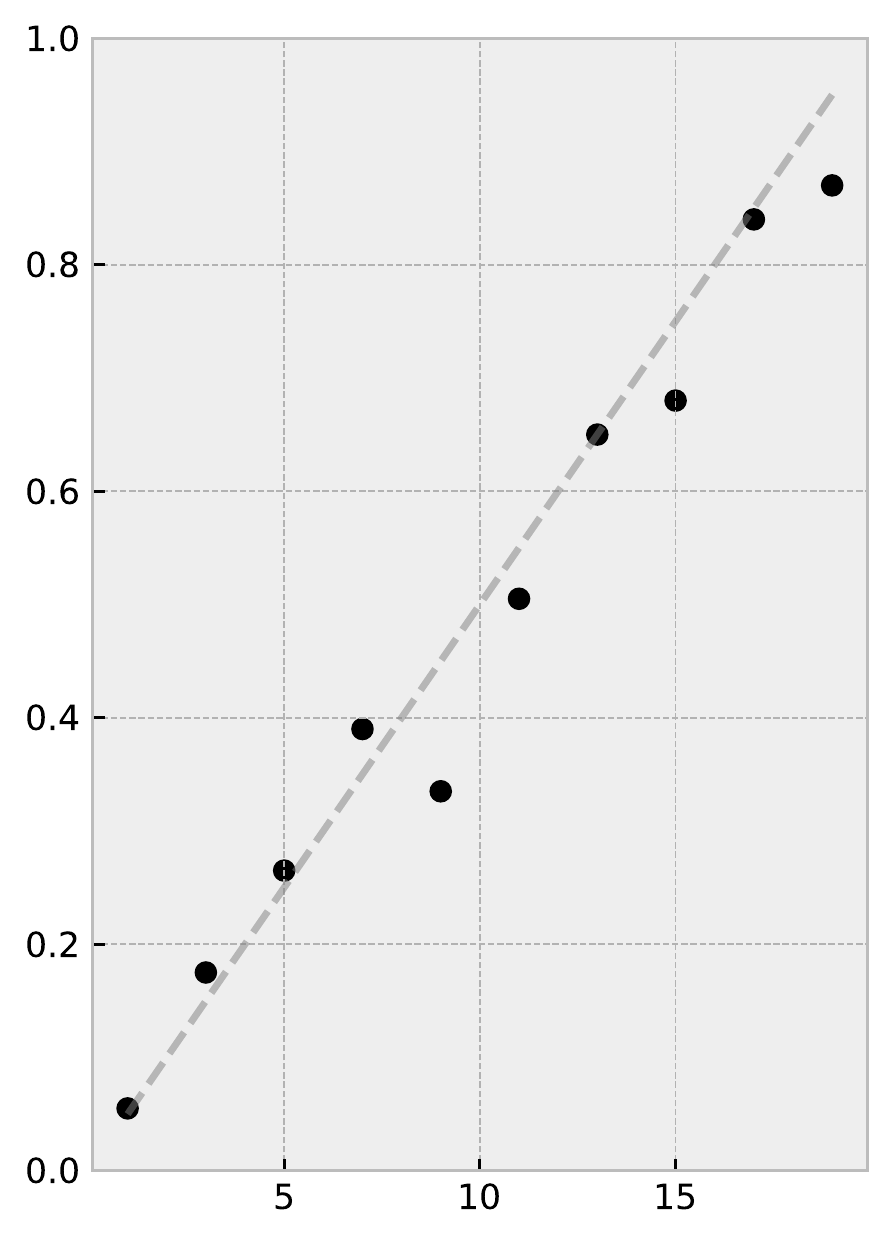}
        \caption{Block 4}
    \end{subfigure}
        \begin{subfigure}[t]{0.18\textwidth}
        \centering
        \includegraphics[width=\textwidth]{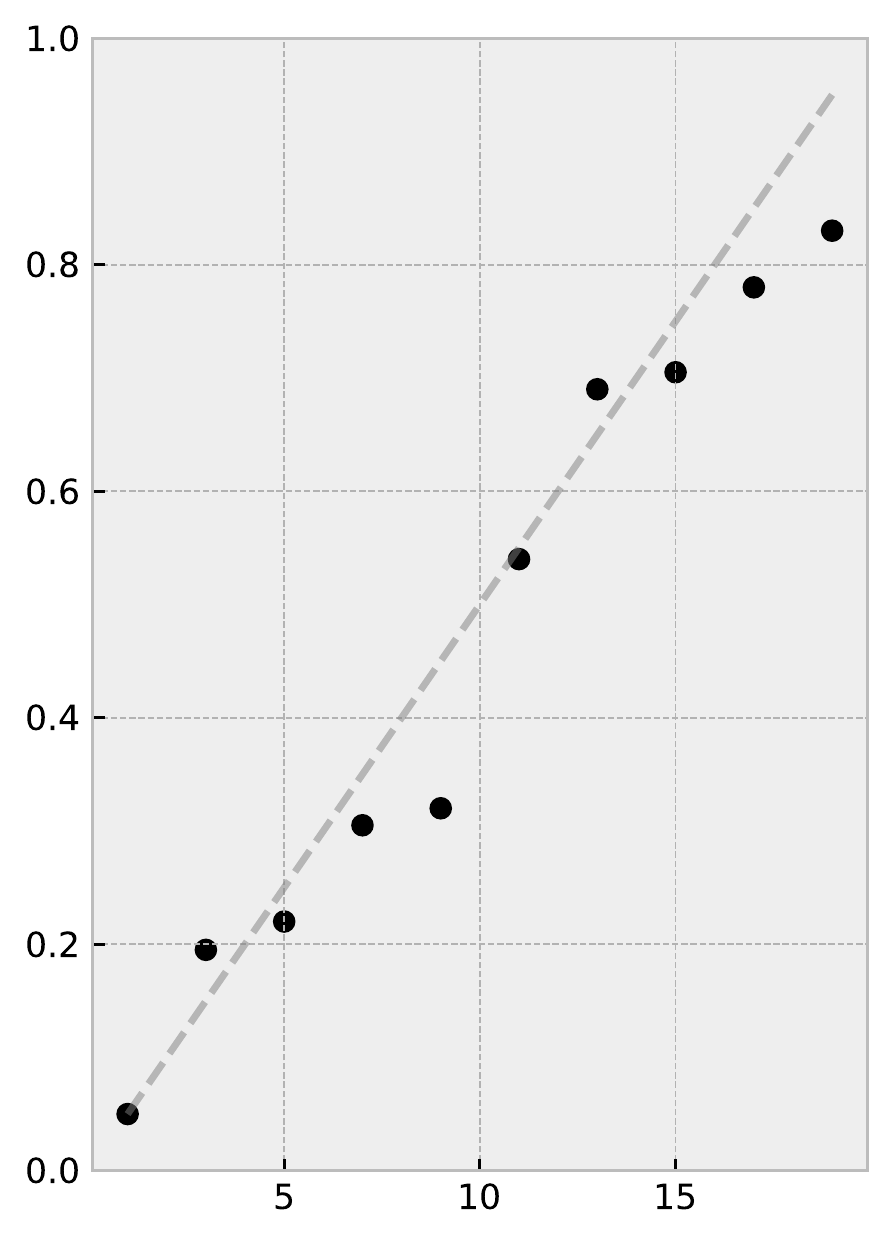}
        \caption{Block 5}
    \end{subfigure}
    \caption{Market Entrance game at various blocks using the experimental data from \cite{sundali1995coordination}. We see over time, the experimental data becomes closer to the equilibrium of perfect attendance (dotted gray line).}
    \label{figMarketEntranceExperimental}
\end{figure*}

\subsection{Beauty Contest}\label{secAppendixBeauty}
\cite{keynes1937general, keynes2018general} originally formulated the beauty contest game as follows. Contestants are asked to vote for the six prettiest faces out of a selection of 100. The winner is the contestant who most closely picks the overall consensus. A naive (level-$0$) strategy is to choose based on personal preference. A level-$1$ strategy is to choose as if everyone is choosing on personal preference, so the player chooses whom they think others will find most desirable. A level-$2$ strategy is then for players to choose whom they think that others will think others will choose, and so on (with Keynes believing there are players who ``practise the fourth, fifth, and higher" levels). The game was originally outlined to highlight how investors are not necessarily driven by fundamentals but rather by anticipating the thoughts of others. 

An extension of the game is the $p$-beauty contest of \cite{moulin1986game}, where contestants are asked to guess fraction $p \in [0\dots1]$ (commonly $p=\frac{2}{3}$) of the average value of the other competitors guesses within the range $[0,\dots100]$. The Nash equilibrium dictate that every player should choose $0$. However, experimentally this is not the case, and players act boundedly rational \citep{nagel1995unraveling}. Such a game shows out-of-equilibrium behaviour and motivates the modelling of such decisions in a finite-depth manner \citep{aumann1992irrationality, stahl1993evolution, binmore1987modeling, binmore1988modeling}. 

\begin{figure}[htb]
\centering
\begin{subfigure}{0.3\textwidth}
  \includegraphics[width=\linewidth]{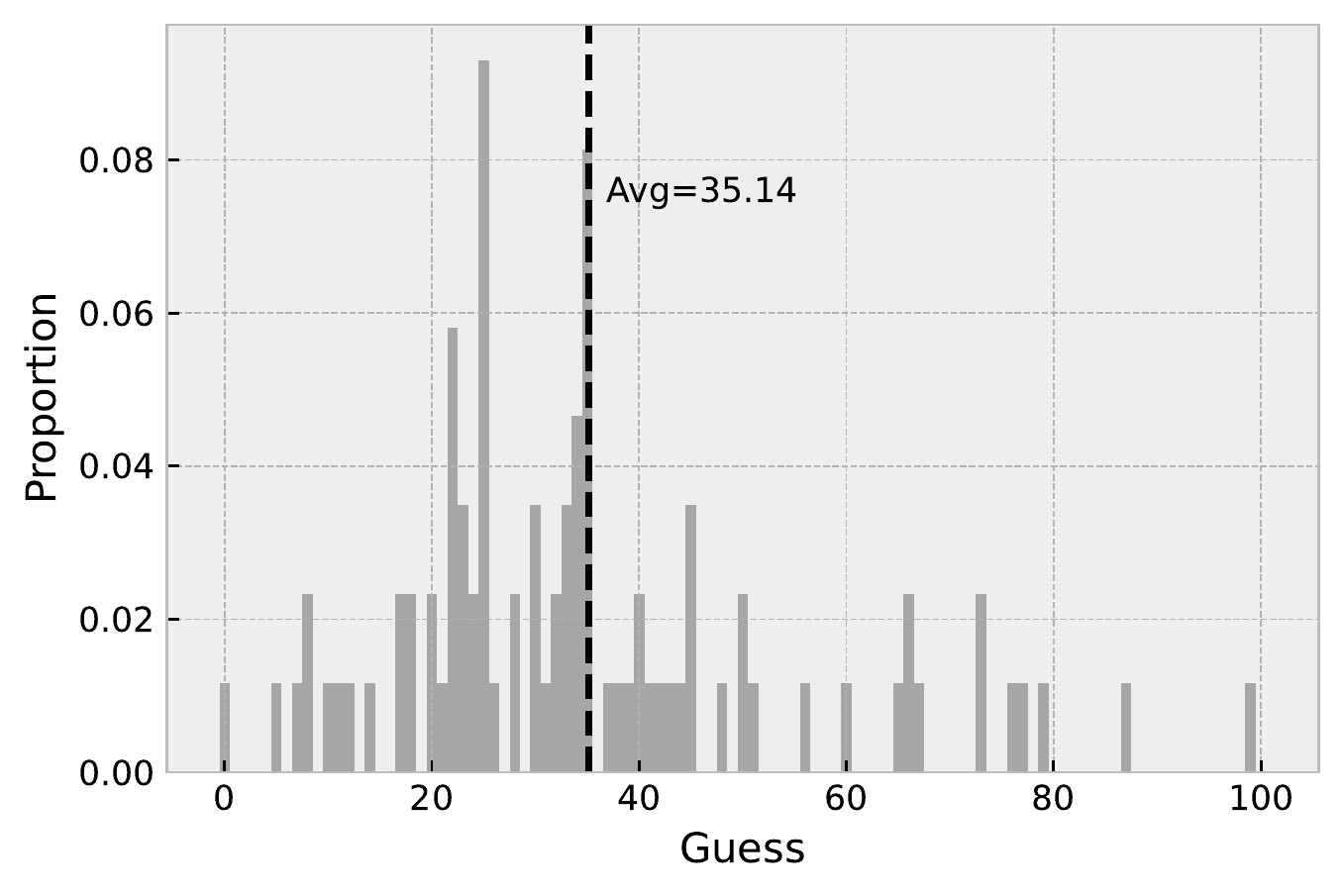}
  \caption{Lab}
\end{subfigure}\hfil
\begin{subfigure}{0.3\textwidth}
  \includegraphics[width=\linewidth]{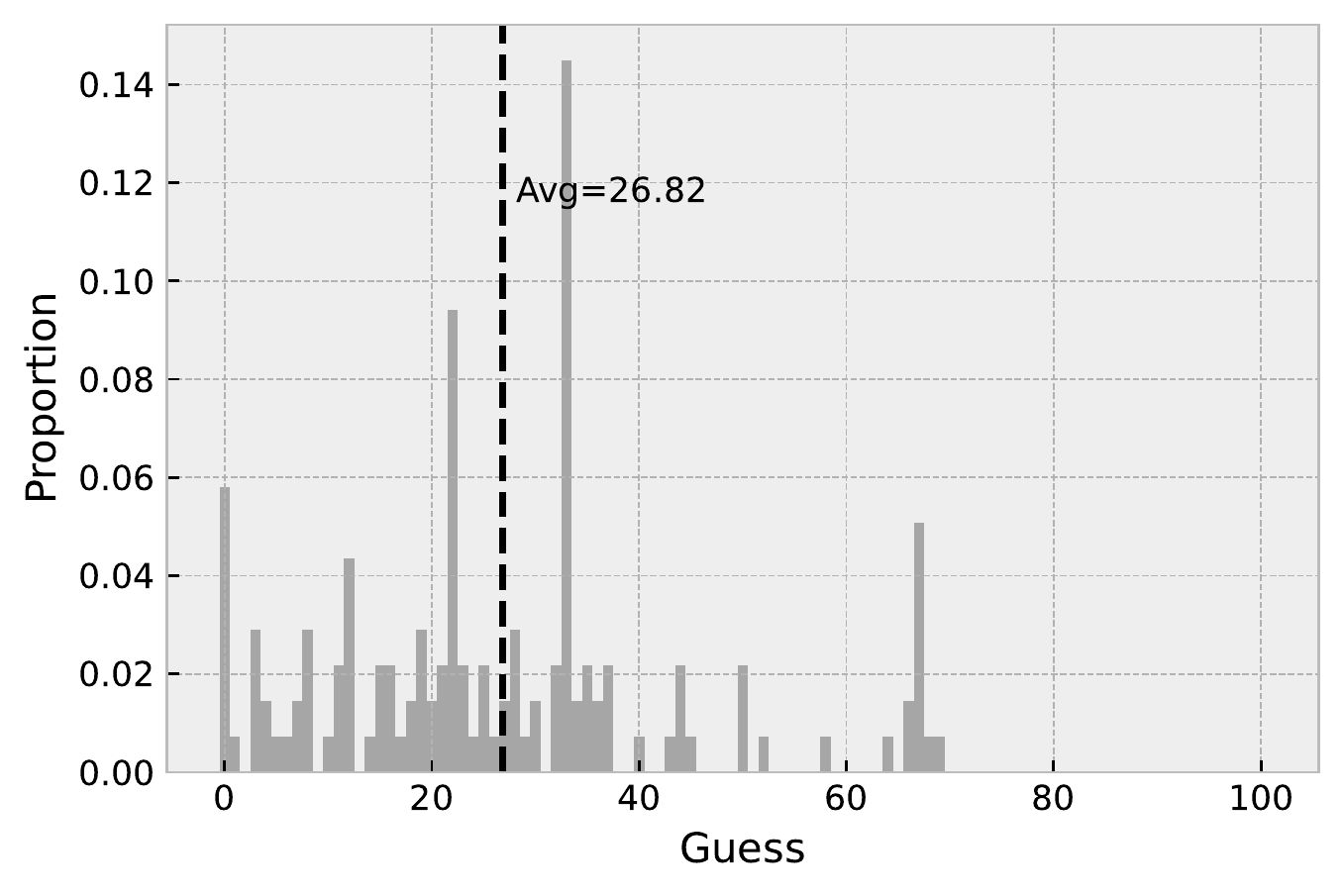}
  \caption{Classroom}
\end{subfigure}\hfil
\begin{subfigure}{0.3\textwidth}
  \includegraphics[width=\linewidth]{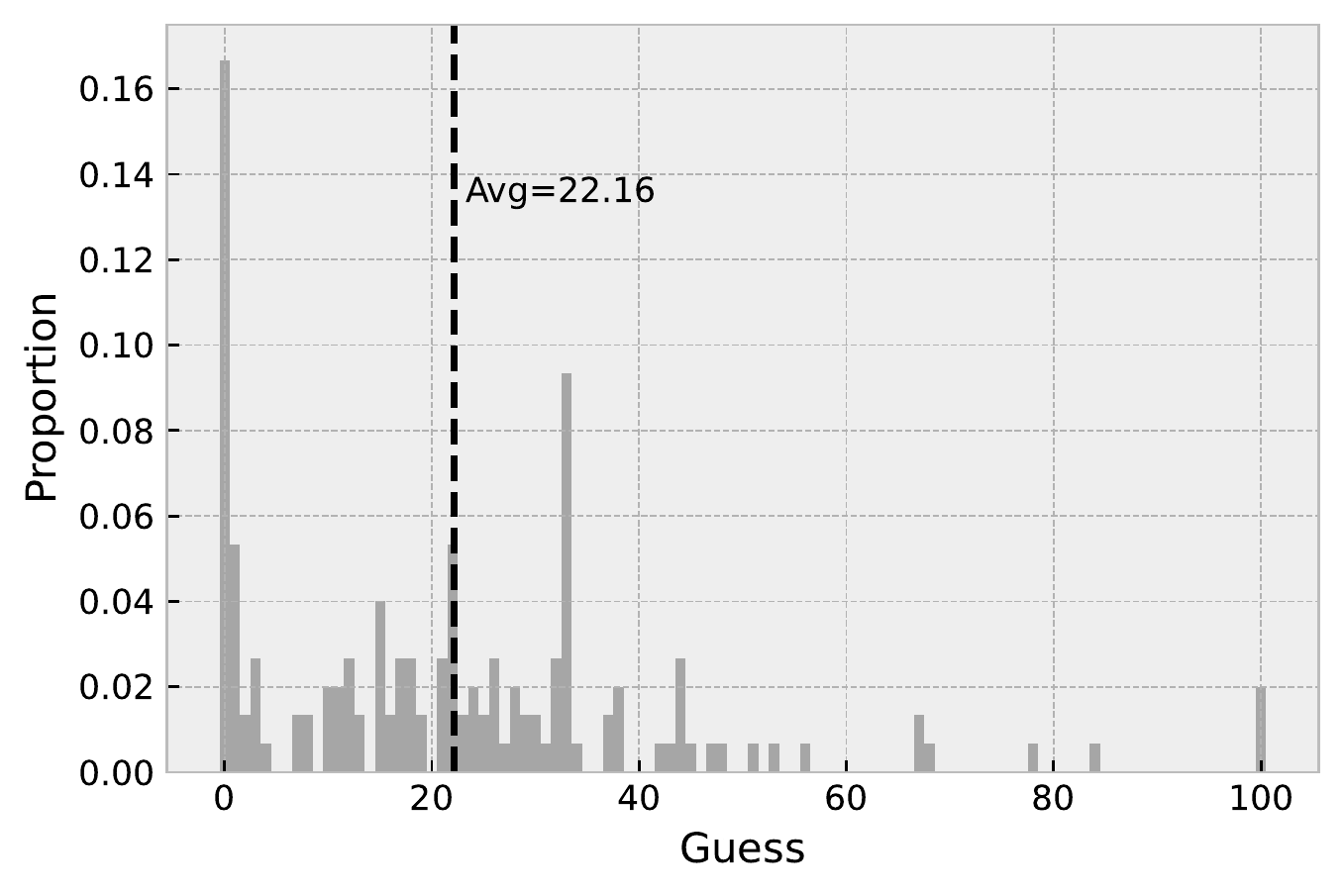}
  \caption{Internet}
\end{subfigure}\hfil
\begin{subfigure}{0.3\textwidth}
  \includegraphics[width=\linewidth]{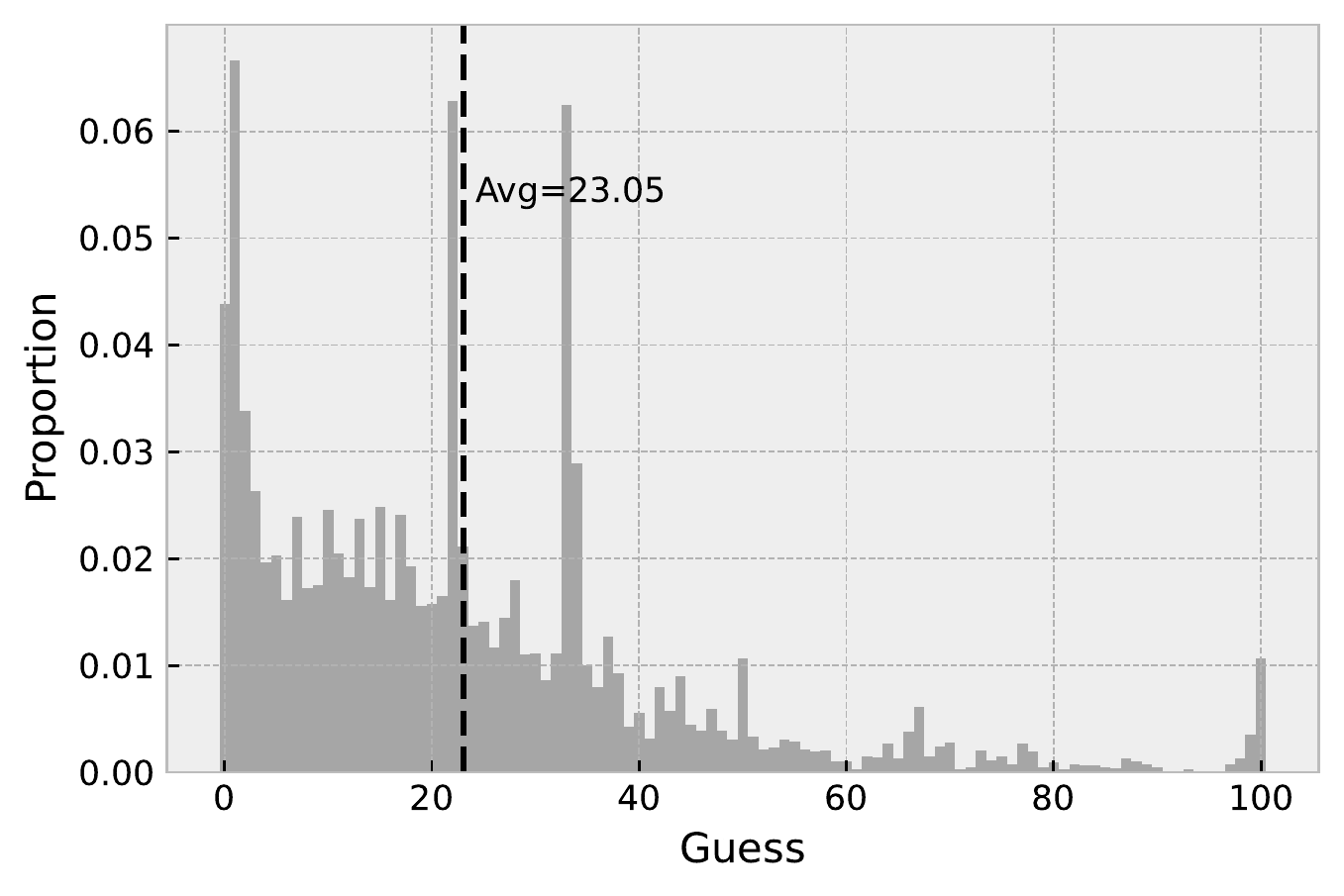}
  \caption{Newspaper}
\end{subfigure}\hfil
\begin{subfigure}{0.3\textwidth}
  \includegraphics[width=\linewidth]{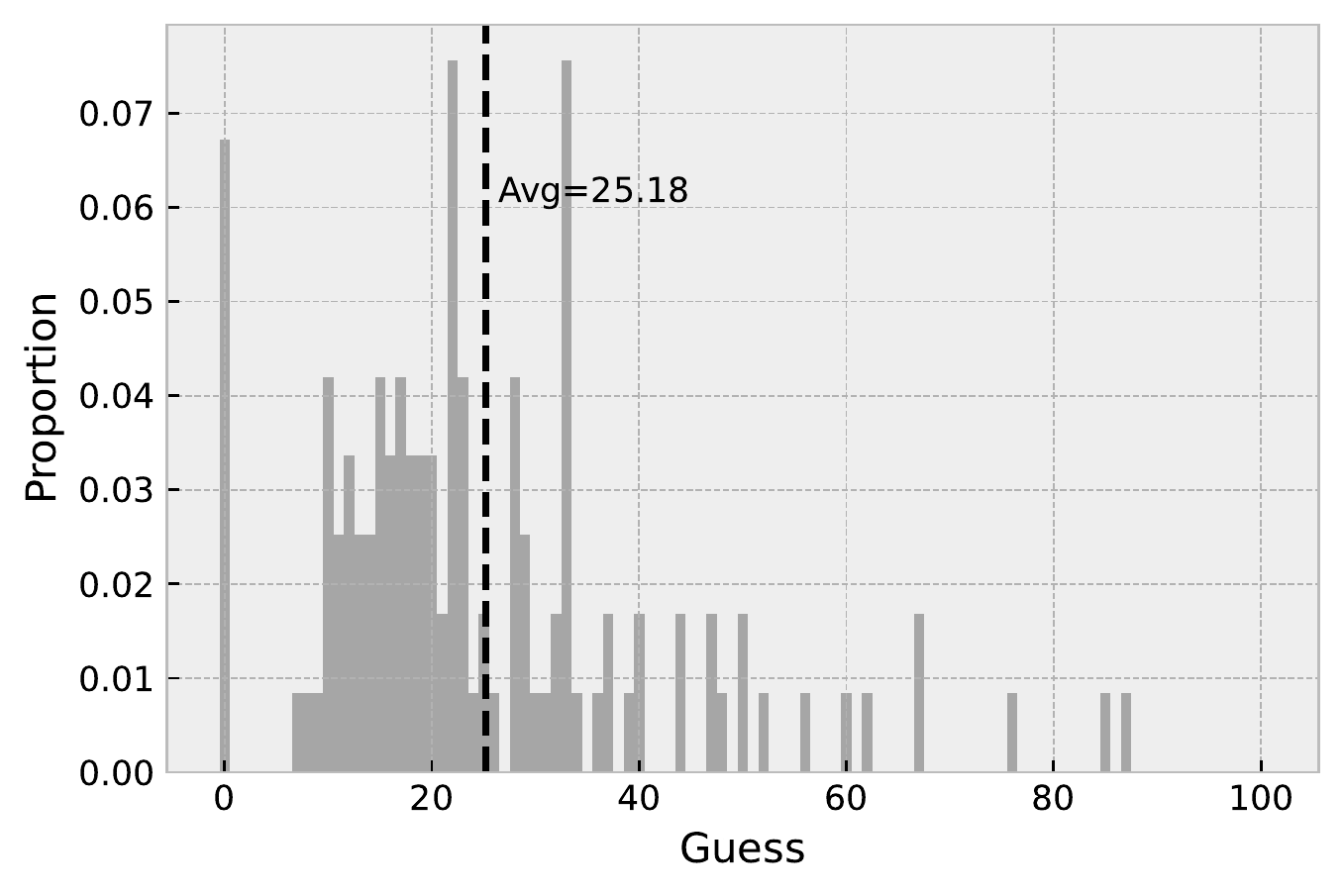}
  \caption{Take Home}
\end{subfigure}\hfil
\begin{subfigure}{0.3\textwidth}
  \includegraphics[width=\linewidth]{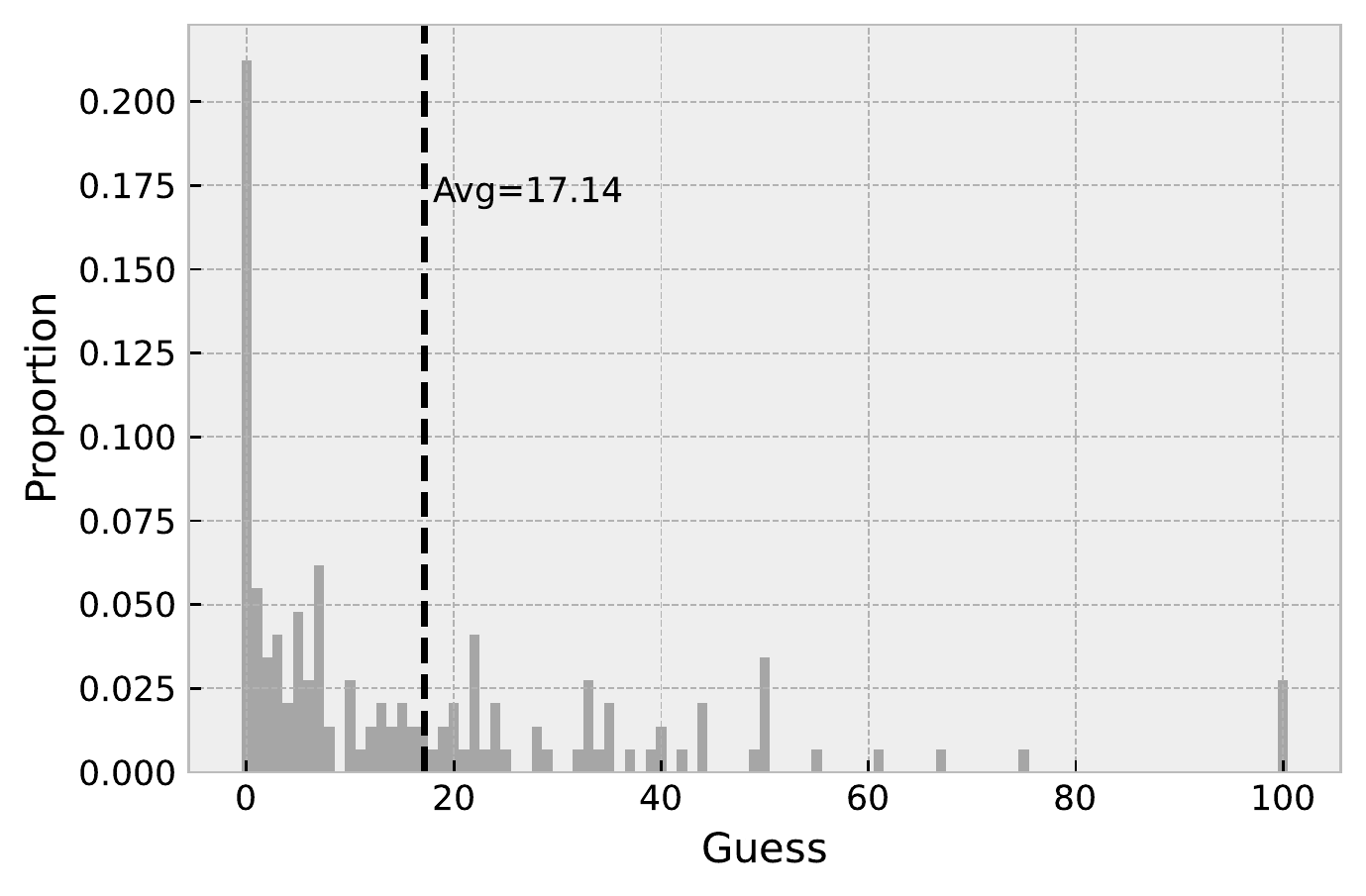}
  \caption{Theorists}
\end{subfigure}\hfil
\caption{Visualisation of various experimental $p$-beauty contests. The dotted vertical line indicates the average for the given dataset. Datasets source: \cite{bosch2002one}.}\label{figBeautyExperiment}
\end{figure}

We use the experimental data provided by \cite{bosch2002one} with $p=\frac{2}{3}$. The resulting guesses are visualised in \cref{figBeautyExperiment}.

The utilities are represented as follows:
\begin{equation}
\begin{split}
g_k &= p \times \frac{\sum_{a_{k+1}} a_{k+1} \times f[a_{k+1}]}{\sum_{a_{k+1}}f[a_{k+1}]} \\ 
U[a_k] &= \lvert a_k - g_k \rvert
\end{split}
\end{equation}
where $g_k$ represents the predicted goal, i.e., $2/3$'s of the average weighted prediction of the lower level thinkers. The utilities for each choice then become the distance to the goal.

\subsubsection{Comparison Methods}

\paragraph{Level-$k$} Level-$0$ competitors are assumed to guess randomly between $[0,100]$ (the same level-$0$ configuration is used for the naive player in the QH model). Level-$1$ players then anticipate this and guess $p \times 50$
(50 being the average from the level-$0$ players), level-$2$ players then guess $p \times (p \times 50)$ and so forth. As the levels increase, the guesses, therefore, approach $0$, coinciding with the perfectly rational choice.

\paragraph{Cognitive Hierarchy} Rather than assuming all players are at $k-1$, the cognitive hierarchy model fits a distribution to these $k$ players, and best responds according to this distribution of lower level thinkers. Again, following the convention of \cite{camerer2004cognitive}, we use the Poisson distribution.

\paragraph{Quantal response equilibrium} For the quantal response equilibrium, we use the logit rule following \cite{breitmoser2012strategic} estimated using fixed point iteration (see also Section F4.1 of the supplementary material for \cite{anufriev2022learning}).

\begin{figure}[htb]
\centering
\begin{subfigure}{0.3\textwidth}
  \includegraphics[width=\linewidth]{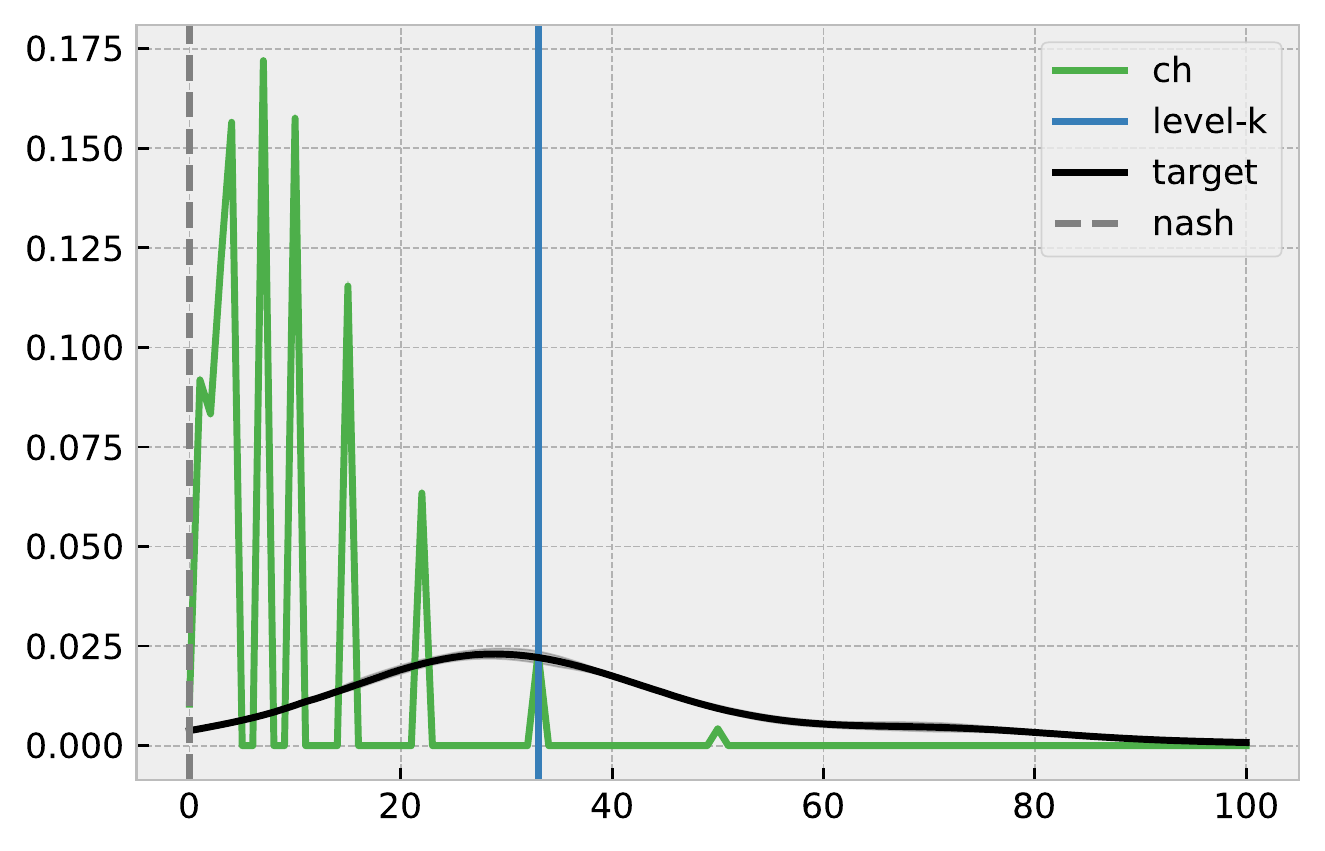}
  \caption{Lab}
\end{subfigure}\hfil
\begin{subfigure}{0.3\textwidth}
  \includegraphics[width=\linewidth]{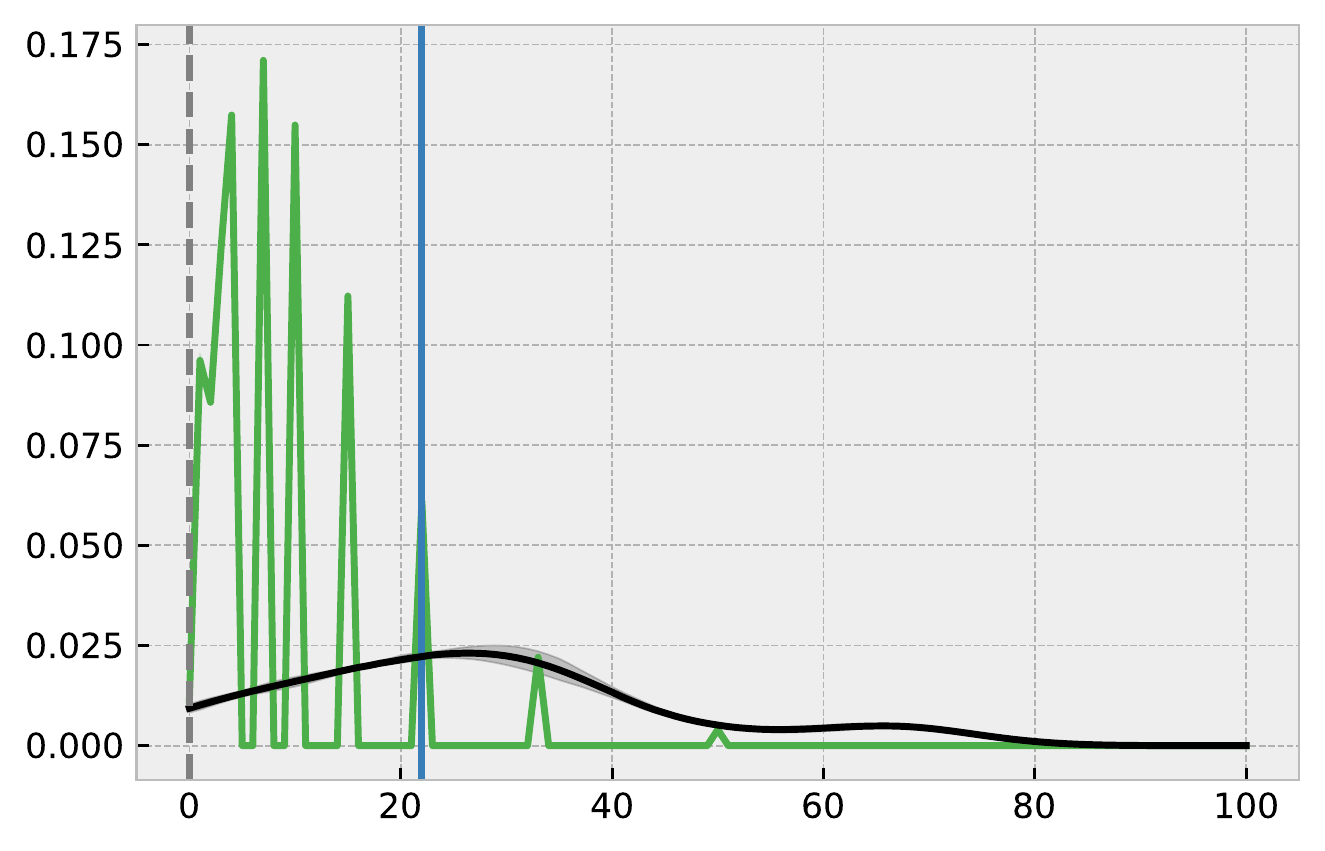}
  \caption{Classroom}
\end{subfigure}\hfil
\begin{subfigure}{0.3\textwidth}
  \includegraphics[width=\linewidth]{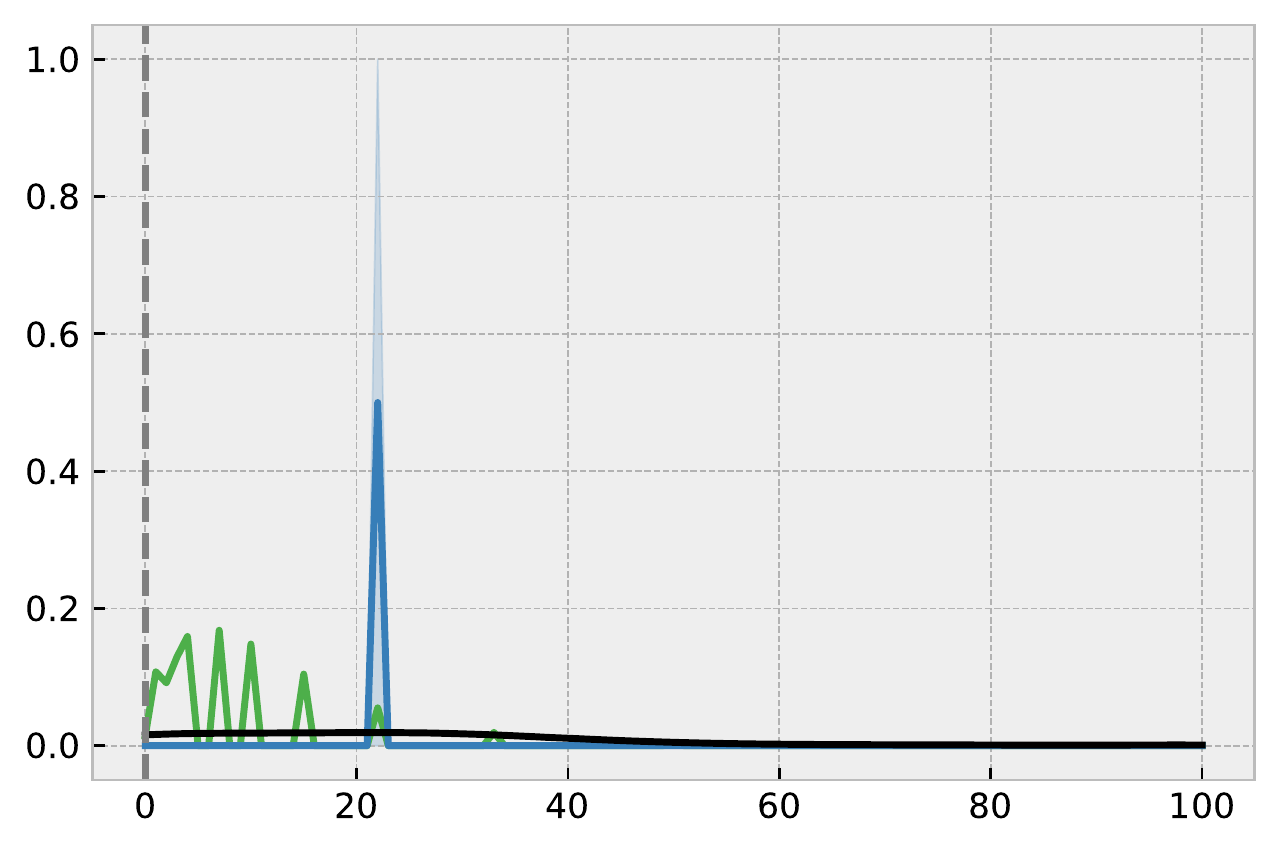}
  \caption{Internet}
\end{subfigure}\hfil
\begin{subfigure}{0.3\textwidth}
  \includegraphics[width=\linewidth]{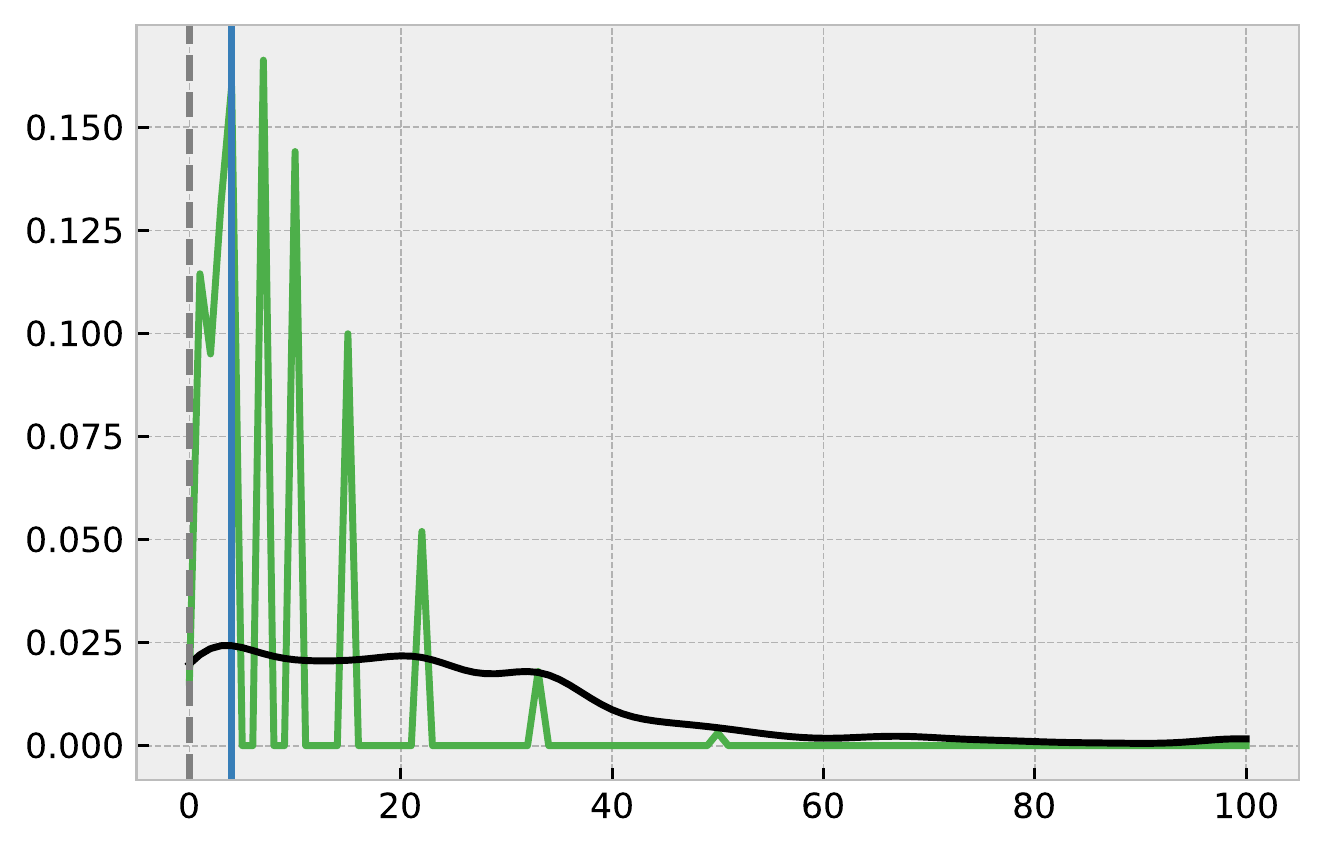}
  \caption{Newspaper}
\end{subfigure}\hfil
\begin{subfigure}{0.3\textwidth}
  \includegraphics[width=\linewidth]{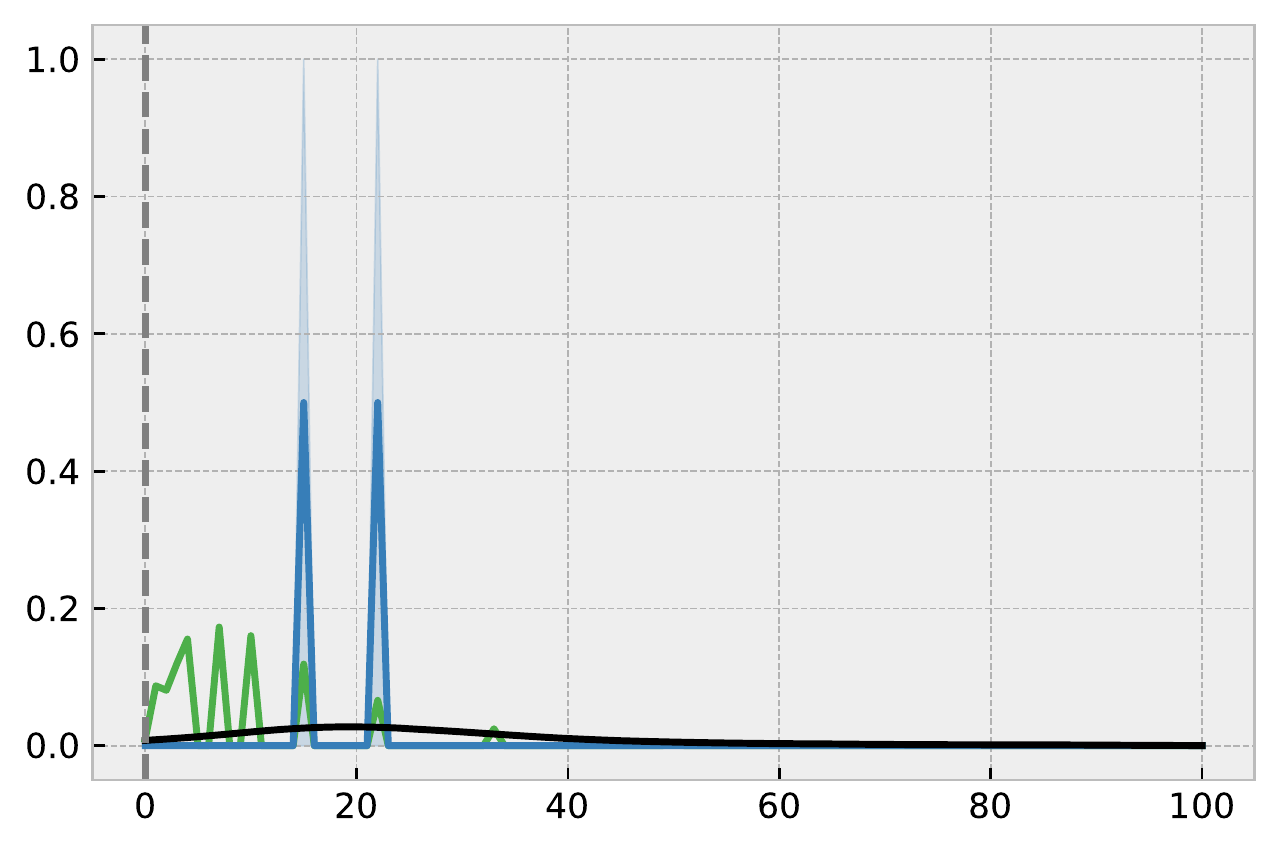}
  \caption{Take Home}
\end{subfigure}\hfil
\begin{subfigure}{0.3\textwidth}
  \includegraphics[width=\linewidth]{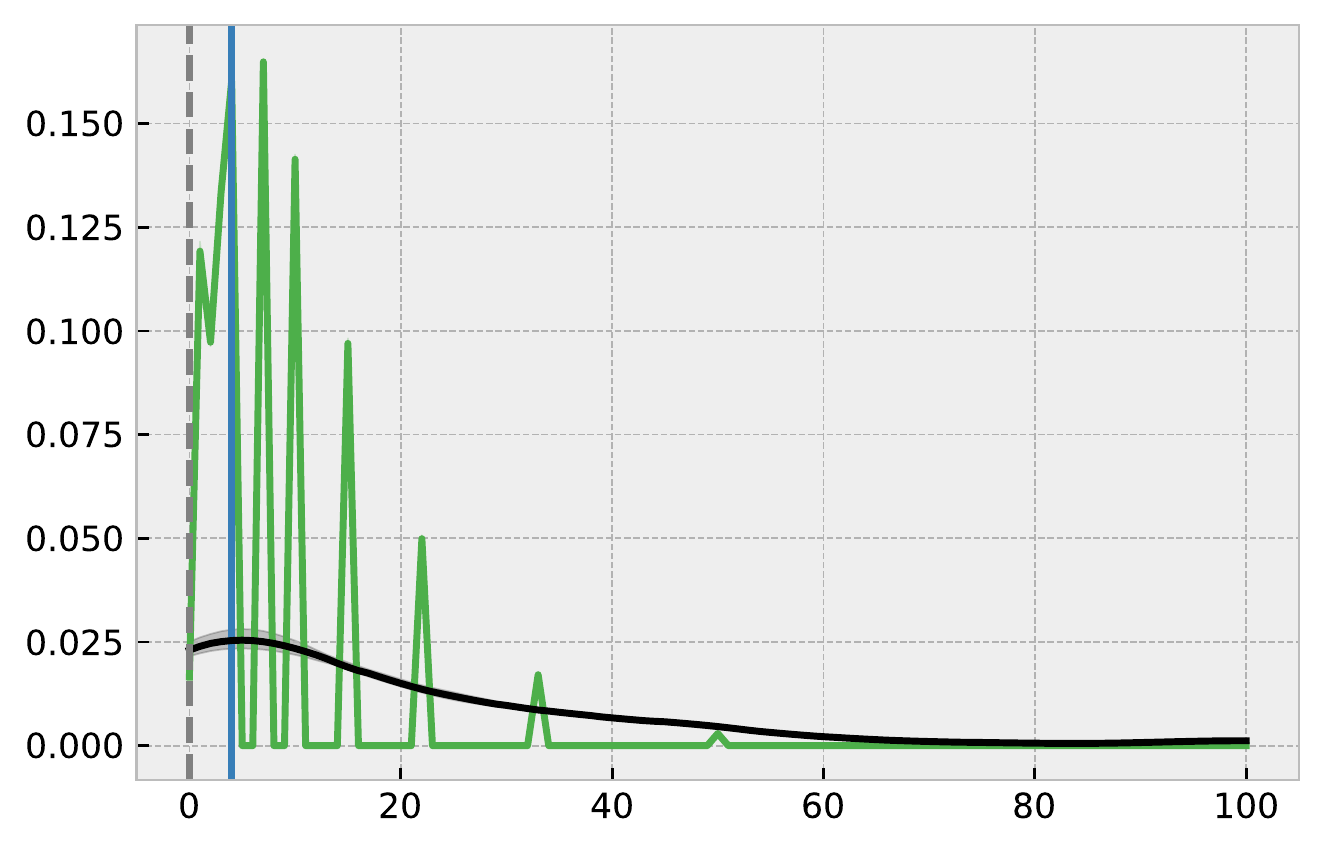}
  \caption{Theorists}
\end{subfigure}\hfil
\caption{Beauty contest games (extended). The darker lines indicate the mean result from 5x2 cross-validation. The shaded regions indicate $\pm$ one standard deviation. The out-of-sample data are shown as the black line. The level-$k$ model is shown as the blue line. The Cognitive Hierarchy model as the green line. The Nash equilibrium solution is indicated as the diagonal dashed grey line.}\label{figBeautyExtended}
\end{figure}


\subsection{Centipede Game}\label{secAppendixCentipede}

\begin{figure}[ht]
    \centering
    \includegraphics[width=\textwidth]{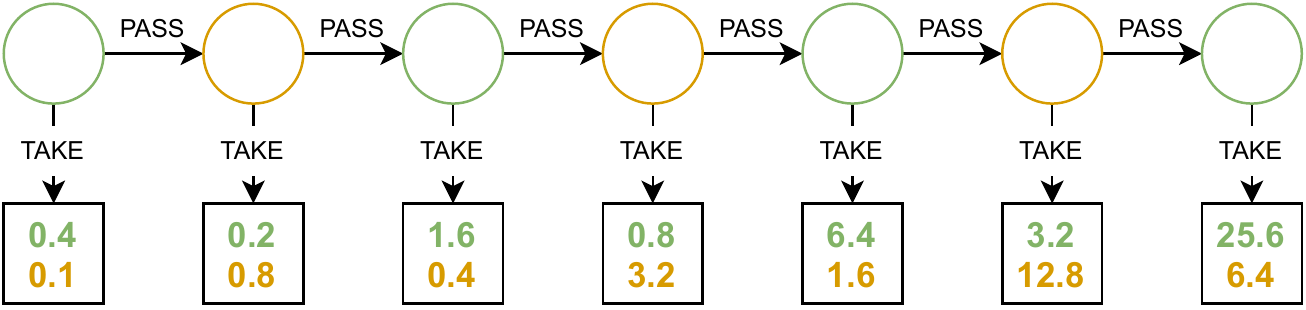}
    \caption{The six move extensive-form centipede Game. Green (orange) circles highlight Player 1(2)'s turn, and the top (bottom) row of the boxes highlights Player 1(2)s payoff. The four-move game is equivalent until the fourth node, however, the payoffs for the fifth node become the ``PASS" payoffs for Node 4. }\label{figCentipedeGame}
\end{figure}

With perfectly rational backward induction, the subgame perfect equilibrium of the centipede game is for each player to immediately take the pot without proceeding to any further rounds. However, this is a poor predictor of what happens experimentally \citep{ho2013dynamic}, where players are shown to ``grow" the money pile by playing for several rounds before taking \citep{ke2019boundedly}. Again, there are multiple reasons proposed to explain players deviation from the predicted unique subgame equilibrium  \citep{kawagoe2012level, krockow2018far, georgalos2020comparing}.

In this work, we use the experimental data of \cite{mckelvey1992experimental} (from their Appendix C) for four and six-level centipede games. The utilities are represented as:

\begin{equation}
\begin{split}
U_1[\text{take}_1] &= 0.4 \\
U_1[\text{pass}_1] &= 0.2 \times f[\text{take}_2] + V_1 \times f[\text{pass}_2] \\
U_2[\text{take}_2] &= 0.8 \\
U_2[\text{pass}_2] &= 0.4 \times f[\text{take}_3] + V_2 \times f[\text{pass}_3] \\
\dots
\end{split}
\end{equation}
where $V_1, V_2$ are derived as the average expected return for the remainder of the moves. These payoffs are also visualised in \cref{figCentipedeGame}. The conditioning on the history of decisions is implicit here, as to take an action $a_n$ at $n > 1$, all previous actions must have been to pass (otherwise the game would have ended).

\begin{figure*}
    \centering
    \begin{subfigure}[t]{0.4\textwidth}
        \centering
        \includegraphics[width=\textwidth]{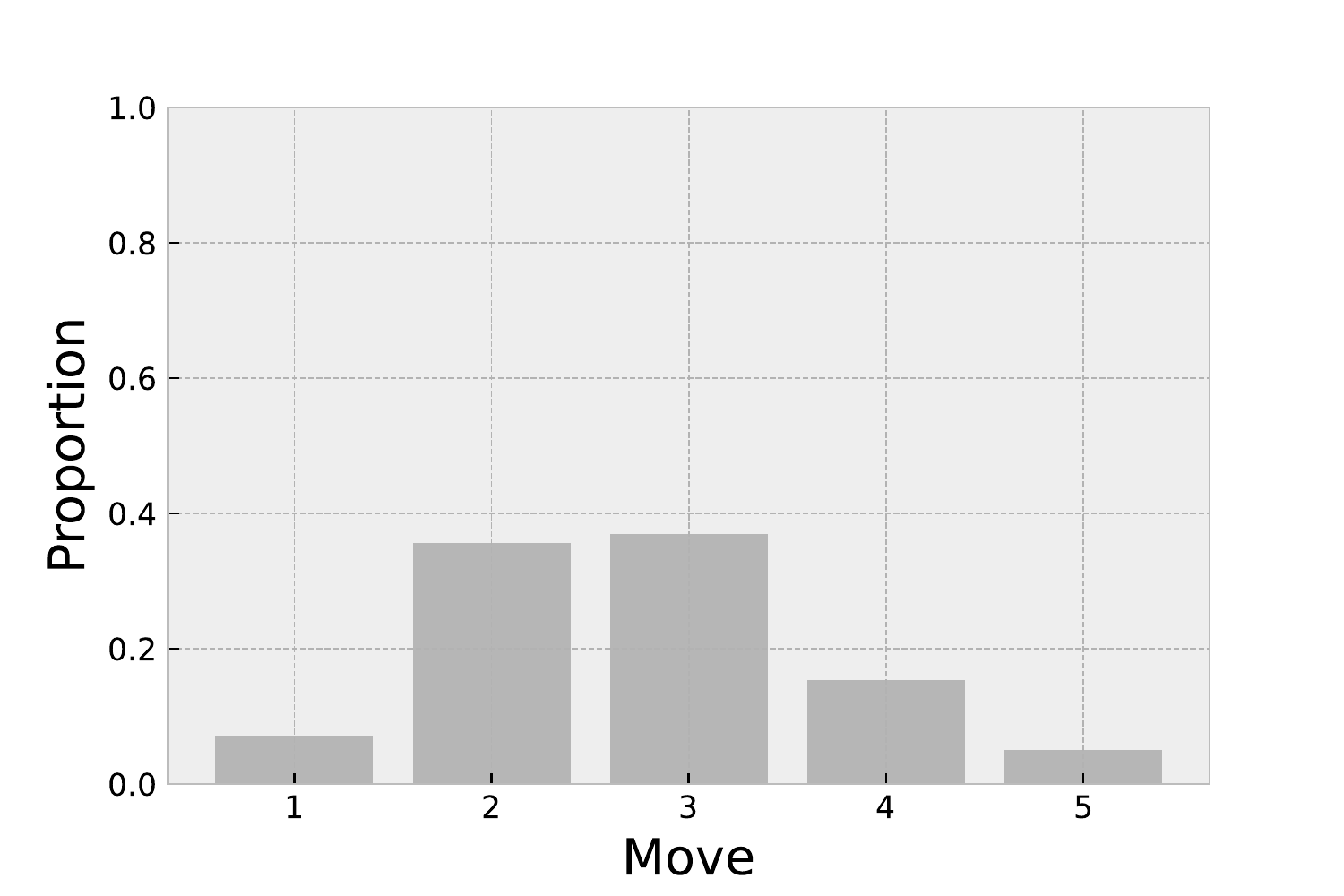}
        \caption{Four Move.}
    \end{subfigure}
    \begin{subfigure}[t]{0.4\textwidth}
        \centering
        \includegraphics[width=\textwidth]{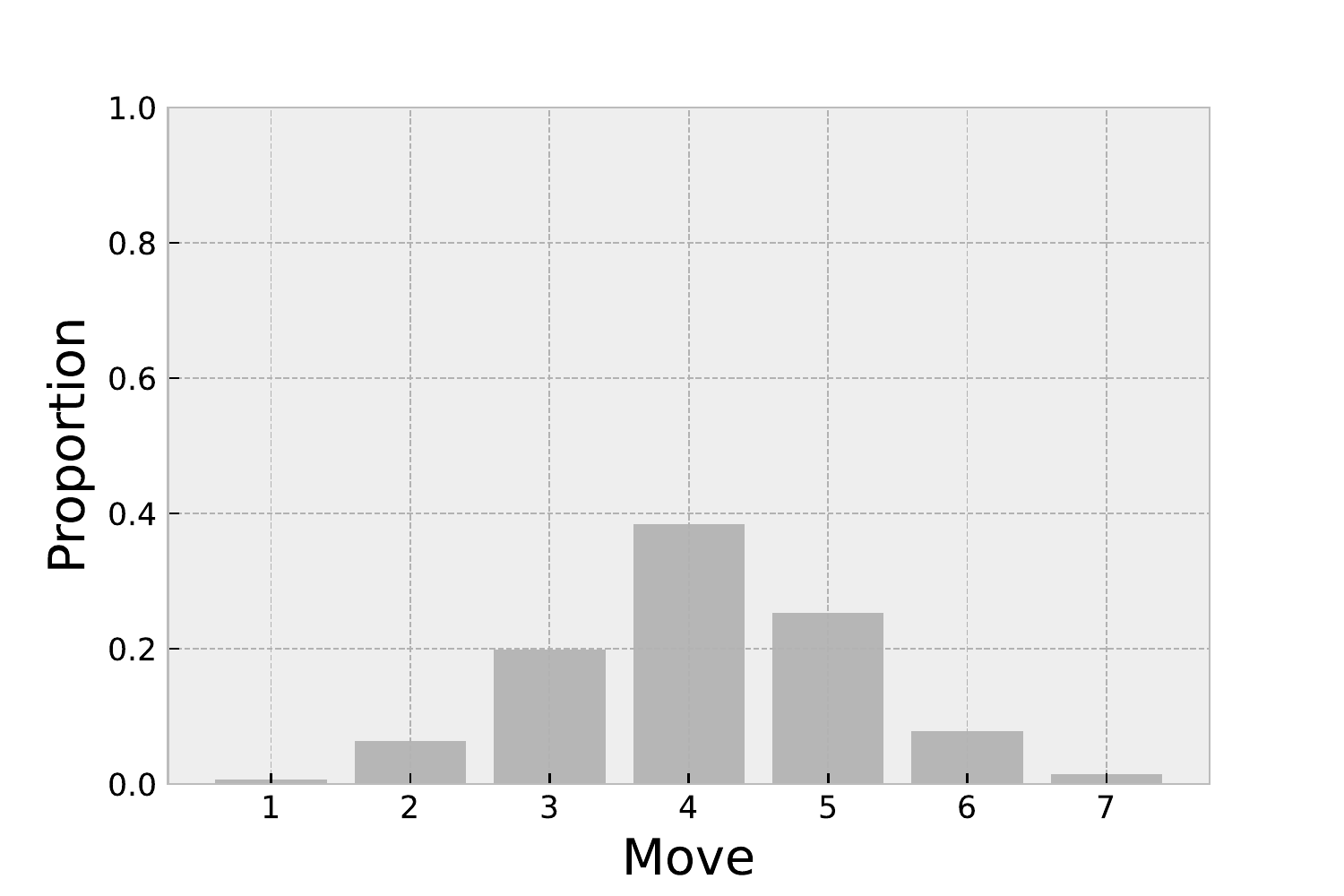}
        \caption{Six Move.}
    \end{subfigure}
    \caption{Four and six-level Centipede Games using the dataset from \cite{mckelvey1992experimental}.}\label{figCentipedeExperimental}
\end{figure*}

\subsubsection{Comparison Methods}

\paragraph{Level-$k$}

Under the level-$k$ framework, we assume a level-$0$ agent is equally as likely to take or pass at each stage of the game (the same configuration is used as the naive player under the proposed Quantal Hierarchy approach). A level-$1$ player then takes at the node which maximises the expected utility subject to this, and so on and so forth.  A full analysis of level-$k$ framework in centipede games is presented in \cite{kawagoe2012level}, but this configuration used (referred to as Random Behavioral strategy (RBS) in \cite{kawagoe2012level}) was shown to be the best specification for matching the experimental data (for both level-$k$ and cognitive hierarchy).

\paragraph{Cognitive Hierarchy} Rather than assuming all players are at $k-1$, the cognitive hierarchy model fits a distribution to these $k$ players, and best responds according to this distribution of lower level thinkers. Again, the Poisson distribution was used, which was shown to be the best experimental fit in \cite{kawagoe2012level}. 

\paragraph{Quantal Response Equilibirum} An agent-form of the QRE \citep{mckelvey1998quantal} is used here , where at each node, the agent choices nosily based on resource parameter $\beta$, and assuming their opponent is also operating under the same resource constraint $\beta$. This is calculated recursively following \citep{mckelvey1998quantal}. 

\subsection{Sequential Bargaining}\label{secAppendixBargaining}

\begin{figure*}[ht]
    \centering
    \begin{subfigure}[t]{0.28\textwidth}
        \centering
        \includegraphics[width=\textwidth]{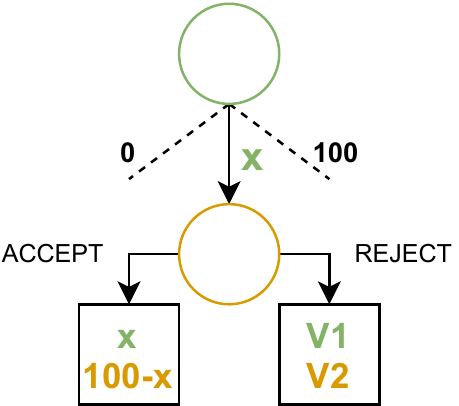}
        \caption{Ultimatum (One-stage)}\label{figUltimatumTree}
    \end{subfigure}
    \begin{subfigure}[t]{0.28\textwidth}
        \centering
        \includegraphics[width=\textwidth]{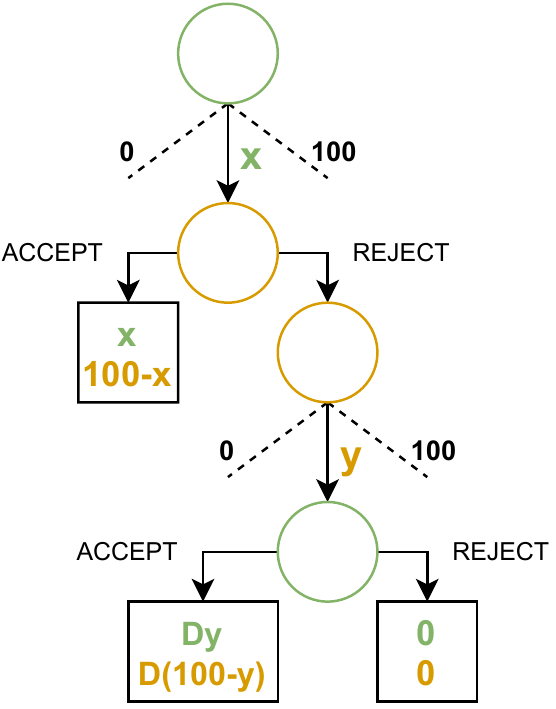}
        \caption{Two-Stage}\label{figTwoStage}
    \end{subfigure}
    \caption{Example extensive-form sequential bargaining games. In the ultimatum game (\cref{figUltimatumTree}), Player 1 makes a request $x \in [0,100]$. If Player 2 accepts the request, Player 2 receives a payoff of $100-x$, and Player 1 receives $x$. If Player 2 declines the request, they each receive the rejection payoff ($V_1$ or $V_2$). In the two-stage game, if Player 2 rejects, they can come back with a counteroffer $y$. Now the process repeats, and it is up to Player 1 to accept or reject. If Player 1 accepts, Player 2 gets a disagreement penalised ($D$) payoff of $D(100-y)$, and Player 1 gets a payoff of $Dy$. However, if both decline they each get a payoff of $0$.}\label{figBargaining}
\end{figure*}

We examine the experimental results of the Ultimatum Game (one-stage) and two-stage alternating-offer bargaining games \citep{binmore2002backward}, which consistently demonstrate violations of backward induction \citep{webster2013note}, even when accounting for ``fairness" in the system  \citep{johnson2002detecting}. 

\subsubsection{Ultimatum Game}
For the ultimatum game, we use the experimental data of Game 1 from \cite{binmore2002backward}. In the ultimatum game, the players are faced with the following payoffs:

\begin{equation}
\begin{split}
U_1[a_1] &= a_1 \times f[\text{accept} \mid a_1] + V_1 \times f[\text{reject} \mid a_1] \\
U_2[\text{accept}_{2} \mid a_1] &= 100 - a_1 \\
U_2[\text{reject}_{2} \mid a_1] &= V_2 \\
\end{split}
\end{equation}
where $V_1, V_2$ are the rejection payoffs for Player 1 and Player 2.

If opponents (Player 2) are rational, then Player 1, being rational, should request no more than their opponent's rejection payoff $V_2$. However, if opponents are not believed to be rational, then there is potential for Player 1 to exploit this fact and request higher (or lower) amounts. That is, it becomes rational for Player 1 to play as if Player 2 is not perfectly rational.

\begin{figure*}[ht]
    \centering
    \begin{subfigure}[t]{0.28\textwidth}
        \centering
        \includegraphics[width=\textwidth]{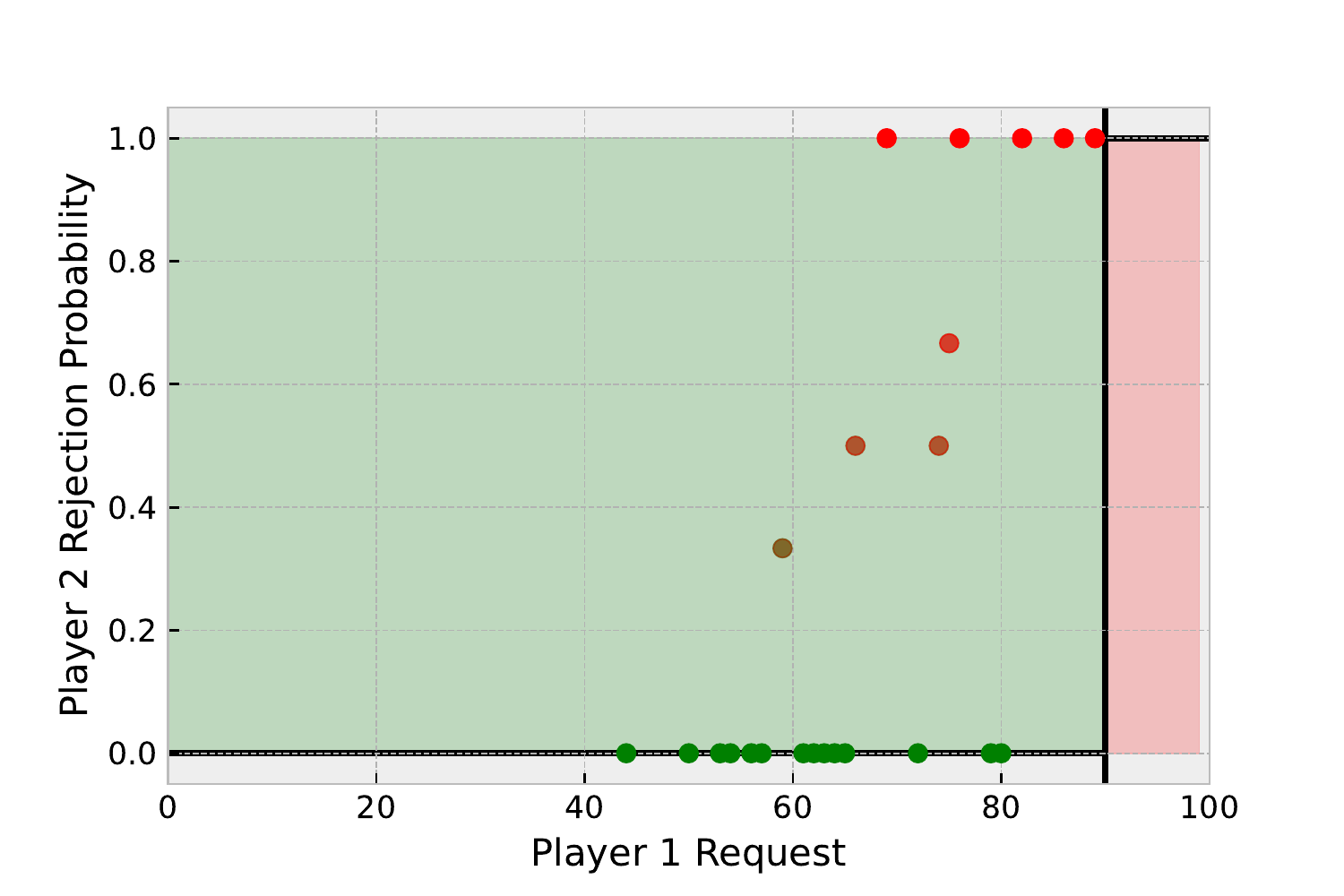}
        \caption{($V_1$=10, $V_2$=10)}
    \end{subfigure}
    \begin{subfigure}[t]{0.28\textwidth}
        \centering
        \includegraphics[width=\textwidth]{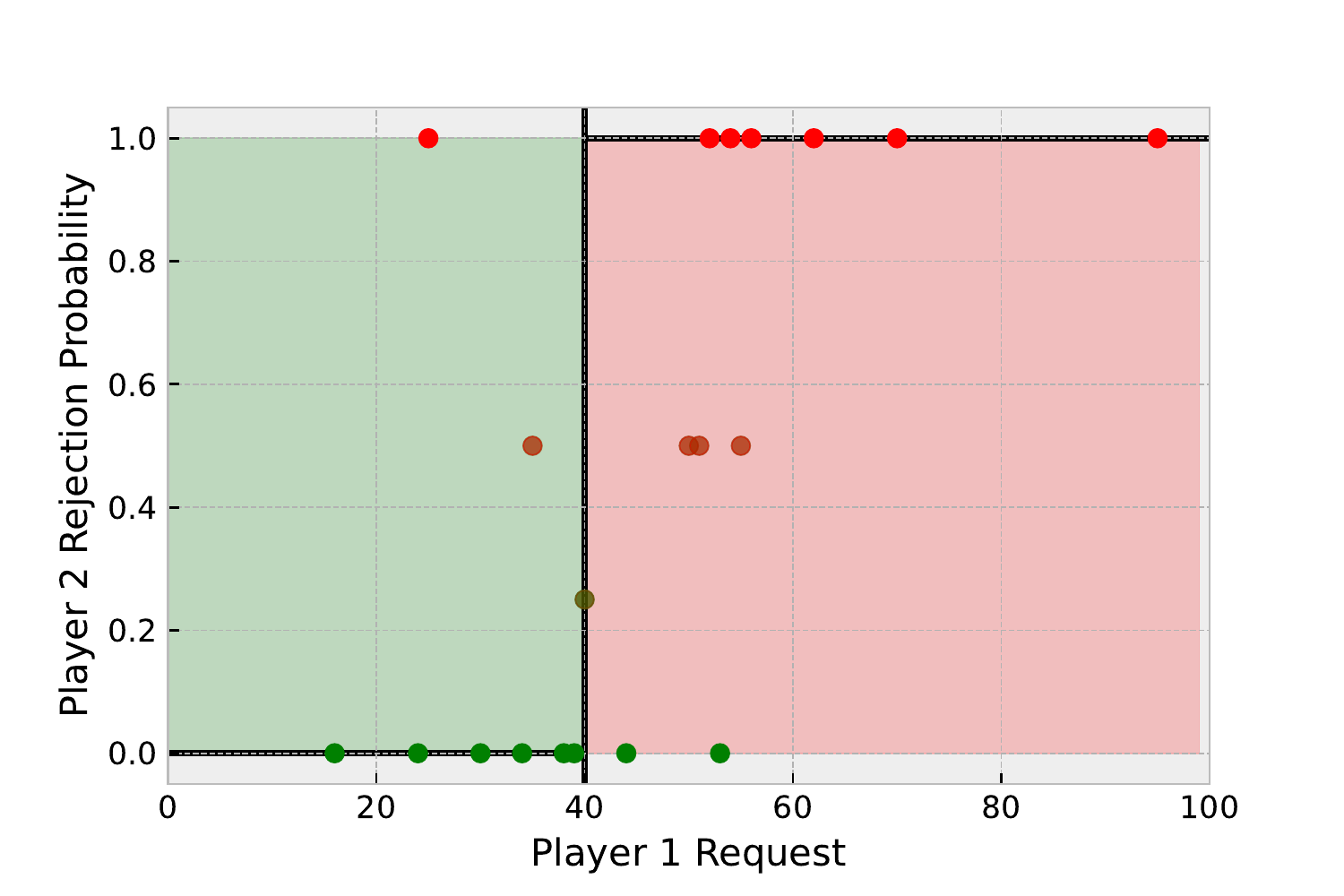}
        \caption{($V_1$=10, $V_2$=60)}
    \end{subfigure}
    \begin{subfigure}[t]{0.28\textwidth}
        \centering
        \includegraphics[width=\textwidth]{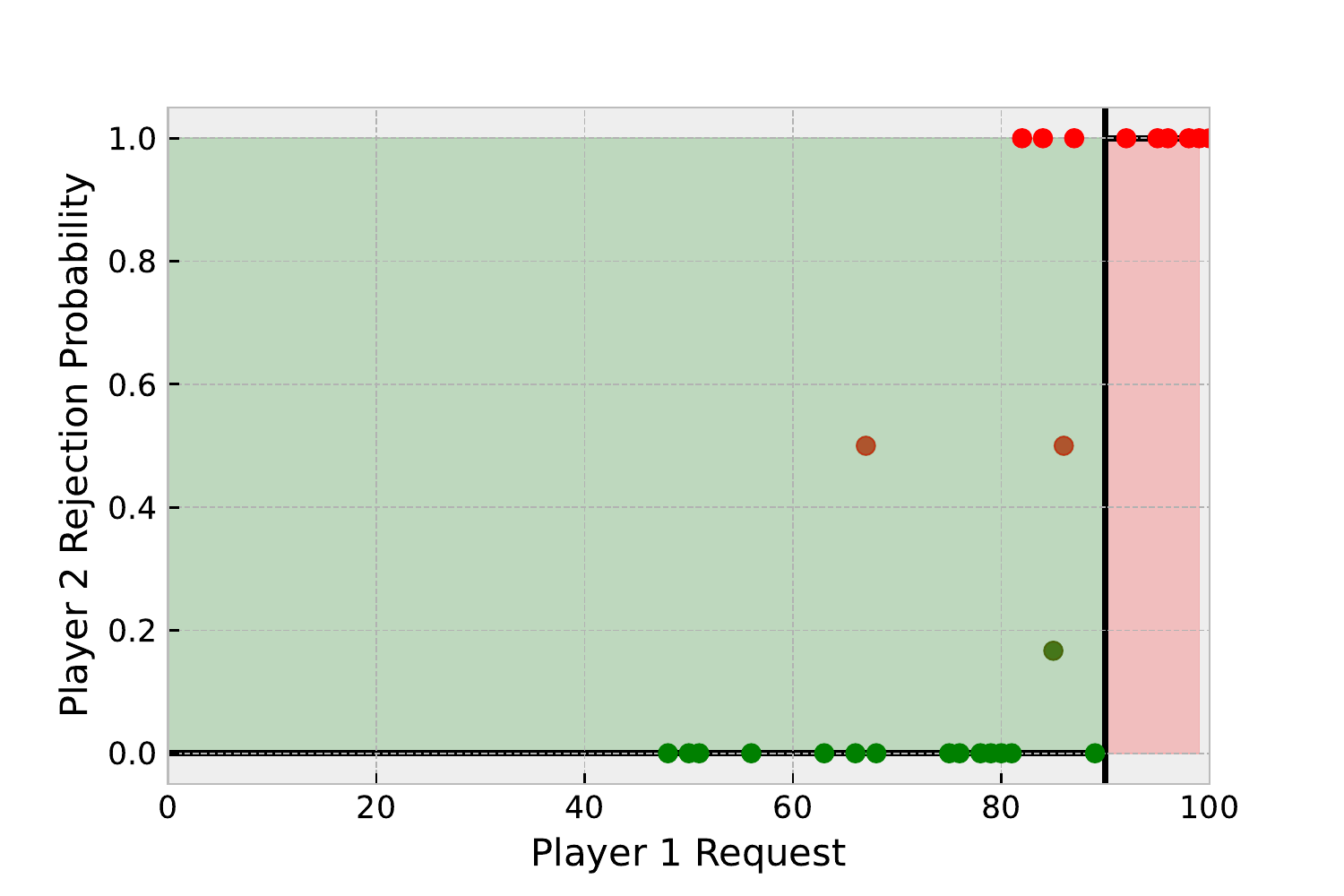}
        \caption{($V_1$=70, $V_2$=10)}
    \end{subfigure}
    \caption{Observed rejection rates from experimental data of \cite{binmore2002backward}. A rational opponent is governed by the step function (black line), where a rational player would reject in the red area and accept in the green area. The observed points show deviations from rationality.
    }\label{figUltimatumStep}
\end{figure*}

A rational opponent implies a step function, where for $a_1$, with payoff $100-a_1 > V_2$ the player accepts with probability $1$, and for payoffs below $V_2$ the player reject with certainty as shown in \cref{figUltimatumStep}. However, from \cref{figUltimatumStep} we can see deviations from rationality in the observed play. This is directly shown by non-deterministic outputs, where the players may or may not accept the request based on the requested value, as well as violations where the players reject or accept with probability $1$ even if a rational actor would do the opposite.

\begin{figure*}[ht]
    \centering
    \begin{subfigure}[t]{0.28\textwidth}
        \centering
        \includegraphics[width=\textwidth]{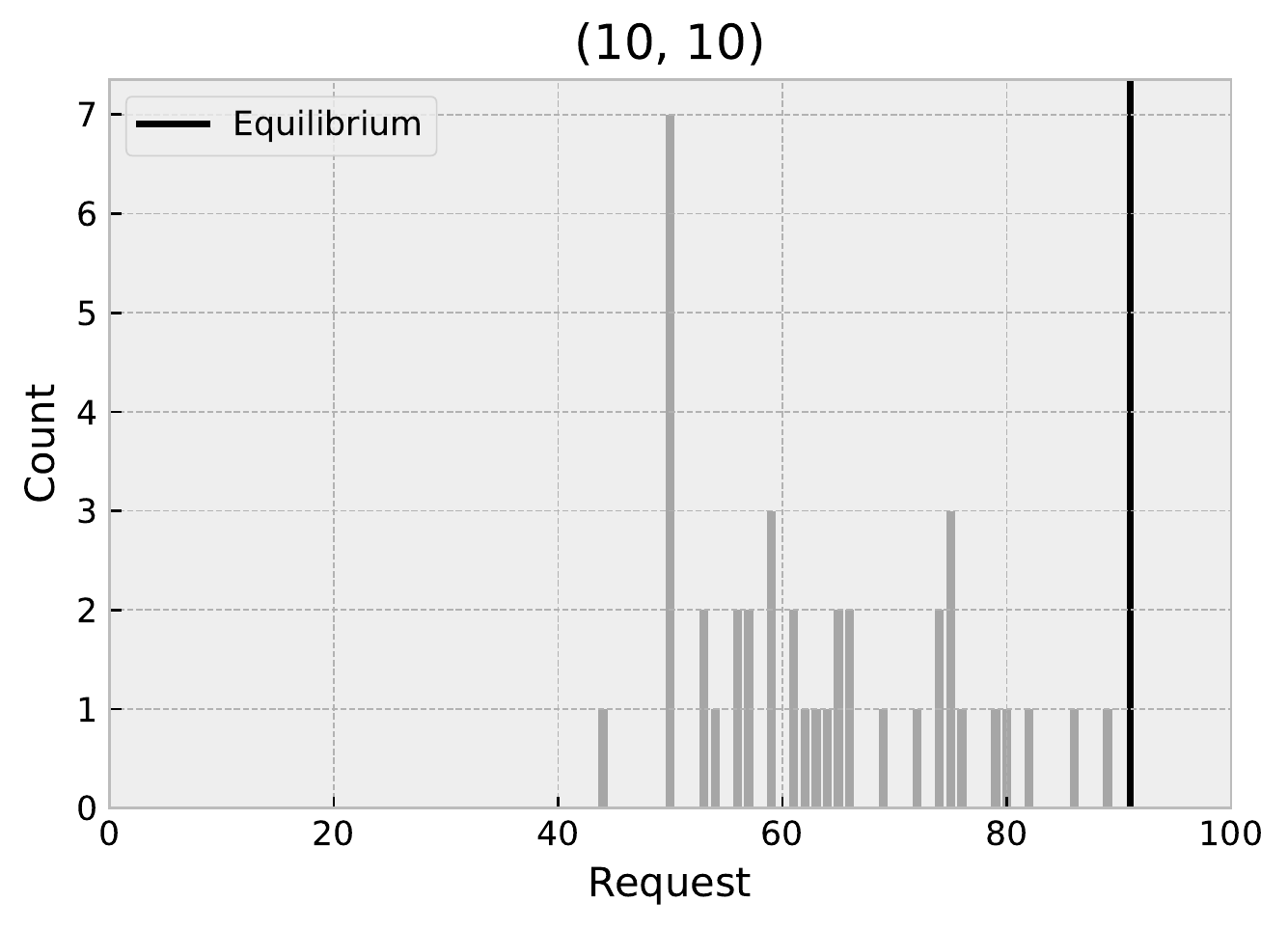}
        \caption{($V_1$=10, $V_2$=10)}
    \end{subfigure}
    \begin{subfigure}[t]{0.28\textwidth}
        \centering
        \includegraphics[width=\textwidth]{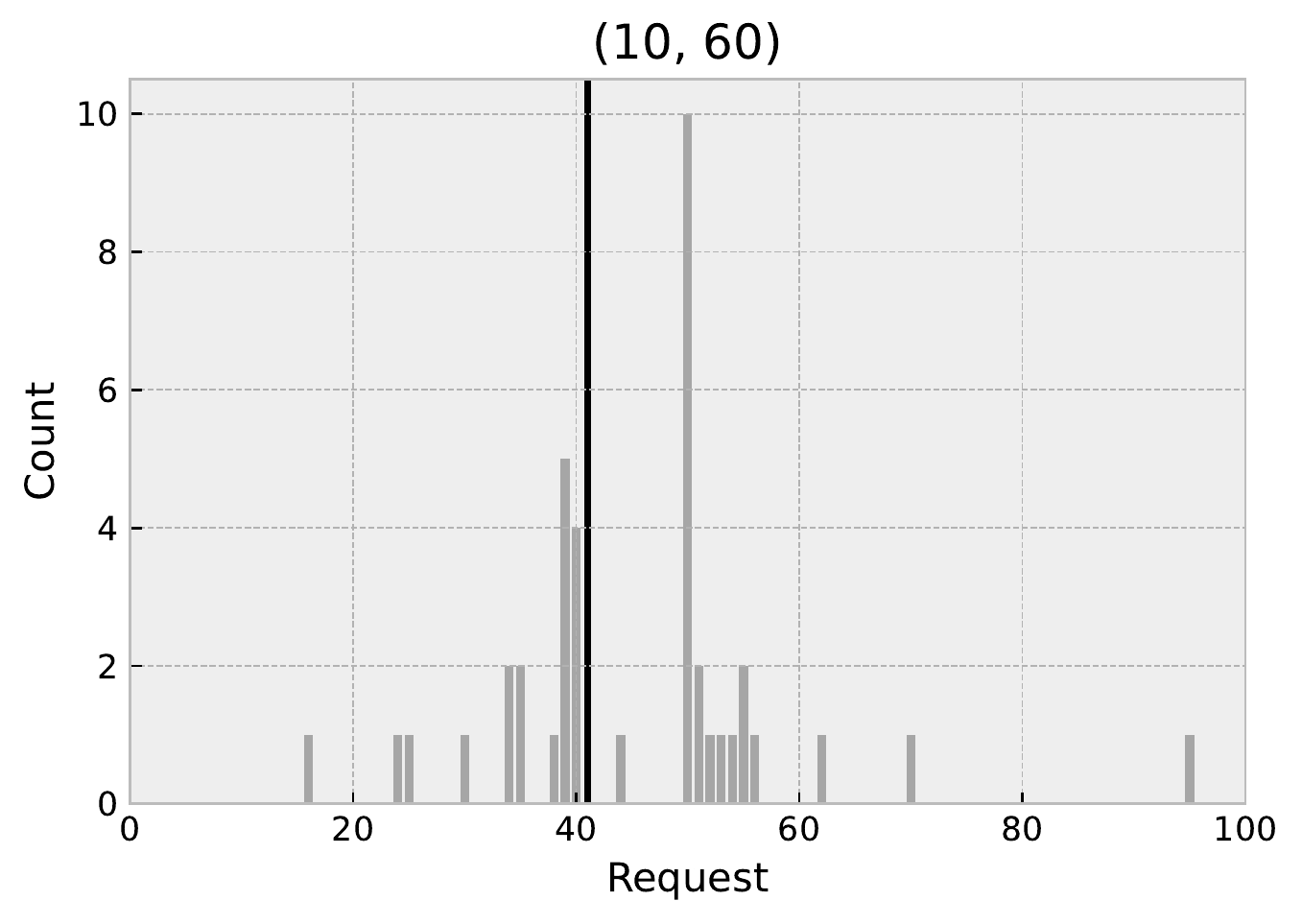}
        \caption{($V_1$=10, $V_2$=60)}
    \end{subfigure}
    \begin{subfigure}[t]{0.28\textwidth}
        \centering
        \includegraphics[width=\textwidth]{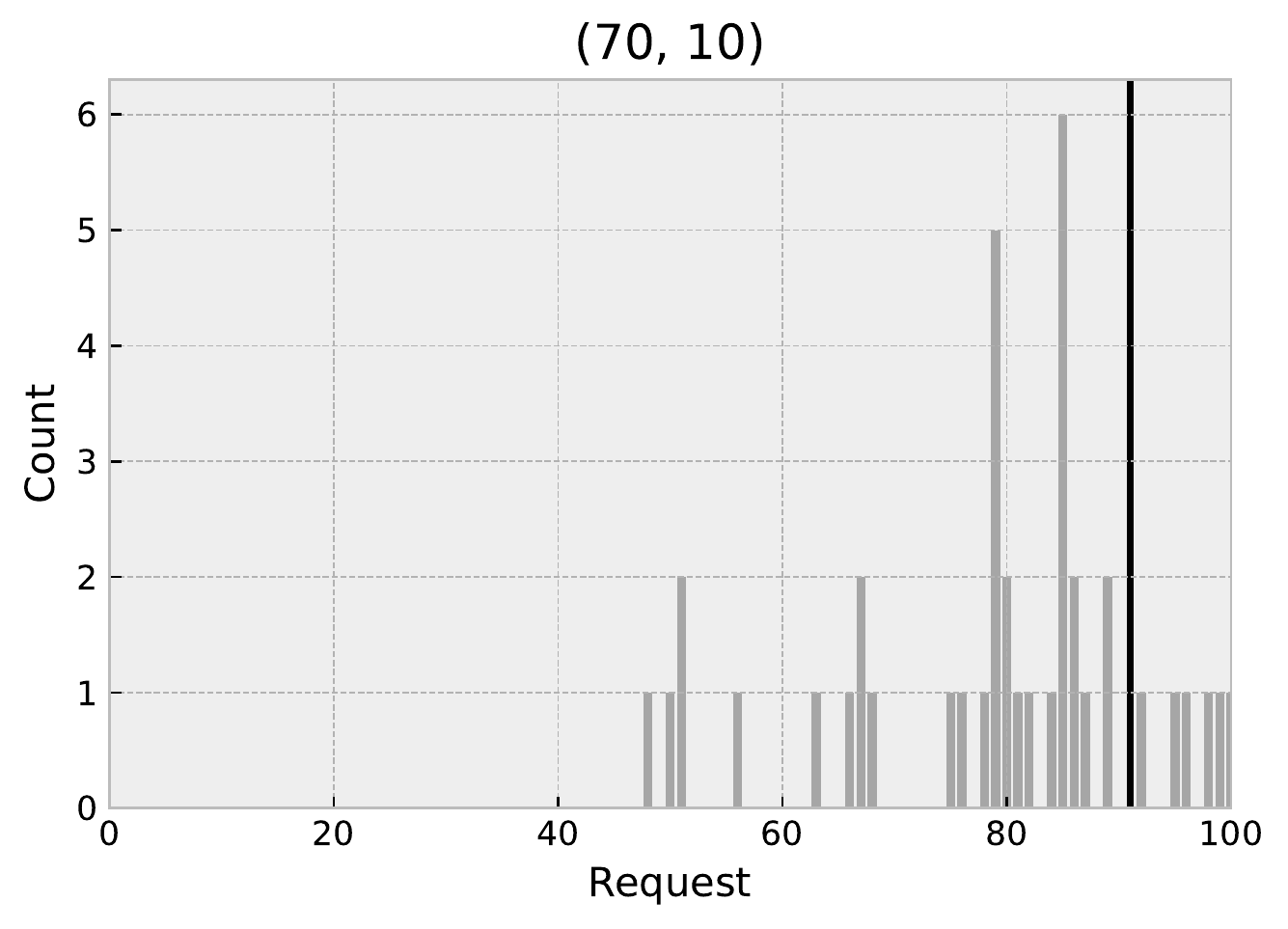}
        \caption{($V_1$=70, $V_2$=10)}
    \end{subfigure}
    \caption{Player 1 requests in the Ultimatum game, with experimental data from \cite{binmore2002backward}. The black line indicates the perfectly rational choice (when assuming opponent is perfectly rational), i.e., a rational opponent would reject any lower request (left of the black line), and would accept any value above their rejection payoff (right of the black line).
     }\label{figUltimatumRequest}
\end{figure*}

Now, knowing the opponent has potential bounds on their rationality, a rational player would respond accordingly.  \cref{figUltimatumRequest} plots the distribution of Player 1 requests. We observe that Player 1's request still deviates from perfect rationality. Perfect rationality would imply a Dirac delta function with the probability mass situated at the optimal request. The observed deviation from rationality may be due to uncertainty in their opponent's abilities (reflected in the probabilities from \cref{figUltimatumStep}), or limitations of Player 1's reasoning.

\subsubsection{Two-stage Bargaining}

Next, we examine a two-stage bargaining game (shown in \cref{figTwoStage}). Now, if Player 2 rejects Player 1's request, they can come back with a counteroffer of their own. If the players can not come to an agreement, they both receive $0$. It is, therefore, in both players best interest to reach an agreement. This is represented with the following utilities:
\begin{equation}
\begin{split}
U_1[a_1] &= a_1 \times  f[\text{accept} \mid a_1] + V_1 \times  f[\text{reject} \mid a_1] \\
U_2[\text{accept}_{2} \mid a_1] &= 100 - a_1 \\
U_2[\text{reject}_{2} \mid a_1] &= V_2 \\
U_2[a_3] &= D(100-a_3) \times  f[\text{accept} \mid a_3]\\
U_1[\text{accept}_{4} \mid a_3] &= D \times a_3 \\
U_1[\text{reject}_{4} \mid a_3] &= 0 \\
\end{split}
\end{equation}
where now $V_1$ and $V_2$ are derived from the expected payoff of the rejection branch. Conditioning on past decisions are excluded from $U_2[a_3]$ as it is implicit that this can only occur when Player $2$ rejects Player $1$'s request. 
We use experimental data of Game 3 from \cite{binmore2002backward}, considering all disagreement penalties $D \in [0.2,0.3,\dots,0.9]$. For discussion sake, here we assume $D=0.9$. With perfect rationality, it is in Player 1's best interest to accept any counteroffer greater than the rejection payoff of $0$ (see \cref{figTwoStage}). Therefore, if prompted, Player 2 should provide a counteroffer of $y=1$, which gives Player 2 a payoff of $D(100-y)=89.1$, and Player 1 a payoff of $Dy=0.9$. Since $Dy > 0$ (the rejection payoff), Player 1 should prefer this to the alternative and accept. With this in mind, Player 1 now knows the payoff for the rejection branch for Player 2 is $89.1$, so if they request $x>10$ (assuming integer requests), Player 2 will reject this request since $100-x < 89.1$ if $x > 10$. Therefore, the rational Player 1 requests $x=10$, maximising their payoff, assuming Player 2 is rational.

\begin{figure}[ht]
    \centering
    \includegraphics[width=.6\textwidth]{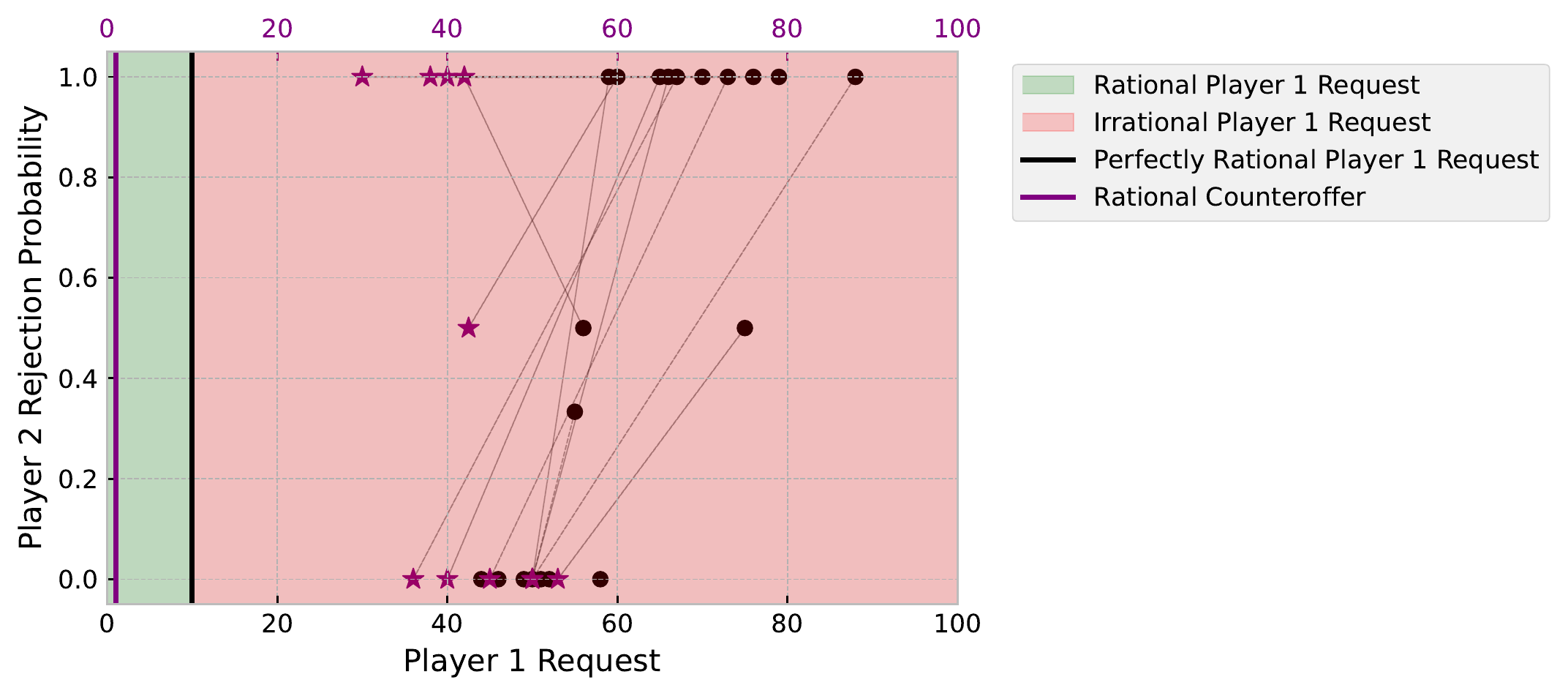}
    \caption{Two-stage Bargaining with disagreement penalty $D=0.9$. Initial requests are shown as black circles. The y-position gives the rejection rate of these requests. For rejected requests, the counteroffers are shown as a purple star (linked to their original request by a line), where again, the y-position shows the rejection rate of the counteroffer. The perfectly rational initial request would be $x=10$ (black line), as any requests in the red region would be rejected by a rational opponent. After a rejection, the perfectly rational counteroffer would be $y=1$ (purple line). Deviations from the subgame perfect equilibrium are clear, with no player performing perfect backward induction. }
    \label{figTwoStageRates}
\end{figure}

\begin{figure*}
    \centering
    \begin{subfigure}[t]{0.4\textwidth}
        \centering
        \includegraphics[width=\textwidth]{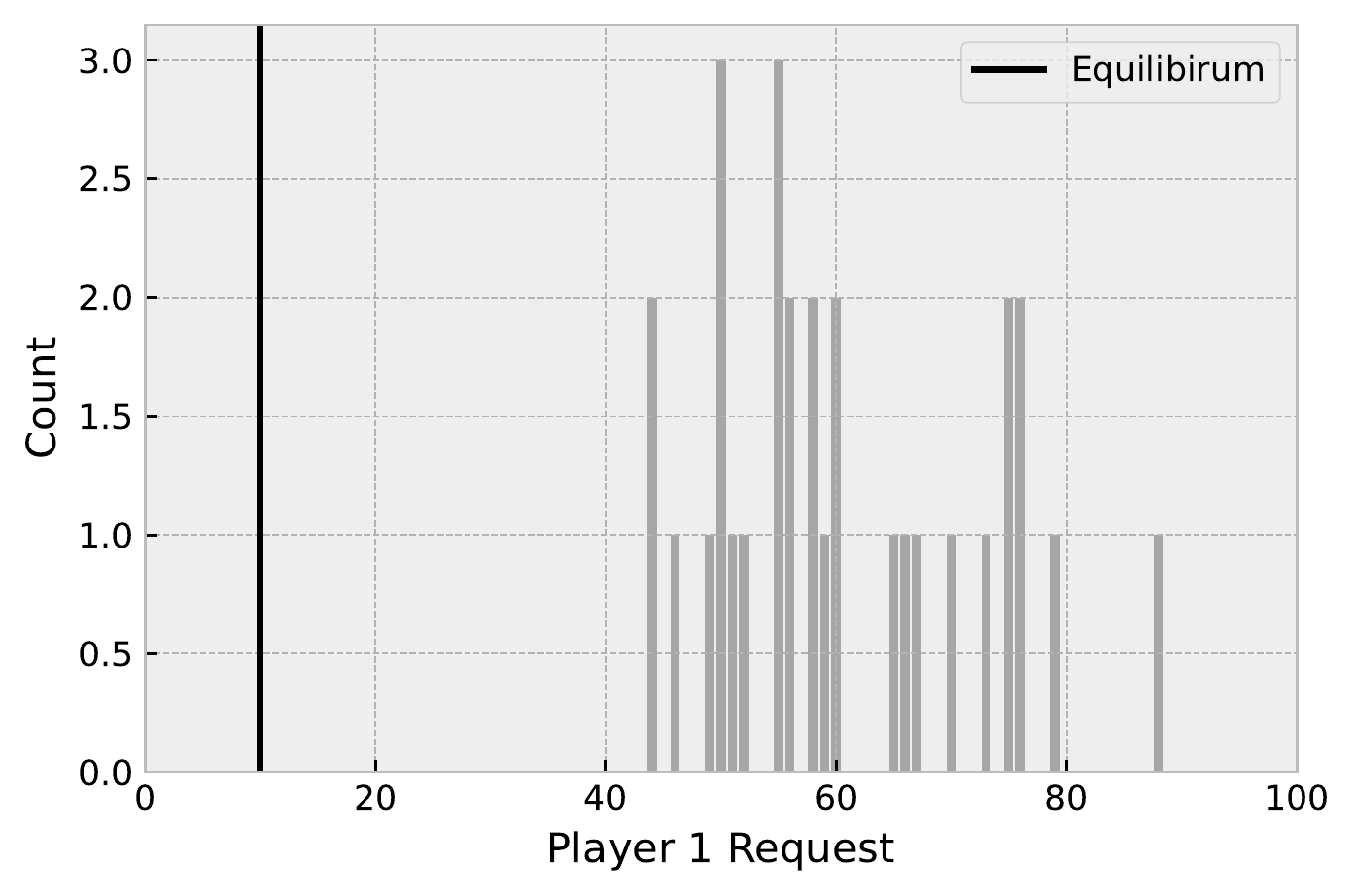}
        \caption{Player 1 Requests}\label{figTwoStageOfferDistributions}
    \end{subfigure}
    \begin{subfigure}[t]{0.4\textwidth}
        \centering
        \includegraphics[width=\textwidth]{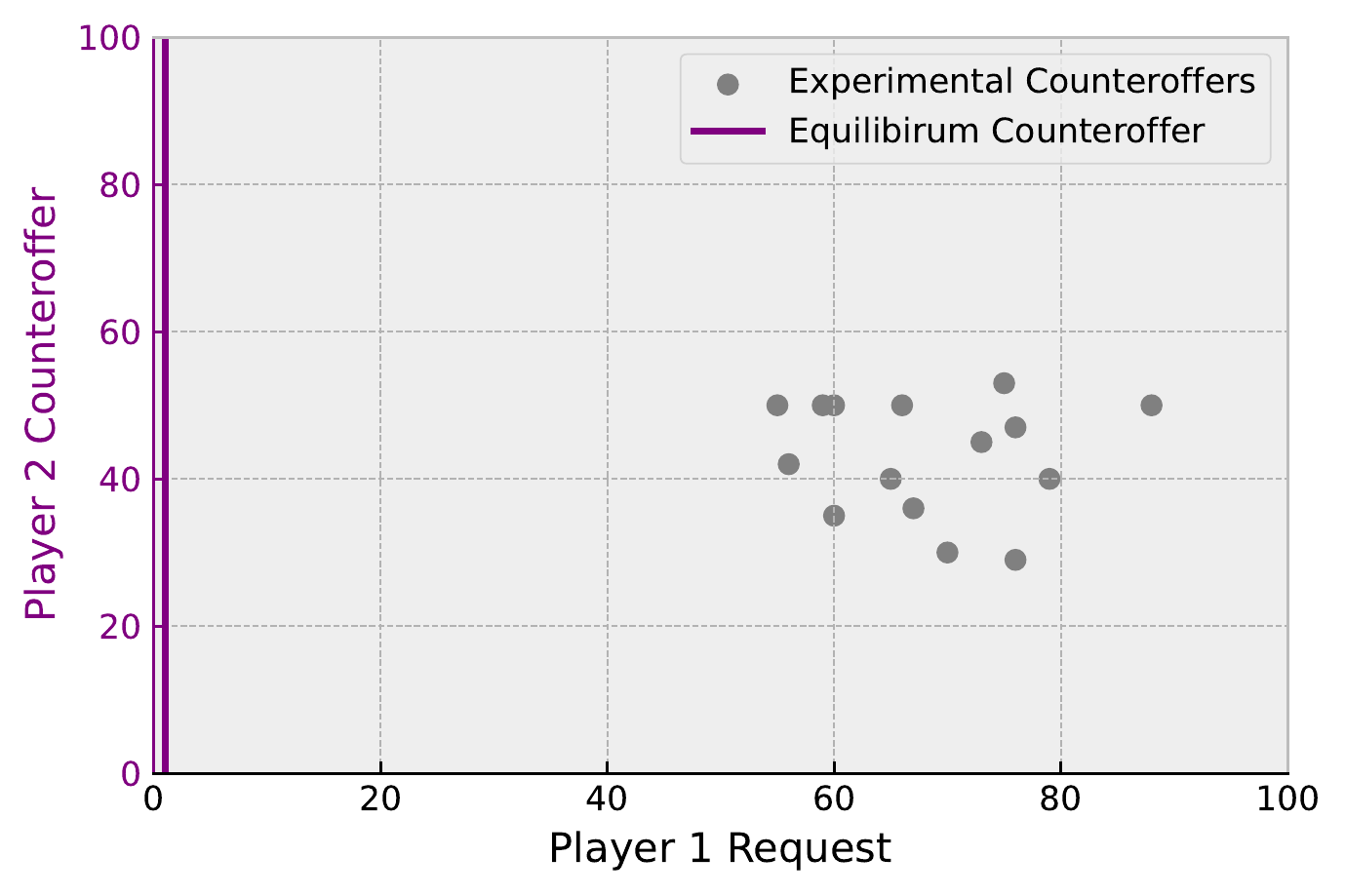}
        \caption{Player 2 Counteroffers}\label{figTwoStageCounterOffers}
    \end{subfigure}
    \caption{Initial requests and counteroffers in the two-stage bargaining game with $D=0.9$.}\label{figTwoStageOffers}
\end{figure*}

However, from \cref{figTwoStageRates} which summarises results from the experiments presented by \cite{binmore2002backward}, we can see substantial deviations from the subgame perfect equilibrium for both players. No Player $1$ requests $x<10$ or the perfectly rational request $x=10$. Furthermore, no Player 2 requests the rational counteroffer of $y=1$. The distribution of initial and counteroffers is visualised in \cref{figTwoStageOffers}.

\subsubsection{Comparison Methods}

\paragraph{Level-$k$} Under the level-$k$ model, level-$0$ players are assumed to be indifferent to all choices, and chooose uniformly. Level-$1$ players exploit this, and choose based on their opponent being a level-$0$ player, and so on.

\paragraph{Cognitive Hierarchy} Similar to the other game classes, again rather than assuming all players are at $k-1$, the cognitive hierarchy model fits a distribution to these $k$ players, and best responds according to this distribution of lower level thinkers. Again, the Poisson distribution is used.

\paragraph{Quantal Response Equilibrium} For the Quantal Response Equilibrium, again, an agent form of QRE is used to account for players noisily responding at each level, which is calculated recursively from the final step.

\section{Sensitivity}
To ensure the method's robustness, we check the sensitivity of the proposed results to various factors that may affect the outcome. Specifically, we carried this testing out with respect to the convergence/termination parameter $\epsilon$ and the fitted density estimates. 

\subsection{Termination parameter}
\begin{figure*}[ht]
\centering
    \begin{subfigure}[t]{0.35\textwidth}
        \centering
        \includegraphics[width=\textwidth]{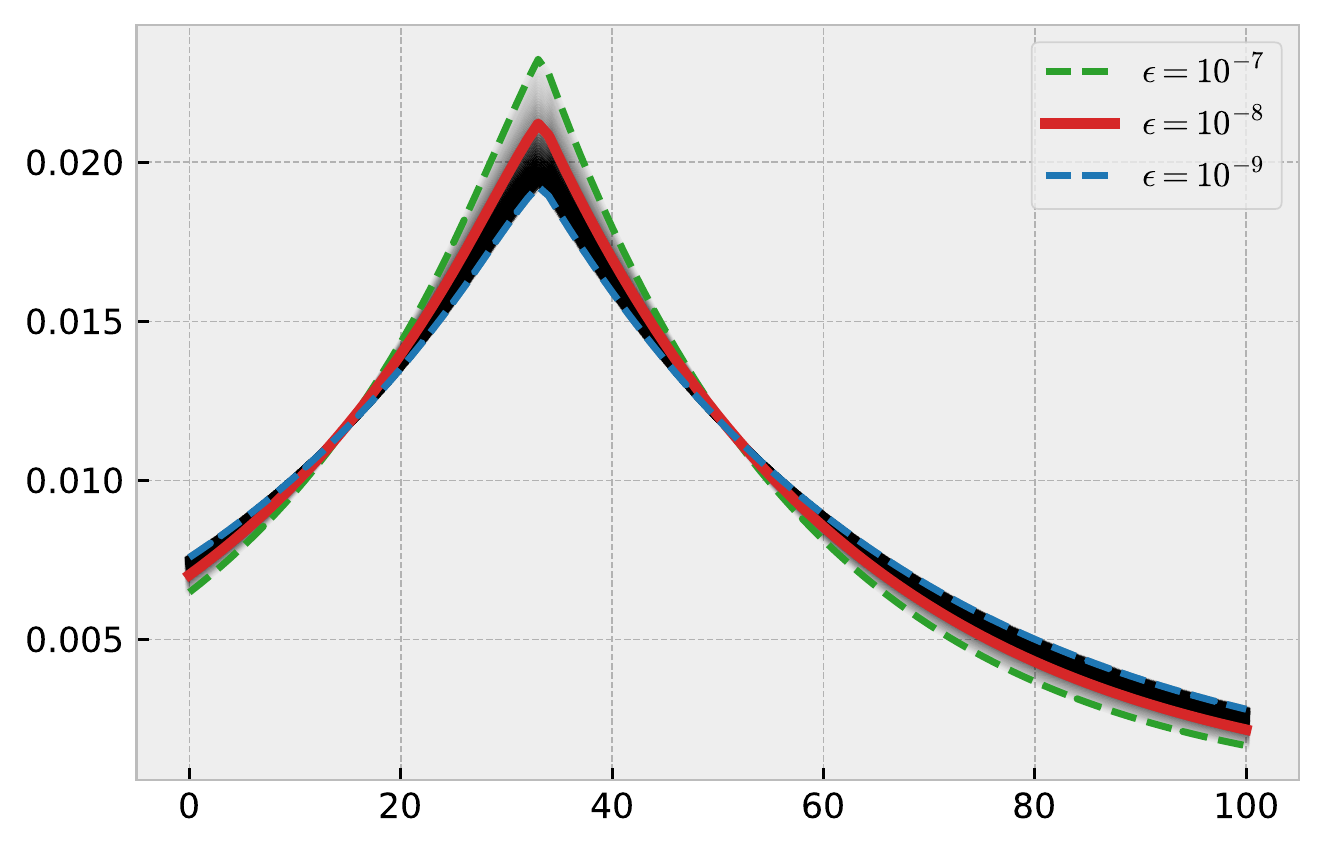}
        \caption{Beauty Contest}
    \end{subfigure}
    \begin{subfigure}[t]{0.35\textwidth}
        \centering
        \includegraphics[width=\textwidth]{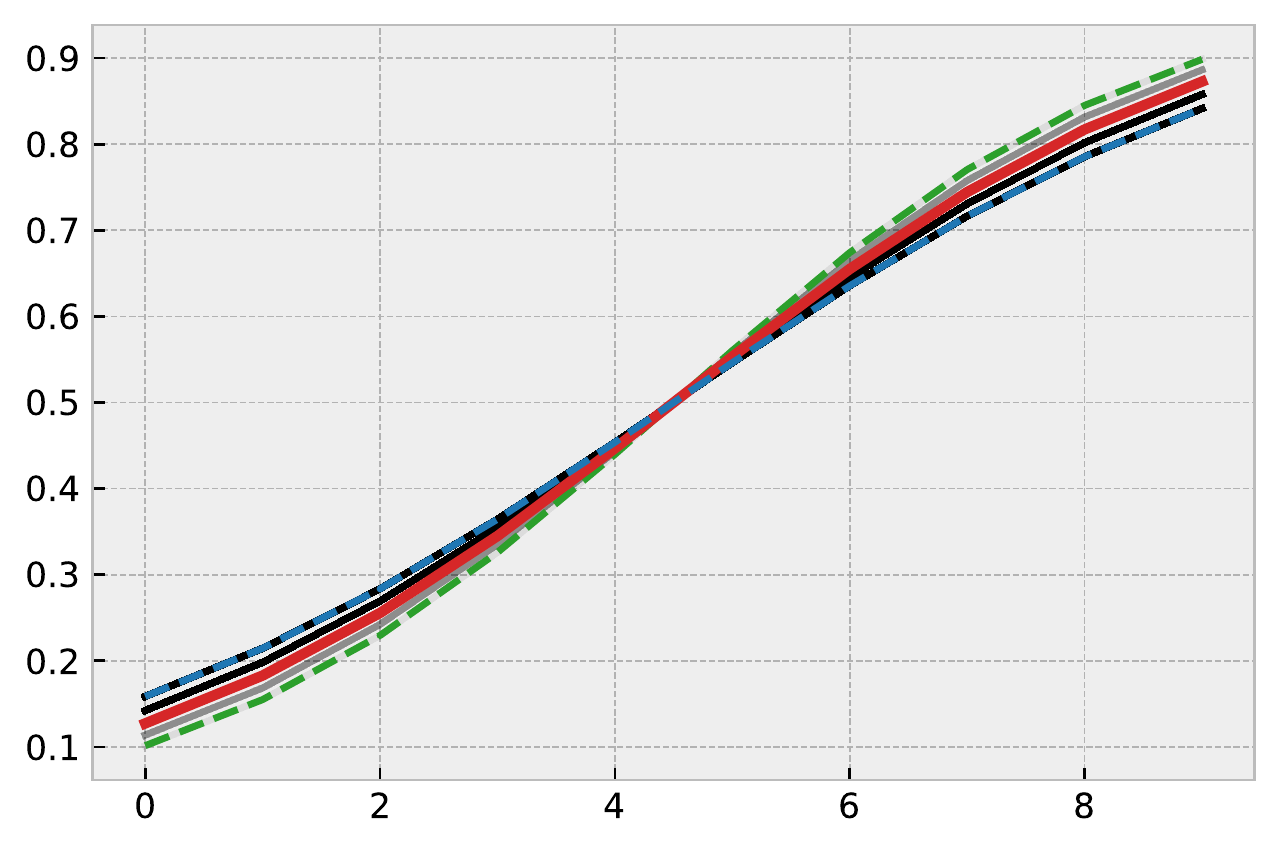}
        \caption{Market}
    \end{subfigure}
    \caption{Termination parameter $\epsilon$ sensitivity. The outcome for the default value is displayed as the red line. The outcome for the upper (lower) threshold is the dashed blue (green) bar. Intermediary values are displayed as light grey lines. }
    \label{figEpsilonSenstivity}
\end{figure*}

For the bargaining and centipede games, the reasoning naturally ends at the end of the extensive-form game. However, for the market entry and beauty contest games (with no defined end point), the reasoning continues until the resources are depleted, i.e., $\beta\gamma^k < \epsilon$. We have used the threshold $\epsilon=10^{-8}$ to determine termination. To check the robustness of the method to $\epsilon$, here we perform sensitivity analysis across the range $10^{-7} < \epsilon < 10^{-9}$, i.e. $\pm$ one order of magnitude from the default value. We sample 1000 points uniformly from this range, presenting the results in \cref{figEpsilonSenstivity}.

In both cases, we can see the approach is robust to these large changes in $\epsilon$, with an order of magnitude change only having slight effects on the resulting outcomes. These results show that $\epsilon$ does not need to be treated as a hyperparameter to optimise (as $\beta$/$\gamma$), but rather as a fixed parameter to determine ``convergence" towards 0 and termination, the choice of which depends on computational/numeric requirements. {We recommend using as small value as possible (e.g. $\epsilon \leq 10^{-8}$) while still achieving reasonable convergence speed.}

\subsection{Density Estimates}\label{appendixDensityEstimate}

We evaluate how the resulting rankings would change with different density estimation methods. Specifically, we analyse the resulting average (out-of-sample) ranks when using Scott's rule \citep{scott2015multivariate} (as presented),  Silverman's rule \citep{silverman2018density}, and the (improved) Sheather \& Jones \citep{botev2010kernel} for automatic bandwidth identification. While this does not provide an exhaustive list, it covers the most common rules used in literature. Density estimates are only used for the bargaining and beauty contest games, so these are the two game classes analysed here.

In \cref{tblRobustRankings}, we see no change in average ranking between Scott's and Silverman's rules for the beauty contest games. However, when using Sheather \& Jones, the rankings between Level-$k$ and Nash change, going from 4 and 5, respectively, to 4.33 and 4.67. This rank change does not alter any of the claims made within the paper, so we can confirm the robustness of the resulting rankings to density estimates for the beauty game.

For the bargaining games, we see slight improvement for the proposed method when comparing Scott's rule with Silverman's and Sheather \& Jones—in both cases, going from 1.27 with Scott's rule to 1.23. At the same time, QREs rank worsens from 1.73 to 1.77. The remaining methods keep the same ranking. These rank changes strengthen the claims made in the paper, showing not only robustness to the rule used but also improvements for the proposed approach when utilising alternative rules for bandwidth estimation.

\begin{table}[!htb]
\caption{Robustness to changes in density estimation. Average rankings are computed using the results from various automatic bandwidth determination methods. Scott's is the method we present in the paper.}\label{tblRobustRankings}
\resizebox{\textwidth}{!}{
\begin{tabular}{@{}lllllll@{}}
\toprule
                                &                   & Quantal Hierarchy & Level-k & Cognitive Hierarchy & QRE & Nash \\ \midrule
\multirow{3}{*}{Beauty Contests} & \textbf{Scott's}   & 1.83 & 4.00 & 3.00 & 1.17 & 5.00 \\ 
                                & Silverman's        & 1.83 & 4.00 & 3.00 & 1.17 & 5.00 \\ 
                                & Sheather \& Jones & 1.83 & 4.33 & 3.00 & 1.17 & 4.67 \\
\midrule
\multirow{3}{*}{Bargaining Games} & \textbf{Scott's} & 1.27 & 3.50 & 3.50 & 1.73 & 5.00 \\
                                & Silverman's        & 1.23                  &   3.50      & 3.50 & 1.77    & 5.00      \\
& Sheather \& Jones & 1.23 & 3.50  & 3.50 &    1.77 & 5.00     \\
\bottomrule
\end{tabular}
}
\end{table}

\end{document}